%% file: main.tex
\documentclass{tlp}
\usepackage{verbatim}
\usepackage{amsmath}
\usepackage{arydshln}
\usepackage{caption}
\usepackage{subcaption}
\let\proof\relax

\usepackage{mathptmx}
\usepackage{url}
\usepackage{amsfonts}
\usepackage{xcolor}
\usepackage{graphicx}
\usepackage{subcaption}
\usepackage{graphics}
\usepackage{threeparttable}
\usepackage{epsfig}
\usepackage{multirow}
\usepackage{xspace}
\usepackage{todonotes}
\usepackage{color}
\usepackage{booktabs}
\usepackage{pgfgantt}
\usepackage{tikz}
\usetikzlibrary{positioning}
\usepackage{pgfplots}
\usepackage{tabularx}
\usepackage{comment}
\usepackage{asplisting}
\usepackage{array}
\usepackage{wrapfig}
\definecolor{darkgreen}{RGB}{0,60,0}
\definecolor{darkgray}{RGB}{80,80,80}
\definecolor{newblue}{rgb}{0.70,0.70,1.00}
\usepackage{amsmath}
\usepackage{graphicx}
\usepackage{multirow}
\newcommand{\Aff}{\operatorname{Aff}}

\lstnewenvironment{asp}
{
	\lstset{
		language=asp,
		showstringspaces=false,
		formfeed=\newpage,
		tabsize=4,
		commentstyle=\color{darkgreen},
		basicstyle=\ttfamily\small,
		keywordstyle=\color{black},
		numbers=left,
		breaklines=true,
        firstnumber=last,
		literate={~} {$\sim$}{1},
	}
}
{}
\newtheorem{definition}{Definition}

\newcommand{\myparagraph}[1]{\vspace{-0.0cm}\paragraph{\textbf{#1}}}

\begin{document}

\lefttitle{Cambridge Author}

\jnlPage{1}{8}
\jnlDoiYr{2021}
\doival{10.1017/xxxxx}

\title[Theory and Practice of Logic Programming]{Long-term Power Grid Planning via Answer Set Programming
}

\begin{authgrp}
\author{\sn{Antonio} \gn{Ielo}, \sn{Francesco} \gn{Doria}}
\affiliation{University of Calabria, Italy}
\author{\sn{Sandra} \gn{Castellanos-Paez}}
\affiliation{University of Huddersfield, UK}
\author{\sn{Marco} \gn{Maratea}}
\affiliation{University of Calabria, Italy}
\author{\sn{Francesco} \gn{Percassi}, \sn{Mauro} \gn{Vallati}}
\affiliation{University of Huddersfield, UK}
\end{authgrp}

\history{\sub{xx xx xxxx;} \rev{xx xx xxxx;} \acc{xx xx xxxx}}

\maketitle

\begin{abstract}
The Power grid is a critical infrastructure underpinning all aspects of modern society and its services. Maintaining its effectiveness requires continuous adaptations. In particular, addressing sustainability targets, demand patterns, and urbanisation trends requires implementing changes to the network. Actual developments can potentially span over a decade, with supply continuity and service quality that must be preserved throughout by ensuring conformance to several topological and combinatorial invariants.
Long-term power grid planning deals with the above process, and
although planning languages could be a natural choice, the kind of properties and invariants needed are cumbersome to express in such languages; on the contrary, they can be elegantly and succinctly encoded in Answer Set Programming (ASP). 
In this paper, we propose the first approach to automate and optimise the long-term power grid planning process using ASP. Experimental evaluations conducted on synthetic and real‑world grid data confirm the expressive power of the proposed ASP‑based approach and demonstrate its effectiveness.
\end{abstract}


\section{Introduction}
Power grids are the backbone of modern societies, and they are currently experiencing significant structural transformations. The increasing penetration of renewable energy sources (RES) at the distribution level, combined with the rapid uptake of electric vehicles, heat pumps, and the wider electrification of end‑uses, is changing operational conditions and load patterns. As distribution networks evolve from passive infrastructures to active systems, their management becomes more complex and uncertain \citep{fangmisra}.
In this context, long‑term planning plays a central role. It refers to the process through which national grid operators define how the current distribution network should evolve over extended time horizons, to meet future demand, integrate new technologies, and comply with national and regional energy transition scenarios \citep{conejo2010decision}. More specifically, the long-term planning problem focuses on identifying the feasible multi‑stage evolution path between the initial and the target desired power grid configuration. The two networks may differ significantly in topology, and many sequences of line additions, removals, or switching actions may connect them. However, considering the time span and the role played by the power grids, supply continuity and service quality must be preserved throughout the duration of the evolution \citep{khator1997power}. 
 {Long-term planning decisions are characterized by their scale and long-lasting impact. For instance, the French distribution network spans approximately 1.4 million kilometres, and the cost of a single power line ranges from 50k€ to 200k€ per kilometre, depending on technology and location. Planning horizons typically extend up to 30 years, while infrastructure lifetimes may reach 60 years. As a result, even small inefficiencies in planning decisions can lead to significant economic consequences.}
At present, long‑term power grid planning remains largely manual. National Distribution System Operators rely on human experts using power‑flow and cost‑analysis tools, but the construction of multi‑stage evolution strategies is still mainly based on expert knowledge and iterative adjustments. This limits reproducibility, scalability, and the systematic exploration of alternative investment trajectories. 
{In particular, the identification of feasible evolution strategies remains a major bottleneck, as it requires exploring a large combinatorial space of possible network transformations while ensuring that all intermediate configurations remain operational.} 
Recent work has called for automated, intelligent decision‑support tools that can complement human expertise and address the growing complexity of distribution networks~\citep{castellanos2022automated,castellanos2023decision}.
Artificial Intelligence methods have a long tradition in power systems, ranging from expert systems for operator support~\citep{wollenberg1987aiops}, to planning‑based approaches for sequential operational tasks~\citep{bell2009role,bertoli2002psr,thiebaux2013mip}, to evolutionary and learning‑based techniques~\citep{arabali2012genetic,glavic2017reinforcement,vazquez2019reinforcement}.  
{
Mixed Integer Linear Programming has been used for network reconfiguration problems focused on power and load flow constraints~\citep{milp-flow-1,milp-flow-2}, typically in static or short-horizon settings where the goal is to compute a feasible configuration rather than model transitions between configurations.
Moreover, they are less adequate to express purely combinatorial properties, such as radiality \citep{milp-radiality}, while being more suitable for flow-based constraints.
}
Planning languages could be a natural choice, but graph properties and invariants (e.g., connectedness) are cumbersome to express and are better handled in knowledge representation and reasoning languages.
Within this landscape, Answer Set Programming offers a suitable approach, providing a declarative and expressive framework for modelling graph‑theoretic constraints, discrete actions, and multi‑stage feasibility requirements. {Answer Set Programming (ASP;~\citep{DBLP:journals/cacm/BrewkaET11}) is a declarative programming language based on stable model semantics~\citep{DBLP:conf/iclp/GelfondL88} of logic programs. Its expressive power and availability of efficient reasoners makes it a popular framework to model knowledge-intensive combinatorial problems. ASP has been successfully applied to academic and industrial problems alike~\citep{DBLP:journals/aim/ErdemGL16,DBLP:journals/ki/FalknerFSTT18}.}

This paper presents the first approach for automating long-term power grid planning that can handle all grid topological constraints. We propose an approach inspired by recent applications of ASP to \emph{reconfiguration}  problems~\citep{DBLP:conf/jelia/YamadaBIS23,DBLP:journals/tplp/KatoBSST24}, where one has to find sequences of \emph{edits} to navigate among solutions of (hard) combinatorial problems. We define an ASP encoding for the long-term power grid planning problem, that allows for both sequential and parallel planning. Results of experiments, using the {\sc clingo} ASP solver, executed on synthetic and real-world instances from the French grid, show that our solution is able to efficiently solve   instances of a size comparable to those that experts deal with in daily operations, when allowing a parallel execution of actions, and to solve larger real-world instances within 30 minutes.

\section{Background}
{This section introduces the required preliminary notions about distribution networks and the long-term power grid planning problem, that will be useful to understand our solution.}

\subsection{Distribution Network Architecture}

\begin{figure}[t]
    \centering
    \includegraphics[scale=0.30]{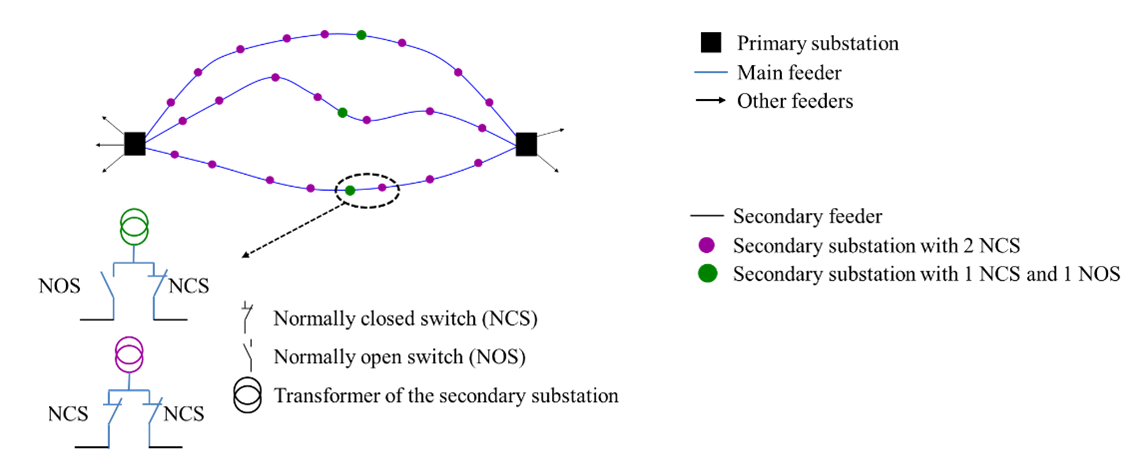}
    \caption{{Medium-voltage distribution network supplied by two primary substations (black squares) feeding interconnected feeders (blue lines) and secondary substations; purple circles represent secondary substations equipped with two normally closed switches (NCS), while green circles denote secondary substations equipped with one NCS and one normally open switch (NOS) used for radial operation and reconfiguration. %
    In this work we treat both types of secondary substations uniformly, and represent operational differences through the status of the network edges (open, close).
    } (Figure adapted from ~\cite{castellanos2023decision}.)}
    \label{fig:secure_feeder}
\end{figure}

Urban electrical distribution networks are commonly designed according to a secured feeder architecture (see Figure \ref{fig:secure_feeder}). Primary substations act as the supply points of the medium-voltage distribution network, injecting power into the system through main feeders that interconnect the substations and supply multiple secondary substations. Along these feeders, secondary substations serve local areas and include distribution transformers that step down the medium-voltage level to low voltage for end users, as well as switching devices that control network operation. {Each secondary substation is connected to a feeder through two switches, providing operational flexibility: normally closed switches (NCS) allow power flow under normal operating conditions, while normally open switches (NOS) are kept open to ensure radial operation, so that each load is supplied by a single primary substation. In the event of a fault, the network can be reconfigured by operating the switches, enabling secondary substations to be resupplied from an alternative primary substation while repair actions are carried out.}

\subsection{Power Grid Planning}\label{power-systems-planning}
Distribution System Operators (DSOs) are responsible for ensuring the safe, reliable, and cost‑effective operation of power grids. This mandate includes maintaining existing infrastructures, guaranteeing continuity and quality of supply, and anticipating future developments in demand and distributed generation \citep{khator1997power,conejo2010decision}. In this context, long‑term power grid planning is a core activity.

Long‑term planning concerns the definition of how a distribution network should evolve over extended horizons, typically spanning multiple years. DSOs construct these plans considering multiple inputs, including consumption forecasts, expected production profiles—including the integration of renewable energy sources—and broader energy transition scenarios. The goal is to identify a future network architecture that remains technically feasible, and operationally secure. %

Infrastructure modifications in distribution grids are expensive and difficult to reverse. The cost of a single power line may range from 50 k€ to 200 k€ per kilometre, depending on technology (overhead vs. underground), cable type, and local conditions (rural, semi‑urban, or urban). At the scale of a national distribution system—such as the French network, which extends over approximately 1.4 million kilometres—even minor changes can result in significant investments. Consequently, DSOs aim to minimise deviations from the existing infrastructure, favouring incremental and targeted interventions \citep{conejo2010decision}.

A distinctive feature of long‑term planning is that it must consider not only the final target network, but also the sequence of intermediate networks that will emerge as investment actions are gradually executed over many years. Each intermediate configuration must remain compliant with operational and topological constraints: such intermediate configurations correspond to networks that will be in service for extended periods. Ensuring their feasibility is therefore critical to maintaining security of supply and service quality over time.

According to the ERDF methodology~\citep{reseauERDF} for the development of distribution network master plans in France, the long-term planning process can be represented as a workflow composed of six main stages 
: system diagnosis, load forecasting, definition of a long-term target network, identification of development strategies, techno-economic scheduling of investments, and assessment of the resulting quality of supply. Among these stages, the identification of development strategies is a key step, as it determines how the transition from the current network to the target configuration is achieved. More informally, given the current network configuration and the target configuration, this step generates the list of interventions to be performed, and the corresponding resulting intermediate networks. The present work focuses on this stage, which remains largely manual and based on expert knowledge in current practice.

To ensure the operationality of intermediate networks during the development strategies step, and for minimising the potential for wasting resources in the long-term power grid planning process, a number of implicit rules are usually followed by grid operators. \cite{castellanos2023decision} extracted the following set of rules via in-depth interviews with experts from French DSOs that are in charge of the development strategies: %

\begin{itemize}

\item[R1]
There is no closed loop between two primary substations; that is, a normally open switch is required to ensure radial operation, meaning that each secondary substation in the network is supplied by exactly one primary substation.

\item[R2] The network must tolerate the loss of any primary node, ensuring that a radial configuration, in which all secondary nodes remain supplied, can always be restored via switching operations (opening and closing lines).

\item[R3] For urban areas, the number of connections of secondary substations are between 2 and 3.

\item[R4]
Only lines belonging to the target network can be created, and only lines not belonging to the target network can be removed.
These operations are irreversible, e.g., a line cannot be removed and later re-added.

\item[R5]
There is only one connection between two nodes (distribution/source).

\end{itemize}

\section{Problem Formalisation}\label{sec:problem}
For our purposes, the power grid is an undirected graph with two types of nodes (primary, secondary) and two types of edges (open, close). 
One acts upon (``reconfigures'') the power grid by changing its graph structure. 
The \emph{long-term planning problem} (LPP) refers to computing a sequence of physical interventions on the power grid infrastructure that results in a specific \emph{target} configuration, 
in such a way that each intermediate state complies with the DSOs rules as described in Section~\ref{power-systems-planning}.
In the remainder of this paper, we use the term LPP to refer to the discrete combinatorial component of the long-term planning process.
In this section, we abstract and formalise the problem and the DSOs requirements into graph-theoretic terms.

\myparagraph{Notation.} In the rest of the section, we adopt the following notation. Let $P, S$ be two finite disjoint sets of \emph{primary} and \emph{secondary} stations, with $V = P \cup S$. 
A \emph{power grid} is an edge-labeled undirected graph $G(V, E, \lambda)$, with $E \subseteq V\times V$ and $\lambda: E \mapsto \{open, close\}$. The labeling function $\lambda$ marks closed and open lines in the power grid. 
We will refer to the graph $G(V, \{e \in E: \lambda(e) = close\})$ as the \emph{closed network} and $G(V, E)$ as the \emph{open network}.
The representation in Figure \ref{fig:secure_feeder} is consistent with this notation: secondary substations are not distinguished at the node level, and their operational differences are encoded through function $\lambda$.

\subsection{Acting upon the power grid}\label{sec:actions}
There are three possible kinds of actions available to reconfigure the power grid: edge addition, edge removal and edge switching. Each action changes the power grid graph. Whenever applying action $a$ over the power grid $G$ results in $G'$, we denote this by $G \vdash_{a} G'$. We describe effects of each possible action on $G$.

\myparagraph{Edge addition.} Edge addition consists of adding a new edge to the open network. Let $x, y \in V$ with $(x,y) \not\in E$. We denote the action of adding $(x,y)$ to $G$ by $add(x,y)$. It results in the power grid $G(V, E', \lambda')$, where $E' = E \cup \{(x,y)\}$ and $\lambda'(e) = \lambda(e)$ for all $e \in E$, and $\lambda'(x,y) = open$. 

\myparagraph{Edge removal.} Edge removal consists of removing an edge from the open network. Let $(x,y) \in E$ such that $\lambda(x,y) = open$. We denote the action of removing $(x,y)$ as  $remove(x,y)$. It results in the power grid $G(V, E', \lambda')$, where $E' = E \setminus \{(x,y)\}$ and $\lambda'(e) = \lambda(e)$ for all $e \in E'$. 

\myparagraph{Edge switching.} Edge switching consists of swapping the state of two edges that are adjacent to the same secondary station. Let $x \in S$, $y, z \in V$ such that $\lambda(x,y) = close$ and $\lambda (x,z) = open$. We denote the switching action by $switch(x,y,z)$. 
It results in the power grid $G(V, E, \lambda')$, with $\lambda(e) = \lambda'(e)$ for all $e \in E \setminus \{(x,y), (x,z)\}$ and $\lambda'(x,y) = open$, $\lambda'(x,z) = close$.

\myparagraph{Non-interfering actions.} Executing an action causes a change in the power grid graph. Intuitively, concurrent execution of actions that involve the same edges leads to ambiguities during plans' execution. {We formalise the notion of ``allowed concurrency'' in our domain, inspired by the notion of \emph{non-interference} in planning~(see, e.g.,~\citep{DBLP:journals/ai/RintanenHN06}}). We denote by $\Aff(a)$ the set of edges that is involved in the execution of action $a$, and in particular we have that $\Aff(p(x,y)) = \{(x,y)\}$ if $p \in \{add, remove\}$ and $\Aff(switch(x,y,z)) = \{(x,y), (x,z)\}$.

We say two actions $a_1$ and $a_2$ are \emph{non-interfering} (or \emph{compatible}) if $\Aff(a_1)$ and $\Aff(a_2)$ are disjoint sets. Intuitively, non-interfering actions are ``compatible'', in the sense that they can be carried out on the graph ``at the same time'' without ambiguities. A set of actions $A$ is non-interfering if all its elements are pairwise non-interfering. Thus, we extend the notion of affected edges to sets of actions, with $\Aff(A) = \bigcup_{a \in A} \Aff(a)$. We say that $G \vdash_A G'$ if we obtain $G'$ by applying all the actions $a \in A$; if $A$ is non-interfering, there is no ambiguity in the resulting graph $G'$ (as no edge is affected by multiple actions).

\subsection{Compliant Networks}

The DSOs' rules described in Section~\ref{power-systems-planning} provide properties that power grids must comply with to be considered valid. These can be succinctly expressed in terms of graph-theoretic properties. We say a power grid $G(P \cup S, E, \lambda)$ is compliant whenever the following properties are satisfied.

\begin{definition}[Radial]\label{def:radial}
A power grid $G(P\cup S, E, \lambda)$ is \emph{radial} if its closed network is acyclic and, in this network, each secondary node is connected to exactly one primary node.
That is, $G$ is a forest with roots in $P$.
\end{definition}

\begin{definition}[Redundantly Connected]\label{def:redundantlyconnected}
Let $G(P\cup S, E, \lambda)$ be a power grid.
Let $s \in S$ be a secondary station connected to $p \in P$ in the closed network.
We say that $s$ is \emph{redundantly connected} if, in the open network, there exists a path from $s$ to some primary $p' \neq p$ whose internal nodes do not include primary nodes.
\end{definition}

\begin{definition}[Reconfigurable]\label{def:reconfigurable}
Let $G(P\cup S, E, \lambda)$ be a power grid. $G$ is \emph{reconfigurable} if all its secondary stations $s \in S$ are redundantly connected.
\end{definition}

\begin{definition}[Degree-compliant]\label{def:degreecompliance}
A power grid $G(P\cup S, E, \lambda)$ is degree-compliant if it holds that $2 \le deg(s) \le 3$ for all $s \in S$.
\end{definition}

\subsection{Long-term Planning Problem}

Finally, given an initial power grid $G(P \cup S, E, \lambda)$ and a target one $G_T(P \cup S, E_T, \lambda_T)$, we define the \emph{long-term planning problem} as computing a sequence of sets of actions $(A_0, \dots, A_n)$ such that $G_i \vdash_{A_i} G_{i+1}$, with $G_0 = G$, $G_n = G_T$, and each intermediate network $G_i$ satisfying the radiality, reconfigurability and degree-compliant properties.
We denote by $B = E_T \setminus E$ the set of \emph{buildable} edges, i.e., edges present in the target network but absent from the initial one, and by $R = E \setminus E_T$ the set of \emph{removable} edges, i.e., edges present in the initial network but not in the target.

\begin{definition}[Long-term Planning Problem (LPP)]\label{def:lpp}
Let $G(P \cup S, E, \lambda)$ and $G_T(P \cup S, E_T, \lambda_T)$ be two power grids.
$LPP(G, G_T)$ is the problem of computing a sequence of compatible sets of actions $(A_0, \dots, A_n)$ such that:
\begin{itemize}[leftmargin=2em]

\item[C1] $G_0 = G$, $G_n = G_T$;

\item[C2] $\forall\, i \in \{0, \dots, n-1\}$, $G_i \vdash_{A_i} G_{i+1}$, with $G_i$ being radial, reconfigurable, and degree-compliant;

\item[C3] If $add(x,y) \in A_i$ then $(x,y) \in B$;

\item[C4] If $remove(x,y) \in A_i$ then $(x,y) \in R$;

\item[C5] No $(x,y)$ such that $add(x,y) \in A_i$ and $remove(x,y) \in A_j$ for some $0 \le i, j \le n$.
\end{itemize}
\end{definition}

\myparagraph{Relationship with DSOs rules.} 
C1 requires that the network evolves from $G$ to $G_T$ by the end of the horizon.
C2 requires every intermediate network $G_i$ to satisfy radiality (R1), fault-tolerance under the loss of any primary node (R2), and the degree compliance (R3).
R4 is captured by C3--C5, which constrain admissible line additions and removals and enforce their irreversibility.\\

In the remainder, we refer to the \emph{bounded} version of LPP whenever we assume an integer upper bound $k \in \mathbb{N}$ on the length of the solutions. We refer to the \emph{sequential} version of the problem whenever we add the requirement that $|A_i| = 1$, and \emph{parallel} otherwise (that is, $|A_i| \ge 1$).

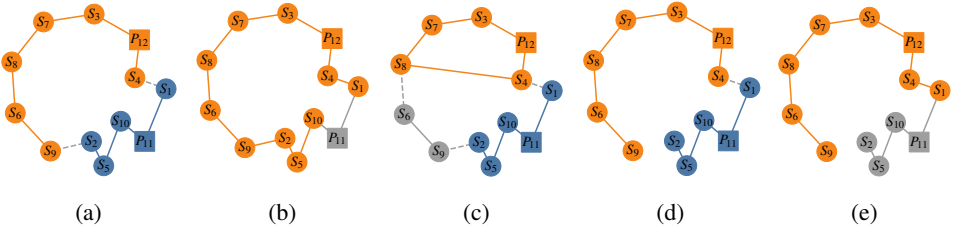
\begin{figure}[t]
\centering

\begin{subfigure}[t]{0.19\linewidth}
    \centering
    \resizebox{\linewidth}{!}{\input{pgf_example/01_single_plot}}
    \caption{}
    \label{fig:ex-compliant}
\end{subfigure}%
\begin{subfigure}[t]{0.19\linewidth}
    \centering
    \resizebox{\linewidth}{!}{\input{pgf_example/02_single_plot}}
    \caption{}
    \label{fig:ex-reconf}
\end{subfigure}%
\begin{subfigure}[t]{0.19\linewidth}
    \centering
    \resizebox{\linewidth}{!}{\input{pgf_example/03_single_plot}}
    \caption{}
    \label{fig:ex-norad}
\end{subfigure}%
\begin{subfigure}[t]{0.19\linewidth}
    \centering
    \resizebox{\linewidth}{!}{\input{pgf_example/04_single_plot}}
    \caption{}
    \label{fig:ex-noconn}
\end{subfigure}%
\begin{subfigure}[t]{0.19\linewidth}
    \centering
    \resizebox{\linewidth}{!}{\input{pgf_example/05_single_plot}}
    \caption{}
    \label{fig:ex-noconn-reconf}
\end{subfigure}

\caption{
{Example graph configurations illustrating the role of constraints.}
}
\label{fig:example}
\end{figure}

\myparagraph{Example.}
{Figure \ref{fig:example} illustrates the constraints considered in this work. 
Figure \ref{fig:ex-compliant} shows a compliant configuration, where radiality, redundant connectivity, and degree constraints are all satisfied.
Figure \ref{fig:ex-reconf} shows the reconfigurability of the graph from Figure \ref{fig:ex-compliant} under the failure of $P_{11}$; in particular, radiality can be restored by closing edges $(S_2,S_9)$ and $(S_1,S_4)$.
Figure \ref{fig:ex-norad} shows a configuration that violates radiality for two distinct reasons: nodes $S_6$ and $S_9$ are isolated, and the feeder associated with $P_{12}$ contains a cycle.
Figure \ref{fig:ex-noconn} shows a graph that is radial but not redundantly connected.
Figure \ref{fig:ex-noconn-reconf} shows how the violation of connectivity in the graph provided in \ref{fig:ex-noconn} implies non-reconfigurability: in particular, in case of failure of $P_{11}$, node $S_1$ can be resupplied by closing $(S_1,S_4)$, whereas $S_2$, $S_5$, and $S_{10}$ remain isolated.}

\section{ASP Encoding}

{In this section, we present our Answer Set Programming (ASP) solution designed to model the LPP problem.  We assume the reader to be familiar with ASP and the input language of the \textsc{clingo} system, and refer them to standard introductory material~\citep{DBLP:books/sp/Lifschitz19,DBLP:series/synthesis/2012Gebser}. 
We start with a high-level description of our approach.}

The idea of the encoding is to model how the \emph{starting network} $G$ evolves subject to a sequence of sets of actions $(A_0, \dots, A_k)$, with the goal of transforming it into a desired \emph{target network} $G_T$ within a given number of steps $k$. The approach is inspired by classical ASP applications in planning~\citep{DBLP:journals/tplp/SonPBS23} and bounded reconfiguration~\citep{DBLP:conf/jelia/YamadaBIS23,DBLP:journals/tplp/KatoBSST24} problems. We first present the data model, and then the proposed ASP encoding. %

\subsection{Data Model}\label{sec:data}

Let $G(P \cup S, E, \lambda)$ and $G_T(P \cup S, E_T, \lambda_T)$ be a pair of power grids. 
We describe the predicates and facts we use to represent inputs and outputs of the $LPP(G,G_T)$ problem.

\myparagraph{Input.} To describe the available stations in the power grid, we use atoms of the form $node(x)$ for $x \in P \cup S$. The atom $node\_attr(x, primary)$ models that node $x$ is a primary node, $x \in P$. Edges that belong to the start configuration are described by the predicate $start/3$. In particular, if $\lambda(x,y) = o$, we have the atom $start(x,y,o)$, with $o \in \{open, close\}$. Similarly, edges in the target configuration are described by the predicate $target/3$, with the atom $target(x,y,o)$ whenever $\lambda_T(x,y) = o$. We state that an edge $(x,y)$ may be built or removed by using the atoms $buildable(x,y)$ and $must\_remove(x,y)$. {As input, we expect an \emph{undirected} representation of the graph, and in particular we assume the fact $start(x,y,o)$ with $x < y$. }

\myparagraph{Output.} A solution $(A_0, \dots, A_k)$ to the $LPP(G, G_T)$ problem is described by means of the $action/2$ predicate, where the atom $action(t,a)$ models that $a \in A_t$. In particular, $a$ will be function terms of the form $add/2$, $remove/2$ and $switch/3$, mirroring how possible actions are defined in Section~\ref{sec:actions}. As an example, $action(3,switch(x,y,z))$ means that the solution includes the action $switch(x,y,z)$ at time $3$, that is $switch(x,y,z) \in A_3$.

\subsection{Encoding of the problem}\label{sec:encoding}

We present the ASP encoding used to model the LPP problem. We describe it incrementally, with each paragraph assuming all the rules that precede it. {We refer to the $i$-th line in the encoding by $r_i$.} We present the encoding to compute a bounded solution of length at most $k$.

\myparagraph{Initial definitions.} We start with definitions that make the encoding more readable. Let $G(P \cup S, E, \lambda)$ be a power grid. We denote which nodes of our network are respectively primary stations or secondary stations by the $primary/1$ and $secondary/1$ predicates, respectively. We will have the atom $primary(x)$ if $x \in P$, and $secondary(x)$ if $x \in S$. The $opposite/2$ predicate expresses the notion that open line and closed lines are opposite concepts. The $network/4$ predicate stores the ``state'', with $network(i,x,y,o)$ stating that $\lambda_i(x,y) = o$ in the $i$-th power grid. We assume $t=-1$ corresponds to the initial state of the network\footnote{The encoding assumes actions at time $t$ get applied at time $t$, and depend on time $t-1$. This is a minor implementation detail, just an index shift wrt the definition of a solution $(A_0, \dots, A_k)$ of the LPP problem.}. The $edge/4$ predicate is a ``symmetric closure'' of $network/4$ over time (that also keeps track of each edge's state), that will be useful to define the effects of the $switch$ action {($r_{13}$--$r_{14}$)}.

\begin{asp}
primary(X) :- node_attr(X, is_primary).
secondary(X) :- node(X), not primary(X).
opposite(open,close).     opposite(close,open).
network(-1,X,Y,A) :- start(X,Y,A).
edge(T, X, Y, A) :- network(T, X, Y, A).
edge(T, X, Y, A) :- network(T, Y, X, A).
\end{asp}

\myparagraph{Generating action sequences.} The following rules generate the search space of possible action sequences. Informally, we first guess ``the plan's length'' {$(r_8)$} by guessing a $time/1$ atom (and propagating backwards existence of previous time-points, up to zero {$(r_9)$}:

\begin{asp}
#const k.   step(0..k-1).   time(0).
{ time(T) } :- step(T).
time(T-1) :- time(T), T > 0.
\end{asp}

Next, the following choice rules impose that edges to be built and removed will be subject to an $add$ {($r_{12}$)} and $remove$ action {($r_{10}$)}, respectively, exactly once within the whole solution. We prohibit the removal of closed edges {($r_{11}$)}. The third choice rule {($r_{13}$--$r_{14}$)} guesses edge switching operations. The last constraint $r_{15}$ imposes that we perform at least one action in each time-point. 

\begin{asp}
{ action(T,remove(X,Y)): time(T) } = 1 :- must_remove(X,Y).
:- action(T, remove(X,Y)), network(T-1,X,Y,close).
{ action(T,add(X,Y)): time(T) } = 1 :- buildable(X,Y). 

{  action(T,switch(X,Y,Z)): edge(T-1, X, Y, O), edge(T-1, X, Z, O'),
  opposite(O,O'), secondary(X) } :- time(T).
:- time(T), not action(T,_).
\end{asp}

The atom $action(t,a)$ encodes that at time $t$ the action $a$ is performed, where the action $a$ is a function term as described in Section~\ref{sec:data}. We assume that actions taken at time $t$ act upon the network as available in the previous time-step $t-1$, and take effect at time $t$. 

The $affected\_by/4$ predicate {($r_{16}$--$r_{21}$)} marks networks' edges that are affected by an action, according to the definition of $\Aff(\cdot)$. In particular, the atom $affected\_by(t,x,y,a)$ states that $(x,y) \in \Aff(a)$, and that $a$ is performed at time $t$. Affected edges will either change their state (i.e., whether they are open or close), or be added/removed at the next time-point.

\begin{asp}
affected_by(T,X,Y,add(X,Y)) :- action(T,add(X,Y)).
affected_by(T,X,Y,remove(X,Y)) :- action(T,remove(X,Y)).
affected_by(T,X,Y,switch(X,Y,Z)) :- action(T,switch(X,Y,Z)).
affected_by(T,Y,X,switch(X,Y,Z)) :- action(T,switch(X,Y,Z)).
affected_by(T,X,Z,switch(X,Y,Z)) :- action(T,switch(X,Y,Z)).
affected_by(T,Z,X,switch(X,Y,Z)) :- action(T,switch(X,Y,Z)).
\end{asp}

{Notice that (see $r_{10}$--$r_{12}$) we can add an edge $(x,y)$ only if it we have the $buildable(x,y)$ fact, and similarly we can remove it only if we have the $must\_remove(x,y)$ fact. Thus, for add and remove actions, we know precisely the affected edge in the ASP representation. On the other hand, the available switch actions are defined on the previous network state ($r_{13}$), and an edge may be affected by switch actions from different endpoints. As an example, edge $(S_6, S_9)$ in Figure~\ref{fig:example} may be affected by $switch(S_6, S_8, S_9)$ but also $switch(S_9, S_6, S_2)$. Thus, the definition of $affected\_by/4$ takes this into account ($r_{18}$--$r_{21}$)}. To ensure our solutions consist of compatible action sets, we forbid to guess non-compatible actions (i.e., actions affecting the same edge).

\begin{asp}
:- affected_by(T,X,Y,A), affected_by(T,X,Y,A'), A < A'.
\end{asp}

\myparagraph{Actions' Effect.} {The following rules apply \emph{the effect} of actions taken at time $t$ on the network at time $t-1$, according to the formalisation of Section~\ref{sec:actions}. They share the use of atoms $network(t-1,\_,\_,\_)$ and $action(t,\_)$---past network state and action to execute---in the body, and have $network(t,\_,\_,\_)$---how the network is affected by the action.}

Actions of the form $add(x,y)$ simply define that the edge $(x,y)$ will exist at the next time-step:

\begin{asp}
network(T,X,Y,open) :- time(T), action(T,add(X,Y)).
\end{asp}

The edge switching $switch(x,y,z)$ operation first identifies the correct pair of edges that is involved in the action, then flips their state in the next time-step:

\begin{asp}
network(T,X,Y,O') :- affected_by(T,X,Y,switch(_,_,_)), network(T-1,X,Y,O), 
    opposite(O,O').
\end{asp}

All edges that are not affected by an action are implicitly carried over to the next time-step, with the same state as before. This is a form of \emph{inertia}. The combination of $remove(x,y)$ affecting $(x,y)$ and intertia implicitly removes edges. 

\begin{asp}
network(T,X,Y,A) :- time(T), network(T-1,X,Y,A), not affected_by(T,X,Y,_).
\end{asp}

\myparagraph{Encoding Radiality I (Connectedness).} To enforce the radiality property, {as by Definition~\ref{def:radial}}, we require the closed network to be a forest with roots in the primary nodes $P$. First, we ensure that $(i)$ all secondary $s \in S$ reach at least one primary $p \in P$ in the closed network, and $(ii)$ there are no loops between primaries over the closed network. We start by defining the standard notion of graph reachability over the closed network by means of the predicate $reach/3$ {($r_{27}$--$r_{30})$}:

\begin{asp}
reach(T,X,Y) :- time(T), network(T,X, Y,close).
reach(T,X,Y) :- time(T), network(T,Y, X,close).
reach(T,X,Y) :- time(T), reach(T,Y,X).
reach(T,X,Z) :- time(T), reach(T,X,Y), reach(T,Y,Z).
\end{asp}

\noindent Then, we use the atom $reaches\_primary(t,x,y)$ to model that node $x$ is connected to the primary station $y$ at time $t$. The constraints model, respectively, that two distinct primaries should not reach each other {($r_{32}$)}, and that all secondary stations should be connected to at least one primary station {($r_{33}$)}.

\begin{asp}
reaches_primary(T,X,Y) :- time(T), reach(T,X,Y), primary(Y).
:- primary(X), primary(Y), time(T), X < Y, reach(T,X,Y).
:- secondary(Y), time(T), not reaches_primary(T,Y,_).
\end{asp}

\myparagraph{Encoding Radiality II (Acyclicity).} Since the problem deals with undirected graphs and the previous rules {($r_{32}$--$r_{33}$)} enforce that each secondary station is connected to exactly one primary station in the closed network (recall that if it were connected to two there would be a loop between primaries, which we forbid by means of a constraint {($r_{32}$}), it is sufficient to enforce that the network at time $t$ has \emph{the correct number of edges} to enforce acyclicity. Since we are looking for $|P|$ disjoint trees and the number of edges in a tree of $q$ nodes is $q-1$, we want our network to have exactly $|V| - |P|$ edges. The predicates $num\_primaries/1$ and $num\_nodes/1$ store the number of primary nodes and overall number of nodes in the network, while the constraint {($r_{36}$--$r_{37}$)} enforces the correct number of overall edges.

\begin{asp}
num_primaries(R) :- R=#count{Z: primary(Z)}.
num_nodes(X) :- X=#count{Z: node(Z)}.
:- num_primaries(R), num_nodes(Q), time(T),
    #count{X, Y: network(T,X,Y,close)} != (Q-R).
\end{asp}

\myparagraph{Encoding Reconfigurability.} Secondary stations must be \emph{redundantly connected} {(Definition~\ref{def:redundantlyconnected})} to an extra primary substation over the open network, distinct from the one they are connected to in the closed network. Recall that a secondary station cannot be connected to two primary stations in the closed network, as this would create a loop between primaries and violate radiality.

\begin{asp}
redundancy_reach(T,X,Y) :- time(T), network(T,X,Y,_).
redundancy_reach(T,X,Y) :- time(T), redundancy_reach(T,Y,X).
redundancy_reach(T,X,Z) :- time(T), redundancy_reach(T,X,Y), 
    redundancy_reach(T,Y,Z), not primary(Y).

redundancy(T,S) :- secondary(S), time(T), redundancy_reach(T, S, X),
    primary(X), not reach(T, S, X).
:- secondary(S), time(T), not redundancy(T,S).
\end{asp}

The predicate $redundancy\_reach/3$ models reachability over the open network {($r_{38}$--$r_{41}$)}, with the atom {$redundancy\_reach(t,x,y)$} modelling that node $x$ is connected to node $y$ over the open network \emph{with a path that does not traverse any primary substation} -- notice the \texttt{not primary(Y)} atom in the body of the recursive definition of $redudancy\_reach/3$ {($r_{41}$)}. This ensures the only way networks' edges involving primaries are taken into account in computing $redundancy\_reach/3$ is via direct edges. Finally, $redundancy(t,s)$ denotes that at time $t$ the secondary station $s$ reaches a primary substation over the open network, and we prevent a secondary station that does not {($r_{44}$)}, {ensuring reconfigurability as by Definition~\ref{def:reconfigurable}, i.e. that all secondary nodes are redundantly connected.}

\myparagraph{Encoding degree-compliance.} {We use aggregates to count the number of adjacent edges to each node, with the atom $degree(t,x,d)$ meaning that node $x$ has degree $d$ in the network at time $t$. The constraints {$(r_{46}$--$r_{47})$} enforce that $d \in [2,3]$, as by {Definition~\ref{def:degreecompliance}}:

\begin{asp}
degree(T,X,D) :- time(T), node(X), D=#count{Y: edge(T,X,Y,_)}.
:- secondary(S), time(T), degree(T,S,D), D < 2.
:- secondary(S), time(T), degree(T,S,D), D > 3.
\end{asp}

\myparagraph{Final configuration.} We encode that the final network obtained by applying the guessed actions sequence is exactly the target network $G_T$, {as in Definition~\ref{def:lpp}}: {it contains the same edges as $G_T$ and in the same state ($r_{49}$--$r_{50}$), and all edges to be removed have been removed ($r_{51}$)}.

\begin{asp}
final(T) :- time(T), T = #max{ T' : time(T') }.
:- final(T), target(X,Y,A), not network(T,X,Y,A).
:- final(T), network(T,X,Y,A), not target(X,Y,A).
:- final(T), must_remove(X,Y), network(T,X,Y,_).
\end{asp}

\myparagraph{Preferred solutions.} 
Answer sets of the above logic program
encode sequences of (sets of) actions that reconfigure $G$ into $G_T$. As customary in ASP applications, multiple answer sets may exist, i.e., multiple ways to reconfigure $G$ into $G_T$. 
ASP optimization (via weak constraints) enables us to single out \emph{preferred solutions}, according to different criteria. 
For the purposes of this paper, we are interested in reducing the overall number of actions executed in the plan, as well as its length (lexicographically). We can express this by the constraints:
\begin{asp}
:~ action(T,switch(X,Y,Z)). [1@2,T,A]
:~ time(T). [1@1,T]
\end{asp}
The weak constraint {$r_{52}$} minimizes the total number of actions\footnote{Edges cannot be added (removed) twice, hence minimizing actions is equivalent to minimizing the number of switches.}, while the weak constraint {$r_{53}$} favours shorter plans.

\begin{figure}[t]
  \centering
  \begin{subfigure}{0.47\textwidth}
    \includegraphics[scale=0.45]{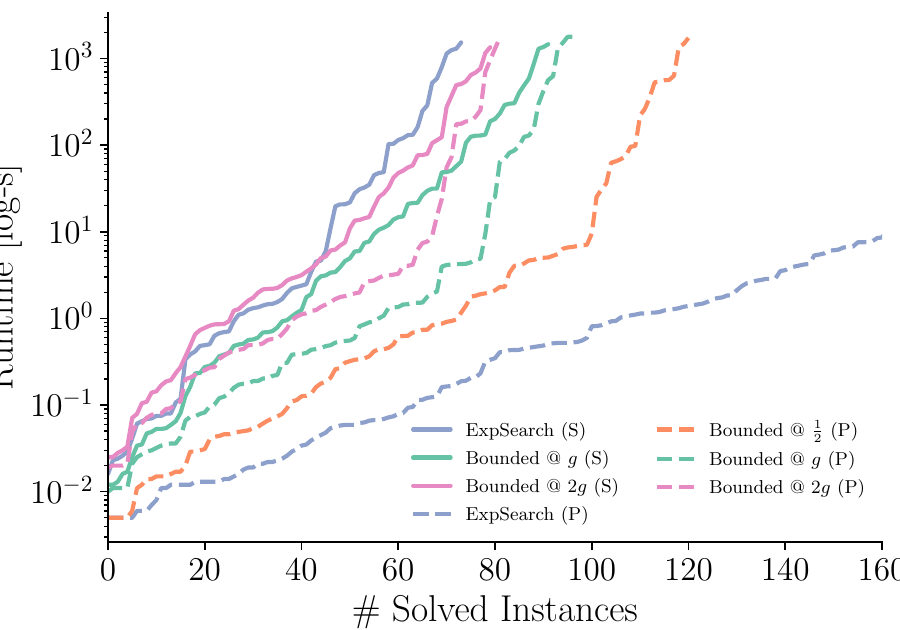}
    \label{fig:exp2-cactus}
  \end{subfigure}
  \hfill
  \begin{subfigure}{0.47\textwidth}
    \includegraphics[scale=0.45]{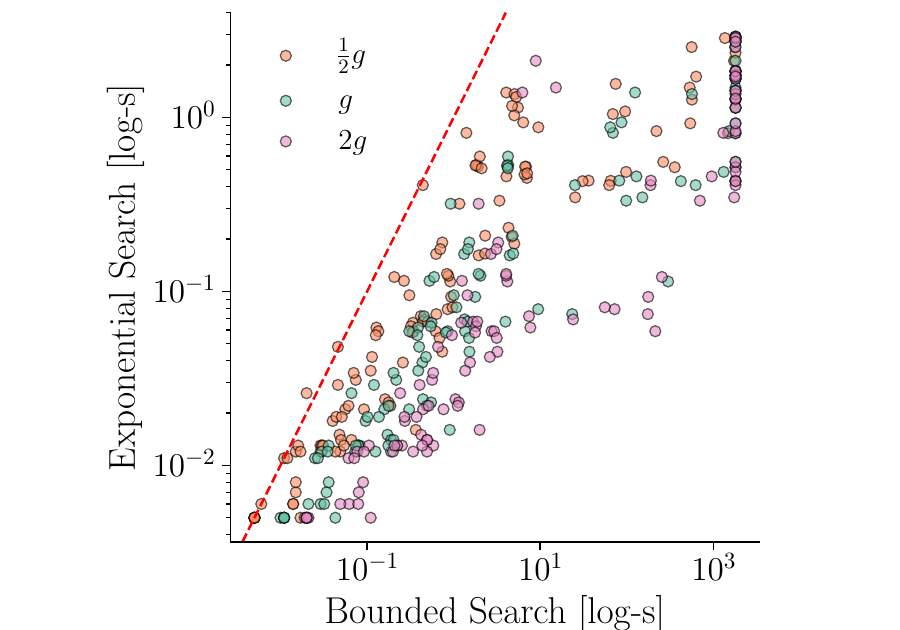}
    \label{fig:exp2-scatt er}
  \end{subfigure}
  \caption{
  (a) A point $(i,t)$ indicates that $i$ instances are solved in $t$ seconds.
  (b) A point $(x,y)$ denotes a synthetic instance solved in $y$ seconds by exponential and in $x$ by bounded.
  Points below the bisector indicate faster exponential search.
  }
  \label{fig:exp}
\end{figure}
\section{Experimental Analysis}

The empirical evaluation aims to assess the capabilities of the proposed approach. We start by describing the experimental settings and, then, in Section~\ref{sec:experiments:results}, we present the results.

\myparagraph{Execution environment.} Experiments were executed on a 64-bit Linux machine equipped with an Intel i7-13700 @ 5.2 GHz with 32GB of RAM and 16 physical cores, running up to 8 concurrent jobs via GNU Parallel. Each run was limited to a timeout of 1800 CPU seconds and 2 GB of memory. We use the \textsc{clingo} solver (version 5.4.1) in default configuration.
\myparagraph{Implementation details.} {In real scenarios, a bound on reconfiguration's length is unknown. We progressively expand such bound by means of multiple solver calls. In particular, we solve the bounded version of the problem over the interval $[2^h,2^{h+1}]$ (i.e., we search for a plan of length at least $2^h$ and at most $2^{h+1}$) starting with $h=0$, and increasing it up to the point where we find a solution (or hit resource limits). To enforce a lower bound $2^h$ on plan's length, we add the fact $time(2^h)$, while to enforce an upper bound $2^{h+1}$ we set the runtime constant $k$ to $2^{h+1}$.}

\myparagraph{Data.}\label{sec:experiments:data} 
Our test set is composed by a collection of synthetic and real-world instances. The 7 real
instances were obtained using 
data from a French urban network consisting of two source stations (primary nodes). This data has been anonymised and used to create smaller subnetworks that comply with a secure feeder architecture, and correspond to French distribution network design practices in terms of topology, feeder capacity constraints, and balancing criteria. The instances range in size from 6 to 40 secondary nodes, and all have 2 primary nodes. %
Further, we generated synthetic instances by varying the number of nodes $|V| \in \{8, 12, 15, 18, 22, {30, 40, 50}\}$, fixing the number of primary nodes to $|P|=2$. This size matches what experts deal with in daily operations.
For each size, we sampled five random compliant power grids. 
Starting from each grid $G$, we constructed a sequential plan by applying valid random actions that preserve compliance, leading to $G_T$. 
In doing so, we avoid trivial redundancies, i.e., addition-removals of the same edge, or ``opposite switches'' in adjacent time-steps. 
Motivated by the empirical observation that, in real instances, solution length is approximately equal to the number of edges, we considered plan depths of the form $\alpha \cdot |E|$ with $\alpha \in \{0.2, 0.6, 1.0, 1.4, 1.8\}$. 
Executing these action sequences yields pairs of compliant configurations $(G, G_T)$, which constitute valid instances of the LPP.
Overall, this process generates {175} synthetic LPP instances.
Moreover, $\alpha \cdot |E|$ is an upper bound for a sequential solution's length.
We will refer to this (instance-wise) upper bound as $g$, and experiments will consider $\frac{1}{2}g$, $g$  and $2g$ as possible bounds.

\begin{table}[t]
  \centering
  \begin{minipage}[c]{0.49\textwidth}
    \centering
    \resizebox{\textwidth}{!}{%
    \begin{tabular}{ccccc|c}
      \toprule
      $|V|$ & Exp. Search & $\frac{1}{2}g$ & $g$ & $2g$ & Total \\
      \midrule
      8 & 25;25 & 0;25 & 25;25 & 25;25 & 75;100 \\
      12 & 21;25 & 0;25 & 25;25 & 25;25 & 71;100 \\
      18 & 11;25 & 0;25 & 19;21 & 16;16 & 46;87 \\
      22 & 7;25 & 0;16 & 13;11 & 9;11 & 29;63 \\
      30 & 5;25 & 0;15 & 5;10 & 5;5 & 15;55 \\
      40 & 5;25 & 0;10 & 5;5 & 0;0 & 10;40 \\
      50 & 0;25 & 0;5 & 0;0 & 0;0 & 0;30 \\
      \midrule
      Total & 74;175 & 0;121 & 92;97 & 80;82 & 246;475 \\
      \bottomrule
    \end{tabular}%
  }

    \caption{{Number of solved instances by graph size and method. Cell content $x;y$ denotes that we solve $x$ instances with sequential planning, and $y$ instances with parallel planning, respectively.}}
    \label{tab:solved-instances-alt}
  \end{minipage}%
  \hfill
  \begin{minipage}[c]{0.47\textwidth}
    \centering
    \includegraphics[width=\textwidth]{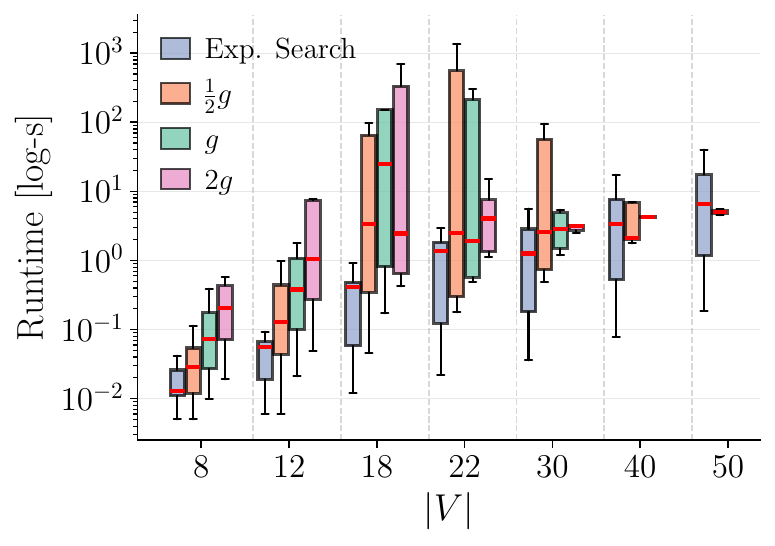}
    \captionof{figure}{{Parallel plans runtime comparison across graph sizes. Red bar is the median runtime.}}
    \label{fig:grouped-boxplots-alt}
  \end{minipage}
\end{table}

\subsection{Results}\label{sec:experiments:results}
Below we present the results of our experiments. All solutions have been validated using an external purposely-designed tool to assess the satisfaction of the DSO rules. 
\myparagraph{Synthetic instances.} In order to select the best approaches, to be employed on real data, we first test our encoding in the bounded and exponential search version, searching for sequential and parallel plans, on the synthetic instances. We report our results in terms of the cactus plot and scatterplot in Figure~\ref{fig:exp}. {First, we can observe in Figure 3a that we are able to solve more problem instances dealing with parallel plans wrt sequential plans. Intuitively, this is due to non-interfering actions ``compressing'' the solutions' length, which translates to dealing with smaller logic programs. In ASP planning, this typically correlates with \emph{easier} problems, given that long plans are a known pain point also for ASP-based planning~(see, e.g.,~\citep{DBLP:journals/tplp/SonPBS23}). Exponential search performs worse than bounded search over sequential plans, as it incurs in overhead due to multiple solver calls that are required to expand the search horizon, while the bounded encoding operates directly on it. {In the parallel setting, the exponential search approach dominates, solving 54 more instances wrt the bounded at horizon $\frac{1}{2}g$, 78 wrt $g$ and 93 wrt $2g$, with all instances being mostly trivial for the parallel approach.} %
The scatterplot in Figure 3b shows that for parallel plans, all instances slow down as bounds increase. In particular, we can observe (top-right corner of the scatter-plot) that these synthetic instances---despite being trivial whenever they admit a short solution---are challenging if we try to solve at a much higher bound: we observe several instances that can be solved within seconds by the exponential search approach, but time-out with larger bounds.}
{
Table~\ref{tab:solved-instances-alt} shows that all techniques solve \emph{fewer} instances as 
$|V|$ increases. 
Parallel exponential search scales better as $|V|$ increases.
This is confirmed by runtimes distributions in Figure~\ref{fig:grouped-boxplots-alt}. 
The largest instances are comparable in size to those handled by human experts, both for training and for the actual long-term planning of power grids. 
Larger grids are generally decomposed into networks of between 20-25 nodes.
In that sense, our approach can readily support the work of experts. 
Overall, these results on synthetic instances motivate our focusing on parallel plans and exponential search approaches to solve real-world scenarios.
}

\begin{table}[t]
\footnotesize
\centering
    \begin{tabular}{ccccccccc}
    \toprule
     & & \multicolumn{3}{c}{Parallel} & \multicolumn{4}{c}{Parallel Optimization} \\
     \cmidrule(lr){3-5} \cmidrule(lr){6-9}
    \multirow{1}{*}{Instance} & \multirow{1}{*}{$\#$ Nodes}  & $r$ & $\Sigma |A_i|$ & $\max |A_i|$ & $r$ & $\Sigma |A_i|$ & $\max |A_i|$ & {Opt?}\\
    \midrule
    P06 & 8 & 0.0 & 12 & 5 & 0.0     & 7 & 4   & $\checkmark$   \\
    P10 & 12 & 0.0 & 20 & 6 & 0.1     & 20 & 6  & $\checkmark$   \\
    P13 & 12 & 0.0 & 19 & 5 & 0.1     & 16 & 7  & $\checkmark$   \\
    P16 & 18 & 1.0 & 30 & 7 & 0.9     & 30 & 7  &                \\
    P22 & 24 & 0.4 & 40 & 12 & 1.8    & 29 & 10 & $\checkmark$   \\
    P30 & 30 & 41.2 & 67 & 10 & 62.2  & 65 & 11 &                \\
    P40 & 42 & --  & -- & -- & 1485.7 & 103 & 14 &                \\
    \bottomrule
    \end{tabular}
    \caption{Runtime ($r$) on real instances of a given size ($\#$ Nodes), with total number of actions ($\Sigma |A_i|$) and maximum number of concurrent actions ($\max |A_i|$). Dashes (--) denote timeout. {The ``Opt?'' column marks instances where we are able to prove optimality of the parallel plan.}}
    \label{tab:real}
\end{table}

{\myparagraph{Real instances.} Table~\ref{tab:real} reports runtime of our exponential search approach, with and without optimization statements (as discussed in Section 4). We are able to solve instances of up to {42} nodes, proving the approach is useful for operation-support in realistic scenarios, and can solve instances of size larger than those experts deal in daily operations. Runtimes are comparable with the ones we obtained over synthetic instances of the same size.  On some solved instances, we are also able to prove optimality of parallel plans. In real-world scenarios, this translate in lower costs and simpler-to-put-in-place operations.}  {
These solutions were validated by an automated tool and visually reviewed by domain experts, who confirmed their consistency with DSO planning rules and their comparability, at the topological level, to solutions typically produced by experts.
}

\section{Conclusion}
The problem of long-term power grid planning requires computing sequences of operations on the power grid infrastructure to \emph{transform} it into a target configuration, while guaranteeing that each intermediate state satisfies certain properties of interest. Enacting these plans may span many years of costly operations. Currently, such plans are devised manually by domain experts. This paper proposes an ASP-based approach to tackle the problem, providing an automated solution inspired by ASP planning and ASP reconfiguration. 
Our experiments on synthetic and real-world data show that the approach can solve complex instances comparable to, or larger than, those currently handled by human operators.
As future work, we plan to further improve scalability by applying, e.g., domain-specific heuristics and decomposition methods, and adapt to infrastructure-specific requirements that may be present in other installations, such as immutability of certain components or budget limitations.

\subsection*{Acknowledgements}
This work has been partially supported by the Italian ``Ministero delle Imprese e Sviluppo Economico'' (MIMIT) under projects ``Ei-TWIN - Energy Digital Twin'' (CUP B29J2400068000); by Regione Calabria under PR Calabria FESR FSE 2021–2027 through the projects ``MOZART - Modelli e tecniche di mOdernizzazione di applicaZioni e data silos a supporto di clienti esterni, field force e impiegati di
backoffice basati su AI GeneRativa e apprendimento auTomatico'' (CUP J49I24001740005), and ``SIGENERA - Large Language Models (LLM) per il controllo di SIstemi informativi, la GENERazione automatica di testo, e l'accesso a basi di
conoscenzA '' (CUP J29I24001770005). Francesco Percassi and Mauro Vallati were supported by a UKRI Future Leaders Fellowship, grant number
MR/Z00005X/1.

\bibliographystyle{tlplike}
\bibliography{refs}

\end{document}

%% file: pgf_example/01_single_plot.tex
\begingroup%
\makeatletter%
\begin{pgfpicture}%
\pgfpathrectangle{\pgfpointorigin}{\pgfqpoint{3.600000in}{3.514317in}}%
\pgfusepath{use as bounding box, clip}%
\begin{pgfscope}%
\pgfsetbuttcap%
\pgfsetmiterjoin%
\definecolor{currentfill}{rgb}{1.000000,1.000000,1.000000}%
\pgfsetfillcolor{currentfill}%
\pgfsetlinewidth{0.000000pt}%
\definecolor{currentstroke}{rgb}{1.000000,1.000000,1.000000}%
\pgfsetstrokecolor{currentstroke}%
\pgfsetdash{}{0pt}%
\pgfpathmoveto{\pgfqpoint{0.000000in}{0.000000in}}%
\pgfpathlineto{\pgfqpoint{3.600000in}{0.000000in}}%
\pgfpathlineto{\pgfqpoint{3.600000in}{3.514317in}}%
\pgfpathlineto{\pgfqpoint{0.000000in}{3.514317in}}%
\pgfpathlineto{\pgfqpoint{0.000000in}{0.000000in}}%
\pgfpathclose%
\pgfusepath{fill}%
\end{pgfscope}%
\begin{pgfscope}%
\pgfpathrectangle{\pgfqpoint{0.100000in}{0.100000in}}{\pgfqpoint{3.400000in}{3.314317in}}%
\pgfusepath{clip}%
\pgfsetbuttcap%
\pgfsetroundjoin%
\pgfsetlinewidth{2.007500pt}%
\definecolor{currentstroke}{rgb}{0.298039,0.470588,0.658824}%
\pgfsetstrokecolor{currentstroke}%
\pgfsetdash{}{0pt}%
\pgfpathmoveto{\pgfqpoint{2.395865in}{1.148125in}}%
\pgfpathlineto{\pgfqpoint{2.841741in}{0.775380in}}%
\pgfusepath{stroke}%
\end{pgfscope}%
\begin{pgfscope}%
\pgfpathrectangle{\pgfqpoint{0.100000in}{0.100000in}}{\pgfqpoint{3.400000in}{3.314317in}}%
\pgfusepath{clip}%
\pgfsetbuttcap%
\pgfsetroundjoin%
\pgfsetlinewidth{2.007500pt}%
\definecolor{currentstroke}{rgb}{0.298039,0.470588,0.658824}%
\pgfsetstrokecolor{currentstroke}%
\pgfsetdash{}{0pt}%
\pgfpathmoveto{\pgfqpoint{2.080565in}{0.322055in}}%
\pgfpathlineto{\pgfqpoint{2.395865in}{1.148125in}}%
\pgfusepath{stroke}%
\end{pgfscope}%
\begin{pgfscope}%
\pgfpathrectangle{\pgfqpoint{0.100000in}{0.100000in}}{\pgfqpoint{3.400000in}{3.314317in}}%
\pgfusepath{clip}%
\pgfsetbuttcap%
\pgfsetroundjoin%
\pgfsetlinewidth{2.007500pt}%
\definecolor{currentstroke}{rgb}{0.298039,0.470588,0.658824}%
\pgfsetstrokecolor{currentstroke}%
\pgfsetdash{}{0pt}%
\pgfpathmoveto{\pgfqpoint{1.853024in}{0.773356in}}%
\pgfpathlineto{\pgfqpoint{2.080565in}{0.322055in}}%
\pgfusepath{stroke}%
\end{pgfscope}%
\begin{pgfscope}%
\pgfpathrectangle{\pgfqpoint{0.100000in}{0.100000in}}{\pgfqpoint{3.400000in}{3.314317in}}%
\pgfusepath{clip}%
\pgfsetbuttcap%
\pgfsetroundjoin%
\pgfsetlinewidth{2.007500pt}%
\definecolor{currentstroke}{rgb}{0.298039,0.470588,0.658824}%
\pgfsetstrokecolor{currentstroke}%
\pgfsetdash{}{0pt}%
\pgfpathmoveto{\pgfqpoint{3.238207in}{1.750135in}}%
\pgfpathlineto{\pgfqpoint{2.841741in}{0.775380in}}%
\pgfusepath{stroke}%
\end{pgfscope}%
\begin{pgfscope}%
\pgfpathrectangle{\pgfqpoint{0.100000in}{0.100000in}}{\pgfqpoint{3.400000in}{3.314317in}}%
\pgfusepath{clip}%
\pgfsetbuttcap%
\pgfsetroundjoin%
\pgfsetlinewidth{2.007500pt}%
\definecolor{currentstroke}{rgb}{0.960784,0.521569,0.094118}%
\pgfsetstrokecolor{currentstroke}%
\pgfsetdash{}{0pt}%
\pgfpathmoveto{\pgfqpoint{0.438926in}{1.300531in}}%
\pgfpathlineto{\pgfqpoint{1.093266in}{0.594842in}}%
\pgfusepath{stroke}%
\end{pgfscope}%
\begin{pgfscope}%
\pgfpathrectangle{\pgfqpoint{0.100000in}{0.100000in}}{\pgfqpoint{3.400000in}{3.314317in}}%
\pgfusepath{clip}%
\pgfsetbuttcap%
\pgfsetroundjoin%
\pgfsetlinewidth{2.007500pt}%
\definecolor{currentstroke}{rgb}{0.960784,0.521569,0.094118}%
\pgfsetstrokecolor{currentstroke}%
\pgfsetdash{}{0pt}%
\pgfpathmoveto{\pgfqpoint{0.438926in}{1.300531in}}%
\pgfpathlineto{\pgfqpoint{0.373099in}{2.249038in}}%
\pgfusepath{stroke}%
\end{pgfscope}%
\begin{pgfscope}%
\pgfpathrectangle{\pgfqpoint{0.100000in}{0.100000in}}{\pgfqpoint{3.400000in}{3.314317in}}%
\pgfusepath{clip}%
\pgfsetbuttcap%
\pgfsetroundjoin%
\pgfsetlinewidth{2.007500pt}%
\definecolor{currentstroke}{rgb}{0.960784,0.521569,0.094118}%
\pgfsetstrokecolor{currentstroke}%
\pgfsetdash{}{0pt}%
\pgfpathmoveto{\pgfqpoint{0.963016in}{2.990663in}}%
\pgfpathlineto{\pgfqpoint{0.373099in}{2.249038in}}%
\pgfusepath{stroke}%
\end{pgfscope}%
\begin{pgfscope}%
\pgfpathrectangle{\pgfqpoint{0.100000in}{0.100000in}}{\pgfqpoint{3.400000in}{3.314317in}}%
\pgfusepath{clip}%
\pgfsetbuttcap%
\pgfsetroundjoin%
\pgfsetlinewidth{2.007500pt}%
\definecolor{currentstroke}{rgb}{0.960784,0.521569,0.094118}%
\pgfsetstrokecolor{currentstroke}%
\pgfsetdash{}{0pt}%
\pgfpathmoveto{\pgfqpoint{1.902667in}{3.136579in}}%
\pgfpathlineto{\pgfqpoint{0.963016in}{2.990663in}}%
\pgfusepath{stroke}%
\end{pgfscope}%
\begin{pgfscope}%
\pgfpathrectangle{\pgfqpoint{0.100000in}{0.100000in}}{\pgfqpoint{3.400000in}{3.314317in}}%
\pgfusepath{clip}%
\pgfsetbuttcap%
\pgfsetroundjoin%
\pgfsetlinewidth{2.007500pt}%
\definecolor{currentstroke}{rgb}{0.960784,0.521569,0.094118}%
\pgfsetstrokecolor{currentstroke}%
\pgfsetdash{}{0pt}%
\pgfpathmoveto{\pgfqpoint{1.902667in}{3.136579in}}%
\pgfpathlineto{\pgfqpoint{2.744162in}{2.658297in}}%
\pgfusepath{stroke}%
\end{pgfscope}%
\begin{pgfscope}%
\pgfpathrectangle{\pgfqpoint{0.100000in}{0.100000in}}{\pgfqpoint{3.400000in}{3.314317in}}%
\pgfusepath{clip}%
\pgfsetbuttcap%
\pgfsetroundjoin%
\pgfsetlinewidth{2.007500pt}%
\definecolor{currentstroke}{rgb}{0.960784,0.521569,0.094118}%
\pgfsetstrokecolor{currentstroke}%
\pgfsetdash{}{0pt}%
\pgfpathmoveto{\pgfqpoint{2.659242in}{1.954744in}}%
\pgfpathlineto{\pgfqpoint{2.744162in}{2.658297in}}%
\pgfusepath{stroke}%
\end{pgfscope}%
\begin{pgfscope}%
\pgfpathrectangle{\pgfqpoint{0.100000in}{0.100000in}}{\pgfqpoint{3.400000in}{3.314317in}}%
\pgfusepath{clip}%
\pgfsetbuttcap%
\pgfsetroundjoin%
\pgfsetlinewidth{2.007500pt}%
\definecolor{currentstroke}{rgb}{0.619608,0.619608,0.619608}%
\pgfsetstrokecolor{currentstroke}%
\pgfsetdash{{7.400000pt}{3.200000pt}}{0.000000pt}%
\pgfpathmoveto{\pgfqpoint{1.853024in}{0.773356in}}%
\pgfpathlineto{\pgfqpoint{1.093266in}{0.594842in}}%
\pgfusepath{stroke}%
\end{pgfscope}%
\begin{pgfscope}%
\pgfpathrectangle{\pgfqpoint{0.100000in}{0.100000in}}{\pgfqpoint{3.400000in}{3.314317in}}%
\pgfusepath{clip}%
\pgfsetbuttcap%
\pgfsetroundjoin%
\pgfsetlinewidth{2.007500pt}%
\definecolor{currentstroke}{rgb}{0.619608,0.619608,0.619608}%
\pgfsetstrokecolor{currentstroke}%
\pgfsetdash{{7.400000pt}{3.200000pt}}{0.000000pt}%
\pgfpathmoveto{\pgfqpoint{3.238207in}{1.750135in}}%
\pgfpathlineto{\pgfqpoint{2.659242in}{1.954744in}}%
\pgfusepath{stroke}%
\end{pgfscope}%
\begin{pgfscope}%
\pgfpathrectangle{\pgfqpoint{0.100000in}{0.100000in}}{\pgfqpoint{3.400000in}{3.314317in}}%
\pgfusepath{clip}%
\pgfsetbuttcap%
\pgfsetroundjoin%
\definecolor{currentfill}{rgb}{0.298039,0.470588,0.658824}%
\pgfsetfillcolor{currentfill}%
\pgfsetlinewidth{1.003750pt}%
\definecolor{currentstroke}{rgb}{0.298039,0.470588,0.658824}%
\pgfsetstrokecolor{currentstroke}%
\pgfsetdash{}{0pt}%
\pgfpathmoveto{\pgfqpoint{3.238207in}{1.566403in}}%
\pgfpathcurveto{\pgfqpoint{3.286934in}{1.566403in}}{\pgfqpoint{3.333671in}{1.585762in}}{\pgfqpoint{3.368126in}{1.620217in}}%
\pgfpathcurveto{\pgfqpoint{3.402581in}{1.654672in}}{\pgfqpoint{3.421940in}{1.701409in}}{\pgfqpoint{3.421940in}{1.750135in}}%
\pgfpathcurveto{\pgfqpoint{3.421940in}{1.798862in}}{\pgfqpoint{3.402581in}{1.845599in}}{\pgfqpoint{3.368126in}{1.880054in}}%
\pgfpathcurveto{\pgfqpoint{3.333671in}{1.914509in}}{\pgfqpoint{3.286934in}{1.933868in}}{\pgfqpoint{3.238207in}{1.933868in}}%
\pgfpathcurveto{\pgfqpoint{3.189481in}{1.933868in}}{\pgfqpoint{3.142744in}{1.914509in}}{\pgfqpoint{3.108289in}{1.880054in}}%
\pgfpathcurveto{\pgfqpoint{3.073834in}{1.845599in}}{\pgfqpoint{3.054475in}{1.798862in}}{\pgfqpoint{3.054475in}{1.750135in}}%
\pgfpathcurveto{\pgfqpoint{3.054475in}{1.701409in}}{\pgfqpoint{3.073834in}{1.654672in}}{\pgfqpoint{3.108289in}{1.620217in}}%
\pgfpathcurveto{\pgfqpoint{3.142744in}{1.585762in}}{\pgfqpoint{3.189481in}{1.566403in}}{\pgfqpoint{3.238207in}{1.566403in}}%
\pgfpathlineto{\pgfqpoint{3.238207in}{1.566403in}}%
\pgfpathclose%
\pgfusepath{stroke,fill}%
\end{pgfscope}%
\begin{pgfscope}%
\pgfpathrectangle{\pgfqpoint{0.100000in}{0.100000in}}{\pgfqpoint{3.400000in}{3.314317in}}%
\pgfusepath{clip}%
\pgfsetbuttcap%
\pgfsetroundjoin%
\definecolor{currentfill}{rgb}{0.298039,0.470588,0.658824}%
\pgfsetfillcolor{currentfill}%
\pgfsetlinewidth{1.003750pt}%
\definecolor{currentstroke}{rgb}{0.298039,0.470588,0.658824}%
\pgfsetstrokecolor{currentstroke}%
\pgfsetdash{}{0pt}%
\pgfpathmoveto{\pgfqpoint{1.853024in}{0.589623in}}%
\pgfpathcurveto{\pgfqpoint{1.901750in}{0.589623in}}{\pgfqpoint{1.948487in}{0.608982in}}{\pgfqpoint{1.982942in}{0.643437in}}%
\pgfpathcurveto{\pgfqpoint{2.017397in}{0.677892in}}{\pgfqpoint{2.036756in}{0.724629in}}{\pgfqpoint{2.036756in}{0.773356in}}%
\pgfpathcurveto{\pgfqpoint{2.036756in}{0.822082in}}{\pgfqpoint{2.017397in}{0.868819in}}{\pgfqpoint{1.982942in}{0.903274in}}%
\pgfpathcurveto{\pgfqpoint{1.948487in}{0.937729in}}{\pgfqpoint{1.901750in}{0.957088in}}{\pgfqpoint{1.853024in}{0.957088in}}%
\pgfpathcurveto{\pgfqpoint{1.804297in}{0.957088in}}{\pgfqpoint{1.757560in}{0.937729in}}{\pgfqpoint{1.723105in}{0.903274in}}%
\pgfpathcurveto{\pgfqpoint{1.688650in}{0.868819in}}{\pgfqpoint{1.669291in}{0.822082in}}{\pgfqpoint{1.669291in}{0.773356in}}%
\pgfpathcurveto{\pgfqpoint{1.669291in}{0.724629in}}{\pgfqpoint{1.688650in}{0.677892in}}{\pgfqpoint{1.723105in}{0.643437in}}%
\pgfpathcurveto{\pgfqpoint{1.757560in}{0.608982in}}{\pgfqpoint{1.804297in}{0.589623in}}{\pgfqpoint{1.853024in}{0.589623in}}%
\pgfpathlineto{\pgfqpoint{1.853024in}{0.589623in}}%
\pgfpathclose%
\pgfusepath{stroke,fill}%
\end{pgfscope}%
\begin{pgfscope}%
\pgfpathrectangle{\pgfqpoint{0.100000in}{0.100000in}}{\pgfqpoint{3.400000in}{3.314317in}}%
\pgfusepath{clip}%
\pgfsetbuttcap%
\pgfsetroundjoin%
\definecolor{currentfill}{rgb}{0.960784,0.521569,0.094118}%
\pgfsetfillcolor{currentfill}%
\pgfsetlinewidth{1.003750pt}%
\definecolor{currentstroke}{rgb}{0.960784,0.521569,0.094118}%
\pgfsetstrokecolor{currentstroke}%
\pgfsetdash{}{0pt}%
\pgfpathmoveto{\pgfqpoint{1.902667in}{2.952846in}}%
\pgfpathcurveto{\pgfqpoint{1.951394in}{2.952846in}}{\pgfqpoint{1.998131in}{2.972206in}}{\pgfqpoint{2.032586in}{3.006661in}}%
\pgfpathcurveto{\pgfqpoint{2.067041in}{3.041115in}}{\pgfqpoint{2.086400in}{3.087853in}}{\pgfqpoint{2.086400in}{3.136579in}}%
\pgfpathcurveto{\pgfqpoint{2.086400in}{3.185306in}}{\pgfqpoint{2.067041in}{3.232043in}}{\pgfqpoint{2.032586in}{3.266498in}}%
\pgfpathcurveto{\pgfqpoint{1.998131in}{3.300953in}}{\pgfqpoint{1.951394in}{3.320312in}}{\pgfqpoint{1.902667in}{3.320312in}}%
\pgfpathcurveto{\pgfqpoint{1.853941in}{3.320312in}}{\pgfqpoint{1.807204in}{3.300953in}}{\pgfqpoint{1.772749in}{3.266498in}}%
\pgfpathcurveto{\pgfqpoint{1.738294in}{3.232043in}}{\pgfqpoint{1.718935in}{3.185306in}}{\pgfqpoint{1.718935in}{3.136579in}}%
\pgfpathcurveto{\pgfqpoint{1.718935in}{3.087853in}}{\pgfqpoint{1.738294in}{3.041115in}}{\pgfqpoint{1.772749in}{3.006661in}}%
\pgfpathcurveto{\pgfqpoint{1.807204in}{2.972206in}}{\pgfqpoint{1.853941in}{2.952846in}}{\pgfqpoint{1.902667in}{2.952846in}}%
\pgfpathlineto{\pgfqpoint{1.902667in}{2.952846in}}%
\pgfpathclose%
\pgfusepath{stroke,fill}%
\end{pgfscope}%
\begin{pgfscope}%
\pgfpathrectangle{\pgfqpoint{0.100000in}{0.100000in}}{\pgfqpoint{3.400000in}{3.314317in}}%
\pgfusepath{clip}%
\pgfsetbuttcap%
\pgfsetroundjoin%
\definecolor{currentfill}{rgb}{0.960784,0.521569,0.094118}%
\pgfsetfillcolor{currentfill}%
\pgfsetlinewidth{1.003750pt}%
\definecolor{currentstroke}{rgb}{0.960784,0.521569,0.094118}%
\pgfsetstrokecolor{currentstroke}%
\pgfsetdash{}{0pt}%
\pgfpathmoveto{\pgfqpoint{2.659242in}{1.771011in}}%
\pgfpathcurveto{\pgfqpoint{2.707969in}{1.771011in}}{\pgfqpoint{2.754706in}{1.790370in}}{\pgfqpoint{2.789161in}{1.824825in}}%
\pgfpathcurveto{\pgfqpoint{2.823616in}{1.859280in}}{\pgfqpoint{2.842975in}{1.906017in}}{\pgfqpoint{2.842975in}{1.954744in}}%
\pgfpathcurveto{\pgfqpoint{2.842975in}{2.003470in}}{\pgfqpoint{2.823616in}{2.050207in}}{\pgfqpoint{2.789161in}{2.084662in}}%
\pgfpathcurveto{\pgfqpoint{2.754706in}{2.119117in}}{\pgfqpoint{2.707969in}{2.138476in}}{\pgfqpoint{2.659242in}{2.138476in}}%
\pgfpathcurveto{\pgfqpoint{2.610516in}{2.138476in}}{\pgfqpoint{2.563779in}{2.119117in}}{\pgfqpoint{2.529324in}{2.084662in}}%
\pgfpathcurveto{\pgfqpoint{2.494869in}{2.050207in}}{\pgfqpoint{2.475510in}{2.003470in}}{\pgfqpoint{2.475510in}{1.954744in}}%
\pgfpathcurveto{\pgfqpoint{2.475510in}{1.906017in}}{\pgfqpoint{2.494869in}{1.859280in}}{\pgfqpoint{2.529324in}{1.824825in}}%
\pgfpathcurveto{\pgfqpoint{2.563779in}{1.790370in}}{\pgfqpoint{2.610516in}{1.771011in}}{\pgfqpoint{2.659242in}{1.771011in}}%
\pgfpathlineto{\pgfqpoint{2.659242in}{1.771011in}}%
\pgfpathclose%
\pgfusepath{stroke,fill}%
\end{pgfscope}%
\begin{pgfscope}%
\pgfpathrectangle{\pgfqpoint{0.100000in}{0.100000in}}{\pgfqpoint{3.400000in}{3.314317in}}%
\pgfusepath{clip}%
\pgfsetbuttcap%
\pgfsetroundjoin%
\definecolor{currentfill}{rgb}{0.298039,0.470588,0.658824}%
\pgfsetfillcolor{currentfill}%
\pgfsetlinewidth{1.003750pt}%
\definecolor{currentstroke}{rgb}{0.298039,0.470588,0.658824}%
\pgfsetstrokecolor{currentstroke}%
\pgfsetdash{}{0pt}%
\pgfpathmoveto{\pgfqpoint{2.080565in}{0.138322in}}%
\pgfpathcurveto{\pgfqpoint{2.129291in}{0.138322in}}{\pgfqpoint{2.176029in}{0.157681in}}{\pgfqpoint{2.210483in}{0.192136in}}%
\pgfpathcurveto{\pgfqpoint{2.244938in}{0.226591in}}{\pgfqpoint{2.264297in}{0.273328in}}{\pgfqpoint{2.264297in}{0.322055in}}%
\pgfpathcurveto{\pgfqpoint{2.264297in}{0.370781in}}{\pgfqpoint{2.244938in}{0.417519in}}{\pgfqpoint{2.210483in}{0.451973in}}%
\pgfpathcurveto{\pgfqpoint{2.176029in}{0.486428in}}{\pgfqpoint{2.129291in}{0.505788in}}{\pgfqpoint{2.080565in}{0.505788in}}%
\pgfpathcurveto{\pgfqpoint{2.031838in}{0.505788in}}{\pgfqpoint{1.985101in}{0.486428in}}{\pgfqpoint{1.950646in}{0.451973in}}%
\pgfpathcurveto{\pgfqpoint{1.916191in}{0.417519in}}{\pgfqpoint{1.896832in}{0.370781in}}{\pgfqpoint{1.896832in}{0.322055in}}%
\pgfpathcurveto{\pgfqpoint{1.896832in}{0.273328in}}{\pgfqpoint{1.916191in}{0.226591in}}{\pgfqpoint{1.950646in}{0.192136in}}%
\pgfpathcurveto{\pgfqpoint{1.985101in}{0.157681in}}{\pgfqpoint{2.031838in}{0.138322in}}{\pgfqpoint{2.080565in}{0.138322in}}%
\pgfpathlineto{\pgfqpoint{2.080565in}{0.138322in}}%
\pgfpathclose%
\pgfusepath{stroke,fill}%
\end{pgfscope}%
\begin{pgfscope}%
\pgfpathrectangle{\pgfqpoint{0.100000in}{0.100000in}}{\pgfqpoint{3.400000in}{3.314317in}}%
\pgfusepath{clip}%
\pgfsetbuttcap%
\pgfsetroundjoin%
\definecolor{currentfill}{rgb}{0.960784,0.521569,0.094118}%
\pgfsetfillcolor{currentfill}%
\pgfsetlinewidth{1.003750pt}%
\definecolor{currentstroke}{rgb}{0.960784,0.521569,0.094118}%
\pgfsetstrokecolor{currentstroke}%
\pgfsetdash{}{0pt}%
\pgfpathmoveto{\pgfqpoint{0.438926in}{1.116798in}}%
\pgfpathcurveto{\pgfqpoint{0.487652in}{1.116798in}}{\pgfqpoint{0.534389in}{1.136157in}}{\pgfqpoint{0.568844in}{1.170612in}}%
\pgfpathcurveto{\pgfqpoint{0.603299in}{1.205067in}}{\pgfqpoint{0.622658in}{1.251804in}}{\pgfqpoint{0.622658in}{1.300531in}}%
\pgfpathcurveto{\pgfqpoint{0.622658in}{1.349257in}}{\pgfqpoint{0.603299in}{1.395994in}}{\pgfqpoint{0.568844in}{1.430449in}}%
\pgfpathcurveto{\pgfqpoint{0.534389in}{1.464904in}}{\pgfqpoint{0.487652in}{1.484263in}}{\pgfqpoint{0.438926in}{1.484263in}}%
\pgfpathcurveto{\pgfqpoint{0.390199in}{1.484263in}}{\pgfqpoint{0.343462in}{1.464904in}}{\pgfqpoint{0.309007in}{1.430449in}}%
\pgfpathcurveto{\pgfqpoint{0.274552in}{1.395994in}}{\pgfqpoint{0.255193in}{1.349257in}}{\pgfqpoint{0.255193in}{1.300531in}}%
\pgfpathcurveto{\pgfqpoint{0.255193in}{1.251804in}}{\pgfqpoint{0.274552in}{1.205067in}}{\pgfqpoint{0.309007in}{1.170612in}}%
\pgfpathcurveto{\pgfqpoint{0.343462in}{1.136157in}}{\pgfqpoint{0.390199in}{1.116798in}}{\pgfqpoint{0.438926in}{1.116798in}}%
\pgfpathlineto{\pgfqpoint{0.438926in}{1.116798in}}%
\pgfpathclose%
\pgfusepath{stroke,fill}%
\end{pgfscope}%
\begin{pgfscope}%
\pgfpathrectangle{\pgfqpoint{0.100000in}{0.100000in}}{\pgfqpoint{3.400000in}{3.314317in}}%
\pgfusepath{clip}%
\pgfsetbuttcap%
\pgfsetroundjoin%
\definecolor{currentfill}{rgb}{0.960784,0.521569,0.094118}%
\pgfsetfillcolor{currentfill}%
\pgfsetlinewidth{1.003750pt}%
\definecolor{currentstroke}{rgb}{0.960784,0.521569,0.094118}%
\pgfsetstrokecolor{currentstroke}%
\pgfsetdash{}{0pt}%
\pgfpathmoveto{\pgfqpoint{0.963016in}{2.806930in}}%
\pgfpathcurveto{\pgfqpoint{1.011742in}{2.806930in}}{\pgfqpoint{1.058479in}{2.826289in}}{\pgfqpoint{1.092934in}{2.860744in}}%
\pgfpathcurveto{\pgfqpoint{1.127389in}{2.895199in}}{\pgfqpoint{1.146748in}{2.941936in}}{\pgfqpoint{1.146748in}{2.990663in}}%
\pgfpathcurveto{\pgfqpoint{1.146748in}{3.039389in}}{\pgfqpoint{1.127389in}{3.086127in}}{\pgfqpoint{1.092934in}{3.120581in}}%
\pgfpathcurveto{\pgfqpoint{1.058479in}{3.155036in}}{\pgfqpoint{1.011742in}{3.174396in}}{\pgfqpoint{0.963016in}{3.174396in}}%
\pgfpathcurveto{\pgfqpoint{0.914289in}{3.174396in}}{\pgfqpoint{0.867552in}{3.155036in}}{\pgfqpoint{0.833097in}{3.120581in}}%
\pgfpathcurveto{\pgfqpoint{0.798642in}{3.086127in}}{\pgfqpoint{0.779283in}{3.039389in}}{\pgfqpoint{0.779283in}{2.990663in}}%
\pgfpathcurveto{\pgfqpoint{0.779283in}{2.941936in}}{\pgfqpoint{0.798642in}{2.895199in}}{\pgfqpoint{0.833097in}{2.860744in}}%
\pgfpathcurveto{\pgfqpoint{0.867552in}{2.826289in}}{\pgfqpoint{0.914289in}{2.806930in}}{\pgfqpoint{0.963016in}{2.806930in}}%
\pgfpathlineto{\pgfqpoint{0.963016in}{2.806930in}}%
\pgfpathclose%
\pgfusepath{stroke,fill}%
\end{pgfscope}%
\begin{pgfscope}%
\pgfpathrectangle{\pgfqpoint{0.100000in}{0.100000in}}{\pgfqpoint{3.400000in}{3.314317in}}%
\pgfusepath{clip}%
\pgfsetbuttcap%
\pgfsetroundjoin%
\definecolor{currentfill}{rgb}{0.960784,0.521569,0.094118}%
\pgfsetfillcolor{currentfill}%
\pgfsetlinewidth{1.003750pt}%
\definecolor{currentstroke}{rgb}{0.960784,0.521569,0.094118}%
\pgfsetstrokecolor{currentstroke}%
\pgfsetdash{}{0pt}%
\pgfpathmoveto{\pgfqpoint{0.373099in}{2.065305in}}%
\pgfpathcurveto{\pgfqpoint{0.421825in}{2.065305in}}{\pgfqpoint{0.468562in}{2.084664in}}{\pgfqpoint{0.503017in}{2.119119in}}%
\pgfpathcurveto{\pgfqpoint{0.537472in}{2.153574in}}{\pgfqpoint{0.556831in}{2.200311in}}{\pgfqpoint{0.556831in}{2.249038in}}%
\pgfpathcurveto{\pgfqpoint{0.556831in}{2.297764in}}{\pgfqpoint{0.537472in}{2.344502in}}{\pgfqpoint{0.503017in}{2.378956in}}%
\pgfpathcurveto{\pgfqpoint{0.468562in}{2.413411in}}{\pgfqpoint{0.421825in}{2.432771in}}{\pgfqpoint{0.373099in}{2.432771in}}%
\pgfpathcurveto{\pgfqpoint{0.324372in}{2.432771in}}{\pgfqpoint{0.277635in}{2.413411in}}{\pgfqpoint{0.243180in}{2.378956in}}%
\pgfpathcurveto{\pgfqpoint{0.208725in}{2.344502in}}{\pgfqpoint{0.189366in}{2.297764in}}{\pgfqpoint{0.189366in}{2.249038in}}%
\pgfpathcurveto{\pgfqpoint{0.189366in}{2.200311in}}{\pgfqpoint{0.208725in}{2.153574in}}{\pgfqpoint{0.243180in}{2.119119in}}%
\pgfpathcurveto{\pgfqpoint{0.277635in}{2.084664in}}{\pgfqpoint{0.324372in}{2.065305in}}{\pgfqpoint{0.373099in}{2.065305in}}%
\pgfpathlineto{\pgfqpoint{0.373099in}{2.065305in}}%
\pgfpathclose%
\pgfusepath{stroke,fill}%
\end{pgfscope}%
\begin{pgfscope}%
\pgfpathrectangle{\pgfqpoint{0.100000in}{0.100000in}}{\pgfqpoint{3.400000in}{3.314317in}}%
\pgfusepath{clip}%
\pgfsetbuttcap%
\pgfsetroundjoin%
\definecolor{currentfill}{rgb}{0.960784,0.521569,0.094118}%
\pgfsetfillcolor{currentfill}%
\pgfsetlinewidth{1.003750pt}%
\definecolor{currentstroke}{rgb}{0.960784,0.521569,0.094118}%
\pgfsetstrokecolor{currentstroke}%
\pgfsetdash{}{0pt}%
\pgfpathmoveto{\pgfqpoint{1.093266in}{0.411109in}}%
\pgfpathcurveto{\pgfqpoint{1.141992in}{0.411109in}}{\pgfqpoint{1.188730in}{0.430468in}}{\pgfqpoint{1.223184in}{0.464923in}}%
\pgfpathcurveto{\pgfqpoint{1.257639in}{0.499378in}}{\pgfqpoint{1.276998in}{0.546115in}}{\pgfqpoint{1.276998in}{0.594842in}}%
\pgfpathcurveto{\pgfqpoint{1.276998in}{0.643568in}}{\pgfqpoint{1.257639in}{0.690306in}}{\pgfqpoint{1.223184in}{0.724761in}}%
\pgfpathcurveto{\pgfqpoint{1.188730in}{0.759215in}}{\pgfqpoint{1.141992in}{0.778575in}}{\pgfqpoint{1.093266in}{0.778575in}}%
\pgfpathcurveto{\pgfqpoint{1.044539in}{0.778575in}}{\pgfqpoint{0.997802in}{0.759215in}}{\pgfqpoint{0.963347in}{0.724761in}}%
\pgfpathcurveto{\pgfqpoint{0.928892in}{0.690306in}}{\pgfqpoint{0.909533in}{0.643568in}}{\pgfqpoint{0.909533in}{0.594842in}}%
\pgfpathcurveto{\pgfqpoint{0.909533in}{0.546115in}}{\pgfqpoint{0.928892in}{0.499378in}}{\pgfqpoint{0.963347in}{0.464923in}}%
\pgfpathcurveto{\pgfqpoint{0.997802in}{0.430468in}}{\pgfqpoint{1.044539in}{0.411109in}}{\pgfqpoint{1.093266in}{0.411109in}}%
\pgfpathlineto{\pgfqpoint{1.093266in}{0.411109in}}%
\pgfpathclose%
\pgfusepath{stroke,fill}%
\end{pgfscope}%
\begin{pgfscope}%
\pgfpathrectangle{\pgfqpoint{0.100000in}{0.100000in}}{\pgfqpoint{3.400000in}{3.314317in}}%
\pgfusepath{clip}%
\pgfsetbuttcap%
\pgfsetroundjoin%
\definecolor{currentfill}{rgb}{0.298039,0.470588,0.658824}%
\pgfsetfillcolor{currentfill}%
\pgfsetlinewidth{1.003750pt}%
\definecolor{currentstroke}{rgb}{0.298039,0.470588,0.658824}%
\pgfsetstrokecolor{currentstroke}%
\pgfsetdash{}{0pt}%
\pgfpathmoveto{\pgfqpoint{2.395865in}{0.964393in}}%
\pgfpathcurveto{\pgfqpoint{2.444592in}{0.964393in}}{\pgfqpoint{2.491329in}{0.983752in}}{\pgfqpoint{2.525784in}{1.018207in}}%
\pgfpathcurveto{\pgfqpoint{2.560239in}{1.052661in}}{\pgfqpoint{2.579598in}{1.099399in}}{\pgfqpoint{2.579598in}{1.148125in}}%
\pgfpathcurveto{\pgfqpoint{2.579598in}{1.196852in}}{\pgfqpoint{2.560239in}{1.243589in}}{\pgfqpoint{2.525784in}{1.278044in}}%
\pgfpathcurveto{\pgfqpoint{2.491329in}{1.312499in}}{\pgfqpoint{2.444592in}{1.331858in}}{\pgfqpoint{2.395865in}{1.331858in}}%
\pgfpathcurveto{\pgfqpoint{2.347139in}{1.331858in}}{\pgfqpoint{2.300402in}{1.312499in}}{\pgfqpoint{2.265947in}{1.278044in}}%
\pgfpathcurveto{\pgfqpoint{2.231492in}{1.243589in}}{\pgfqpoint{2.212133in}{1.196852in}}{\pgfqpoint{2.212133in}{1.148125in}}%
\pgfpathcurveto{\pgfqpoint{2.212133in}{1.099399in}}{\pgfqpoint{2.231492in}{1.052661in}}{\pgfqpoint{2.265947in}{1.018207in}}%
\pgfpathcurveto{\pgfqpoint{2.300402in}{0.983752in}}{\pgfqpoint{2.347139in}{0.964393in}}{\pgfqpoint{2.395865in}{0.964393in}}%
\pgfpathlineto{\pgfqpoint{2.395865in}{0.964393in}}%
\pgfpathclose%
\pgfusepath{stroke,fill}%
\end{pgfscope}%
\begin{pgfscope}%
\pgfpathrectangle{\pgfqpoint{0.100000in}{0.100000in}}{\pgfqpoint{3.400000in}{3.314317in}}%
\pgfusepath{clip}%
\pgfsetbuttcap%
\pgfsetroundjoin%
\definecolor{currentfill}{rgb}{0.298039,0.470588,0.658824}%
\pgfsetfillcolor{currentfill}%
\pgfsetlinewidth{1.003750pt}%
\definecolor{currentstroke}{rgb}{0.298039,0.470588,0.658824}%
\pgfsetstrokecolor{currentstroke}%
\pgfsetdash{}{0pt}%
\pgfpathmoveto{\pgfqpoint{2.658008in}{0.591648in}}%
\pgfpathlineto{\pgfqpoint{3.025473in}{0.591648in}}%
\pgfpathlineto{\pgfqpoint{3.025473in}{0.959113in}}%
\pgfpathlineto{\pgfqpoint{2.658008in}{0.959113in}}%
\pgfpathlineto{\pgfqpoint{2.658008in}{0.591648in}}%
\pgfpathclose%
\pgfusepath{stroke,fill}%
\end{pgfscope}%
\begin{pgfscope}%
\pgfpathrectangle{\pgfqpoint{0.100000in}{0.100000in}}{\pgfqpoint{3.400000in}{3.314317in}}%
\pgfusepath{clip}%
\pgfsetbuttcap%
\pgfsetroundjoin%
\definecolor{currentfill}{rgb}{0.960784,0.521569,0.094118}%
\pgfsetfillcolor{currentfill}%
\pgfsetlinewidth{1.003750pt}%
\definecolor{currentstroke}{rgb}{0.960784,0.521569,0.094118}%
\pgfsetstrokecolor{currentstroke}%
\pgfsetdash{}{0pt}%
\pgfpathmoveto{\pgfqpoint{2.560430in}{2.474565in}}%
\pgfpathlineto{\pgfqpoint{2.927895in}{2.474565in}}%
\pgfpathlineto{\pgfqpoint{2.927895in}{2.842030in}}%
\pgfpathlineto{\pgfqpoint{2.560430in}{2.842030in}}%
\pgfpathlineto{\pgfqpoint{2.560430in}{2.474565in}}%
\pgfpathclose%
\pgfusepath{stroke,fill}%
\end{pgfscope}%
\begin{pgfscope}%
\definecolor{textcolor}{rgb}{0.000000,0.000000,0.000000}%
\pgfsetstrokecolor{textcolor}%
\pgfsetfillcolor{textcolor}%
\pgftext[x=3.238207in,y=1.750135in,,]{\color{textcolor}\sffamily\fontsize{19.000000}{22.800000}\selectfont \(\displaystyle S_{1}\)}%
\end{pgfscope}%
\begin{pgfscope}%
\definecolor{textcolor}{rgb}{0.000000,0.000000,0.000000}%
\pgfsetstrokecolor{textcolor}%
\pgfsetfillcolor{textcolor}%
\pgftext[x=1.853024in,y=0.773356in,,]{\color{textcolor}\sffamily\fontsize{19.000000}{22.800000}\selectfont \(\displaystyle S_{2}\)}%
\end{pgfscope}%
\begin{pgfscope}%
\definecolor{textcolor}{rgb}{0.000000,0.000000,0.000000}%
\pgfsetstrokecolor{textcolor}%
\pgfsetfillcolor{textcolor}%
\pgftext[x=1.902667in,y=3.136579in,,]{\color{textcolor}\sffamily\fontsize{19.000000}{22.800000}\selectfont \(\displaystyle S_{3}\)}%
\end{pgfscope}%
\begin{pgfscope}%
\definecolor{textcolor}{rgb}{0.000000,0.000000,0.000000}%
\pgfsetstrokecolor{textcolor}%
\pgfsetfillcolor{textcolor}%
\pgftext[x=2.659242in,y=1.954744in,,]{\color{textcolor}\sffamily\fontsize{19.000000}{22.800000}\selectfont \(\displaystyle S_{4}\)}%
\end{pgfscope}%
\begin{pgfscope}%
\definecolor{textcolor}{rgb}{0.000000,0.000000,0.000000}%
\pgfsetstrokecolor{textcolor}%
\pgfsetfillcolor{textcolor}%
\pgftext[x=2.080565in,y=0.322055in,,]{\color{textcolor}\sffamily\fontsize{19.000000}{22.800000}\selectfont \(\displaystyle S_{5}\)}%
\end{pgfscope}%
\begin{pgfscope}%
\definecolor{textcolor}{rgb}{0.000000,0.000000,0.000000}%
\pgfsetstrokecolor{textcolor}%
\pgfsetfillcolor{textcolor}%
\pgftext[x=0.438926in,y=1.300531in,,]{\color{textcolor}\sffamily\fontsize{19.000000}{22.800000}\selectfont \(\displaystyle S_{6}\)}%
\end{pgfscope}%
\begin{pgfscope}%
\definecolor{textcolor}{rgb}{0.000000,0.000000,0.000000}%
\pgfsetstrokecolor{textcolor}%
\pgfsetfillcolor{textcolor}%
\pgftext[x=0.963016in,y=2.990663in,,]{\color{textcolor}\sffamily\fontsize{19.000000}{22.800000}\selectfont \(\displaystyle S_{7}\)}%
\end{pgfscope}%
\begin{pgfscope}%
\definecolor{textcolor}{rgb}{0.000000,0.000000,0.000000}%
\pgfsetstrokecolor{textcolor}%
\pgfsetfillcolor{textcolor}%
\pgftext[x=0.373099in,y=2.249038in,,]{\color{textcolor}\sffamily\fontsize{19.000000}{22.800000}\selectfont \(\displaystyle S_{8}\)}%
\end{pgfscope}%
\begin{pgfscope}%
\definecolor{textcolor}{rgb}{0.000000,0.000000,0.000000}%
\pgfsetstrokecolor{textcolor}%
\pgfsetfillcolor{textcolor}%
\pgftext[x=1.093266in,y=0.594842in,,]{\color{textcolor}\sffamily\fontsize{19.000000}{22.800000}\selectfont \(\displaystyle S_{9}\)}%
\end{pgfscope}%
\begin{pgfscope}%
\definecolor{textcolor}{rgb}{0.000000,0.000000,0.000000}%
\pgfsetstrokecolor{textcolor}%
\pgfsetfillcolor{textcolor}%
\pgftext[x=2.395865in,y=1.148125in,,]{\color{textcolor}\sffamily\fontsize{19.000000}{22.800000}\selectfont \(\displaystyle S_{10}\)}%
\end{pgfscope}%
\begin{pgfscope}%
\definecolor{textcolor}{rgb}{0.000000,0.000000,0.000000}%
\pgfsetstrokecolor{textcolor}%
\pgfsetfillcolor{textcolor}%
\pgftext[x=2.841741in,y=0.775380in,,]{\color{textcolor}\sffamily\fontsize{19.000000}{22.800000}\selectfont \(\displaystyle P_{11}\)}%
\end{pgfscope}%
\begin{pgfscope}%
\definecolor{textcolor}{rgb}{0.000000,0.000000,0.000000}%
\pgfsetstrokecolor{textcolor}%
\pgfsetfillcolor{textcolor}%
\pgftext[x=2.744162in,y=2.658297in,,]{\color{textcolor}\sffamily\fontsize{19.000000}{22.800000}\selectfont \(\displaystyle P_{12}\)}%
\end{pgfscope}%
\end{pgfpicture}%
\makeatother%
\endgroup%

%% file: pgf_example/02_single_plot.tex
\begingroup%
\makeatletter%
\begin{pgfpicture}%
\pgfpathrectangle{\pgfpointorigin}{\pgfqpoint{3.600000in}{3.539972in}}%
\pgfusepath{use as bounding box, clip}%
\begin{pgfscope}%
\pgfsetbuttcap%
\pgfsetmiterjoin%
\definecolor{currentfill}{rgb}{1.000000,1.000000,1.000000}%
\pgfsetfillcolor{currentfill}%
\pgfsetlinewidth{0.000000pt}%
\definecolor{currentstroke}{rgb}{1.000000,1.000000,1.000000}%
\pgfsetstrokecolor{currentstroke}%
\pgfsetdash{}{0pt}%
\pgfpathmoveto{\pgfqpoint{0.000000in}{0.000000in}}%
\pgfpathlineto{\pgfqpoint{3.600000in}{0.000000in}}%
\pgfpathlineto{\pgfqpoint{3.600000in}{3.539972in}}%
\pgfpathlineto{\pgfqpoint{0.000000in}{3.539972in}}%
\pgfpathlineto{\pgfqpoint{0.000000in}{0.000000in}}%
\pgfpathclose%
\pgfusepath{fill}%
\end{pgfscope}%
\begin{pgfscope}%
\pgfpathrectangle{\pgfqpoint{0.100000in}{0.100000in}}{\pgfqpoint{3.400000in}{3.339972in}}%
\pgfusepath{clip}%
\pgfsetbuttcap%
\pgfsetroundjoin%
\pgfsetlinewidth{2.007500pt}%
\definecolor{currentstroke}{rgb}{0.619608,0.619608,0.619608}%
\pgfsetstrokecolor{currentstroke}%
\pgfsetdash{}{0pt}%
\pgfpathmoveto{\pgfqpoint{2.378843in}{1.199990in}}%
\pgfpathlineto{\pgfqpoint{2.816129in}{0.834425in}}%
\pgfusepath{stroke}%
\end{pgfscope}%
\begin{pgfscope}%
\pgfpathrectangle{\pgfqpoint{0.100000in}{0.100000in}}{\pgfqpoint{3.400000in}{3.339972in}}%
\pgfusepath{clip}%
\pgfsetbuttcap%
\pgfsetroundjoin%
\pgfsetlinewidth{2.007500pt}%
\definecolor{currentstroke}{rgb}{0.619608,0.619608,0.619608}%
\pgfsetstrokecolor{currentstroke}%
\pgfsetdash{}{0pt}%
\pgfpathmoveto{\pgfqpoint{3.204959in}{1.790403in}}%
\pgfpathlineto{\pgfqpoint{2.816129in}{0.834425in}}%
\pgfusepath{stroke}%
\end{pgfscope}%
\begin{pgfscope}%
\pgfpathrectangle{\pgfqpoint{0.100000in}{0.100000in}}{\pgfqpoint{3.400000in}{3.339972in}}%
\pgfusepath{clip}%
\pgfsetbuttcap%
\pgfsetroundjoin%
\pgfsetlinewidth{2.007500pt}%
\definecolor{currentstroke}{rgb}{0.960784,0.521569,0.094118}%
\pgfsetstrokecolor{currentstroke}%
\pgfsetdash{}{0pt}%
\pgfpathmoveto{\pgfqpoint{2.069616in}{0.389832in}}%
\pgfpathlineto{\pgfqpoint{2.378843in}{1.199990in}}%
\pgfusepath{stroke}%
\end{pgfscope}%
\begin{pgfscope}%
\pgfpathrectangle{\pgfqpoint{0.100000in}{0.100000in}}{\pgfqpoint{3.400000in}{3.339972in}}%
\pgfusepath{clip}%
\pgfsetbuttcap%
\pgfsetroundjoin%
\pgfsetlinewidth{2.007500pt}%
\definecolor{currentstroke}{rgb}{0.960784,0.521569,0.094118}%
\pgfsetstrokecolor{currentstroke}%
\pgfsetdash{}{0pt}%
\pgfpathmoveto{\pgfqpoint{1.846458in}{0.832440in}}%
\pgfpathlineto{\pgfqpoint{2.069616in}{0.389832in}}%
\pgfusepath{stroke}%
\end{pgfscope}%
\begin{pgfscope}%
\pgfpathrectangle{\pgfqpoint{0.100000in}{0.100000in}}{\pgfqpoint{3.400000in}{3.339972in}}%
\pgfusepath{clip}%
\pgfsetbuttcap%
\pgfsetroundjoin%
\pgfsetlinewidth{2.007500pt}%
\definecolor{currentstroke}{rgb}{0.960784,0.521569,0.094118}%
\pgfsetstrokecolor{currentstroke}%
\pgfsetdash{}{0pt}%
\pgfpathmoveto{\pgfqpoint{1.846458in}{0.832440in}}%
\pgfpathlineto{\pgfqpoint{1.101336in}{0.657365in}}%
\pgfusepath{stroke}%
\end{pgfscope}%
\begin{pgfscope}%
\pgfpathrectangle{\pgfqpoint{0.100000in}{0.100000in}}{\pgfqpoint{3.400000in}{3.339972in}}%
\pgfusepath{clip}%
\pgfsetbuttcap%
\pgfsetroundjoin%
\pgfsetlinewidth{2.007500pt}%
\definecolor{currentstroke}{rgb}{0.960784,0.521569,0.094118}%
\pgfsetstrokecolor{currentstroke}%
\pgfsetdash{}{0pt}%
\pgfpathmoveto{\pgfqpoint{0.459600in}{1.349459in}}%
\pgfpathlineto{\pgfqpoint{1.101336in}{0.657365in}}%
\pgfusepath{stroke}%
\end{pgfscope}%
\begin{pgfscope}%
\pgfpathrectangle{\pgfqpoint{0.100000in}{0.100000in}}{\pgfqpoint{3.400000in}{3.339972in}}%
\pgfusepath{clip}%
\pgfsetbuttcap%
\pgfsetroundjoin%
\pgfsetlinewidth{2.007500pt}%
\definecolor{currentstroke}{rgb}{0.960784,0.521569,0.094118}%
\pgfsetstrokecolor{currentstroke}%
\pgfsetdash{}{0pt}%
\pgfpathmoveto{\pgfqpoint{0.459600in}{1.349459in}}%
\pgfpathlineto{\pgfqpoint{0.395041in}{2.279695in}}%
\pgfusepath{stroke}%
\end{pgfscope}%
\begin{pgfscope}%
\pgfpathrectangle{\pgfqpoint{0.100000in}{0.100000in}}{\pgfqpoint{3.400000in}{3.339972in}}%
\pgfusepath{clip}%
\pgfsetbuttcap%
\pgfsetroundjoin%
\pgfsetlinewidth{2.007500pt}%
\definecolor{currentstroke}{rgb}{0.960784,0.521569,0.094118}%
\pgfsetstrokecolor{currentstroke}%
\pgfsetdash{}{0pt}%
\pgfpathmoveto{\pgfqpoint{0.973595in}{3.007034in}}%
\pgfpathlineto{\pgfqpoint{0.395041in}{2.279695in}}%
\pgfusepath{stroke}%
\end{pgfscope}%
\begin{pgfscope}%
\pgfpathrectangle{\pgfqpoint{0.100000in}{0.100000in}}{\pgfqpoint{3.400000in}{3.339972in}}%
\pgfusepath{clip}%
\pgfsetbuttcap%
\pgfsetroundjoin%
\pgfsetlinewidth{2.007500pt}%
\definecolor{currentstroke}{rgb}{0.960784,0.521569,0.094118}%
\pgfsetstrokecolor{currentstroke}%
\pgfsetdash{}{0pt}%
\pgfpathmoveto{\pgfqpoint{1.895146in}{3.150140in}}%
\pgfpathlineto{\pgfqpoint{0.973595in}{3.007034in}}%
\pgfusepath{stroke}%
\end{pgfscope}%
\begin{pgfscope}%
\pgfpathrectangle{\pgfqpoint{0.100000in}{0.100000in}}{\pgfqpoint{3.400000in}{3.339972in}}%
\pgfusepath{clip}%
\pgfsetbuttcap%
\pgfsetroundjoin%
\pgfsetlinewidth{2.007500pt}%
\definecolor{currentstroke}{rgb}{0.960784,0.521569,0.094118}%
\pgfsetstrokecolor{currentstroke}%
\pgfsetdash{}{0pt}%
\pgfpathmoveto{\pgfqpoint{1.895146in}{3.150140in}}%
\pgfpathlineto{\pgfqpoint{2.720431in}{2.681071in}}%
\pgfusepath{stroke}%
\end{pgfscope}%
\begin{pgfscope}%
\pgfpathrectangle{\pgfqpoint{0.100000in}{0.100000in}}{\pgfqpoint{3.400000in}{3.339972in}}%
\pgfusepath{clip}%
\pgfsetbuttcap%
\pgfsetroundjoin%
\pgfsetlinewidth{2.007500pt}%
\definecolor{currentstroke}{rgb}{0.960784,0.521569,0.094118}%
\pgfsetstrokecolor{currentstroke}%
\pgfsetdash{}{0pt}%
\pgfpathmoveto{\pgfqpoint{3.204959in}{1.790403in}}%
\pgfpathlineto{\pgfqpoint{2.637146in}{1.991070in}}%
\pgfusepath{stroke}%
\end{pgfscope}%
\begin{pgfscope}%
\pgfpathrectangle{\pgfqpoint{0.100000in}{0.100000in}}{\pgfqpoint{3.400000in}{3.339972in}}%
\pgfusepath{clip}%
\pgfsetbuttcap%
\pgfsetroundjoin%
\pgfsetlinewidth{2.007500pt}%
\definecolor{currentstroke}{rgb}{0.960784,0.521569,0.094118}%
\pgfsetstrokecolor{currentstroke}%
\pgfsetdash{}{0pt}%
\pgfpathmoveto{\pgfqpoint{2.637146in}{1.991070in}}%
\pgfpathlineto{\pgfqpoint{2.720431in}{2.681071in}}%
\pgfusepath{stroke}%
\end{pgfscope}%
\begin{pgfscope}%
\pgfpathrectangle{\pgfqpoint{0.100000in}{0.100000in}}{\pgfqpoint{3.400000in}{3.339972in}}%
\pgfusepath{clip}%
\pgfsetbuttcap%
\pgfsetroundjoin%
\definecolor{currentfill}{rgb}{0.960784,0.521569,0.094118}%
\pgfsetfillcolor{currentfill}%
\pgfsetlinewidth{1.003750pt}%
\definecolor{currentstroke}{rgb}{0.960784,0.521569,0.094118}%
\pgfsetstrokecolor{currentstroke}%
\pgfsetdash{}{0pt}%
\pgfpathmoveto{\pgfqpoint{3.204959in}{1.606671in}}%
\pgfpathcurveto{\pgfqpoint{3.253685in}{1.606671in}}{\pgfqpoint{3.300423in}{1.626030in}}{\pgfqpoint{3.334877in}{1.660485in}}%
\pgfpathcurveto{\pgfqpoint{3.369332in}{1.694939in}}{\pgfqpoint{3.388691in}{1.741677in}}{\pgfqpoint{3.388691in}{1.790403in}}%
\pgfpathcurveto{\pgfqpoint{3.388691in}{1.839130in}}{\pgfqpoint{3.369332in}{1.885867in}}{\pgfqpoint{3.334877in}{1.920322in}}%
\pgfpathcurveto{\pgfqpoint{3.300423in}{1.954777in}}{\pgfqpoint{3.253685in}{1.974136in}}{\pgfqpoint{3.204959in}{1.974136in}}%
\pgfpathcurveto{\pgfqpoint{3.156232in}{1.974136in}}{\pgfqpoint{3.109495in}{1.954777in}}{\pgfqpoint{3.075040in}{1.920322in}}%
\pgfpathcurveto{\pgfqpoint{3.040585in}{1.885867in}}{\pgfqpoint{3.021226in}{1.839130in}}{\pgfqpoint{3.021226in}{1.790403in}}%
\pgfpathcurveto{\pgfqpoint{3.021226in}{1.741677in}}{\pgfqpoint{3.040585in}{1.694939in}}{\pgfqpoint{3.075040in}{1.660485in}}%
\pgfpathcurveto{\pgfqpoint{3.109495in}{1.626030in}}{\pgfqpoint{3.156232in}{1.606671in}}{\pgfqpoint{3.204959in}{1.606671in}}%
\pgfpathlineto{\pgfqpoint{3.204959in}{1.606671in}}%
\pgfpathclose%
\pgfusepath{stroke,fill}%
\end{pgfscope}%
\begin{pgfscope}%
\pgfpathrectangle{\pgfqpoint{0.100000in}{0.100000in}}{\pgfqpoint{3.400000in}{3.339972in}}%
\pgfusepath{clip}%
\pgfsetbuttcap%
\pgfsetroundjoin%
\definecolor{currentfill}{rgb}{0.960784,0.521569,0.094118}%
\pgfsetfillcolor{currentfill}%
\pgfsetlinewidth{1.003750pt}%
\definecolor{currentstroke}{rgb}{0.960784,0.521569,0.094118}%
\pgfsetstrokecolor{currentstroke}%
\pgfsetdash{}{0pt}%
\pgfpathmoveto{\pgfqpoint{1.846458in}{0.648707in}}%
\pgfpathcurveto{\pgfqpoint{1.895185in}{0.648707in}}{\pgfqpoint{1.941922in}{0.668066in}}{\pgfqpoint{1.976377in}{0.702521in}}%
\pgfpathcurveto{\pgfqpoint{2.010832in}{0.736976in}}{\pgfqpoint{2.030191in}{0.783713in}}{\pgfqpoint{2.030191in}{0.832440in}}%
\pgfpathcurveto{\pgfqpoint{2.030191in}{0.881166in}}{\pgfqpoint{2.010832in}{0.927903in}}{\pgfqpoint{1.976377in}{0.962358in}}%
\pgfpathcurveto{\pgfqpoint{1.941922in}{0.996813in}}{\pgfqpoint{1.895185in}{1.016172in}}{\pgfqpoint{1.846458in}{1.016172in}}%
\pgfpathcurveto{\pgfqpoint{1.797732in}{1.016172in}}{\pgfqpoint{1.750994in}{0.996813in}}{\pgfqpoint{1.716539in}{0.962358in}}%
\pgfpathcurveto{\pgfqpoint{1.682085in}{0.927903in}}{\pgfqpoint{1.662725in}{0.881166in}}{\pgfqpoint{1.662725in}{0.832440in}}%
\pgfpathcurveto{\pgfqpoint{1.662725in}{0.783713in}}{\pgfqpoint{1.682085in}{0.736976in}}{\pgfqpoint{1.716539in}{0.702521in}}%
\pgfpathcurveto{\pgfqpoint{1.750994in}{0.668066in}}{\pgfqpoint{1.797732in}{0.648707in}}{\pgfqpoint{1.846458in}{0.648707in}}%
\pgfpathlineto{\pgfqpoint{1.846458in}{0.648707in}}%
\pgfpathclose%
\pgfusepath{stroke,fill}%
\end{pgfscope}%
\begin{pgfscope}%
\pgfpathrectangle{\pgfqpoint{0.100000in}{0.100000in}}{\pgfqpoint{3.400000in}{3.339972in}}%
\pgfusepath{clip}%
\pgfsetbuttcap%
\pgfsetroundjoin%
\definecolor{currentfill}{rgb}{0.960784,0.521569,0.094118}%
\pgfsetfillcolor{currentfill}%
\pgfsetlinewidth{1.003750pt}%
\definecolor{currentstroke}{rgb}{0.960784,0.521569,0.094118}%
\pgfsetstrokecolor{currentstroke}%
\pgfsetdash{}{0pt}%
\pgfpathmoveto{\pgfqpoint{1.895146in}{2.966407in}}%
\pgfpathcurveto{\pgfqpoint{1.943872in}{2.966407in}}{\pgfqpoint{1.990609in}{2.985766in}}{\pgfqpoint{2.025064in}{3.020221in}}%
\pgfpathcurveto{\pgfqpoint{2.059519in}{3.054676in}}{\pgfqpoint{2.078878in}{3.101413in}}{\pgfqpoint{2.078878in}{3.150140in}}%
\pgfpathcurveto{\pgfqpoint{2.078878in}{3.198866in}}{\pgfqpoint{2.059519in}{3.245603in}}{\pgfqpoint{2.025064in}{3.280058in}}%
\pgfpathcurveto{\pgfqpoint{1.990609in}{3.314513in}}{\pgfqpoint{1.943872in}{3.333872in}}{\pgfqpoint{1.895146in}{3.333872in}}%
\pgfpathcurveto{\pgfqpoint{1.846419in}{3.333872in}}{\pgfqpoint{1.799682in}{3.314513in}}{\pgfqpoint{1.765227in}{3.280058in}}%
\pgfpathcurveto{\pgfqpoint{1.730772in}{3.245603in}}{\pgfqpoint{1.711413in}{3.198866in}}{\pgfqpoint{1.711413in}{3.150140in}}%
\pgfpathcurveto{\pgfqpoint{1.711413in}{3.101413in}}{\pgfqpoint{1.730772in}{3.054676in}}{\pgfqpoint{1.765227in}{3.020221in}}%
\pgfpathcurveto{\pgfqpoint{1.799682in}{2.985766in}}{\pgfqpoint{1.846419in}{2.966407in}}{\pgfqpoint{1.895146in}{2.966407in}}%
\pgfpathlineto{\pgfqpoint{1.895146in}{2.966407in}}%
\pgfpathclose%
\pgfusepath{stroke,fill}%
\end{pgfscope}%
\begin{pgfscope}%
\pgfpathrectangle{\pgfqpoint{0.100000in}{0.100000in}}{\pgfqpoint{3.400000in}{3.339972in}}%
\pgfusepath{clip}%
\pgfsetbuttcap%
\pgfsetroundjoin%
\definecolor{currentfill}{rgb}{0.960784,0.521569,0.094118}%
\pgfsetfillcolor{currentfill}%
\pgfsetlinewidth{1.003750pt}%
\definecolor{currentstroke}{rgb}{0.960784,0.521569,0.094118}%
\pgfsetstrokecolor{currentstroke}%
\pgfsetdash{}{0pt}%
\pgfpathmoveto{\pgfqpoint{2.637146in}{1.807337in}}%
\pgfpathcurveto{\pgfqpoint{2.685873in}{1.807337in}}{\pgfqpoint{2.732610in}{1.826697in}}{\pgfqpoint{2.767065in}{1.861151in}}%
\pgfpathcurveto{\pgfqpoint{2.801520in}{1.895606in}}{\pgfqpoint{2.820879in}{1.942344in}}{\pgfqpoint{2.820879in}{1.991070in}}%
\pgfpathcurveto{\pgfqpoint{2.820879in}{2.039797in}}{\pgfqpoint{2.801520in}{2.086534in}}{\pgfqpoint{2.767065in}{2.120989in}}%
\pgfpathcurveto{\pgfqpoint{2.732610in}{2.155444in}}{\pgfqpoint{2.685873in}{2.174803in}}{\pgfqpoint{2.637146in}{2.174803in}}%
\pgfpathcurveto{\pgfqpoint{2.588420in}{2.174803in}}{\pgfqpoint{2.541683in}{2.155444in}}{\pgfqpoint{2.507228in}{2.120989in}}%
\pgfpathcurveto{\pgfqpoint{2.472773in}{2.086534in}}{\pgfqpoint{2.453414in}{2.039797in}}{\pgfqpoint{2.453414in}{1.991070in}}%
\pgfpathcurveto{\pgfqpoint{2.453414in}{1.942344in}}{\pgfqpoint{2.472773in}{1.895606in}}{\pgfqpoint{2.507228in}{1.861151in}}%
\pgfpathcurveto{\pgfqpoint{2.541683in}{1.826697in}}{\pgfqpoint{2.588420in}{1.807337in}}{\pgfqpoint{2.637146in}{1.807337in}}%
\pgfpathlineto{\pgfqpoint{2.637146in}{1.807337in}}%
\pgfpathclose%
\pgfusepath{stroke,fill}%
\end{pgfscope}%
\begin{pgfscope}%
\pgfpathrectangle{\pgfqpoint{0.100000in}{0.100000in}}{\pgfqpoint{3.400000in}{3.339972in}}%
\pgfusepath{clip}%
\pgfsetbuttcap%
\pgfsetroundjoin%
\definecolor{currentfill}{rgb}{0.960784,0.521569,0.094118}%
\pgfsetfillcolor{currentfill}%
\pgfsetlinewidth{1.003750pt}%
\definecolor{currentstroke}{rgb}{0.960784,0.521569,0.094118}%
\pgfsetstrokecolor{currentstroke}%
\pgfsetdash{}{0pt}%
\pgfpathmoveto{\pgfqpoint{2.069616in}{0.206100in}}%
\pgfpathcurveto{\pgfqpoint{2.118342in}{0.206100in}}{\pgfqpoint{2.165080in}{0.225459in}}{\pgfqpoint{2.199535in}{0.259914in}}%
\pgfpathcurveto{\pgfqpoint{2.233989in}{0.294368in}}{\pgfqpoint{2.253349in}{0.341106in}}{\pgfqpoint{2.253349in}{0.389832in}}%
\pgfpathcurveto{\pgfqpoint{2.253349in}{0.438559in}}{\pgfqpoint{2.233989in}{0.485296in}}{\pgfqpoint{2.199535in}{0.519751in}}%
\pgfpathcurveto{\pgfqpoint{2.165080in}{0.554206in}}{\pgfqpoint{2.118342in}{0.573565in}}{\pgfqpoint{2.069616in}{0.573565in}}%
\pgfpathcurveto{\pgfqpoint{2.020889in}{0.573565in}}{\pgfqpoint{1.974152in}{0.554206in}}{\pgfqpoint{1.939697in}{0.519751in}}%
\pgfpathcurveto{\pgfqpoint{1.905242in}{0.485296in}}{\pgfqpoint{1.885883in}{0.438559in}}{\pgfqpoint{1.885883in}{0.389832in}}%
\pgfpathcurveto{\pgfqpoint{1.885883in}{0.341106in}}{\pgfqpoint{1.905242in}{0.294368in}}{\pgfqpoint{1.939697in}{0.259914in}}%
\pgfpathcurveto{\pgfqpoint{1.974152in}{0.225459in}}{\pgfqpoint{2.020889in}{0.206100in}}{\pgfqpoint{2.069616in}{0.206100in}}%
\pgfpathlineto{\pgfqpoint{2.069616in}{0.206100in}}%
\pgfpathclose%
\pgfusepath{stroke,fill}%
\end{pgfscope}%
\begin{pgfscope}%
\pgfpathrectangle{\pgfqpoint{0.100000in}{0.100000in}}{\pgfqpoint{3.400000in}{3.339972in}}%
\pgfusepath{clip}%
\pgfsetbuttcap%
\pgfsetroundjoin%
\definecolor{currentfill}{rgb}{0.960784,0.521569,0.094118}%
\pgfsetfillcolor{currentfill}%
\pgfsetlinewidth{1.003750pt}%
\definecolor{currentstroke}{rgb}{0.960784,0.521569,0.094118}%
\pgfsetstrokecolor{currentstroke}%
\pgfsetdash{}{0pt}%
\pgfpathmoveto{\pgfqpoint{0.459600in}{1.165727in}}%
\pgfpathcurveto{\pgfqpoint{0.508327in}{1.165727in}}{\pgfqpoint{0.555064in}{1.185086in}}{\pgfqpoint{0.589519in}{1.219541in}}%
\pgfpathcurveto{\pgfqpoint{0.623974in}{1.253996in}}{\pgfqpoint{0.643333in}{1.300733in}}{\pgfqpoint{0.643333in}{1.349459in}}%
\pgfpathcurveto{\pgfqpoint{0.643333in}{1.398186in}}{\pgfqpoint{0.623974in}{1.444923in}}{\pgfqpoint{0.589519in}{1.479378in}}%
\pgfpathcurveto{\pgfqpoint{0.555064in}{1.513833in}}{\pgfqpoint{0.508327in}{1.533192in}}{\pgfqpoint{0.459600in}{1.533192in}}%
\pgfpathcurveto{\pgfqpoint{0.410874in}{1.533192in}}{\pgfqpoint{0.364136in}{1.513833in}}{\pgfqpoint{0.329682in}{1.479378in}}%
\pgfpathcurveto{\pgfqpoint{0.295227in}{1.444923in}}{\pgfqpoint{0.275868in}{1.398186in}}{\pgfqpoint{0.275868in}{1.349459in}}%
\pgfpathcurveto{\pgfqpoint{0.275868in}{1.300733in}}{\pgfqpoint{0.295227in}{1.253996in}}{\pgfqpoint{0.329682in}{1.219541in}}%
\pgfpathcurveto{\pgfqpoint{0.364136in}{1.185086in}}{\pgfqpoint{0.410874in}{1.165727in}}{\pgfqpoint{0.459600in}{1.165727in}}%
\pgfpathlineto{\pgfqpoint{0.459600in}{1.165727in}}%
\pgfpathclose%
\pgfusepath{stroke,fill}%
\end{pgfscope}%
\begin{pgfscope}%
\pgfpathrectangle{\pgfqpoint{0.100000in}{0.100000in}}{\pgfqpoint{3.400000in}{3.339972in}}%
\pgfusepath{clip}%
\pgfsetbuttcap%
\pgfsetroundjoin%
\definecolor{currentfill}{rgb}{0.960784,0.521569,0.094118}%
\pgfsetfillcolor{currentfill}%
\pgfsetlinewidth{1.003750pt}%
\definecolor{currentstroke}{rgb}{0.960784,0.521569,0.094118}%
\pgfsetstrokecolor{currentstroke}%
\pgfsetdash{}{0pt}%
\pgfpathmoveto{\pgfqpoint{0.973595in}{2.823301in}}%
\pgfpathcurveto{\pgfqpoint{1.022321in}{2.823301in}}{\pgfqpoint{1.069058in}{2.842661in}}{\pgfqpoint{1.103513in}{2.877115in}}%
\pgfpathcurveto{\pgfqpoint{1.137968in}{2.911570in}}{\pgfqpoint{1.157327in}{2.958308in}}{\pgfqpoint{1.157327in}{3.007034in}}%
\pgfpathcurveto{\pgfqpoint{1.157327in}{3.055761in}}{\pgfqpoint{1.137968in}{3.102498in}}{\pgfqpoint{1.103513in}{3.136953in}}%
\pgfpathcurveto{\pgfqpoint{1.069058in}{3.171408in}}{\pgfqpoint{1.022321in}{3.190767in}}{\pgfqpoint{0.973595in}{3.190767in}}%
\pgfpathcurveto{\pgfqpoint{0.924868in}{3.190767in}}{\pgfqpoint{0.878131in}{3.171408in}}{\pgfqpoint{0.843676in}{3.136953in}}%
\pgfpathcurveto{\pgfqpoint{0.809221in}{3.102498in}}{\pgfqpoint{0.789862in}{3.055761in}}{\pgfqpoint{0.789862in}{3.007034in}}%
\pgfpathcurveto{\pgfqpoint{0.789862in}{2.958308in}}{\pgfqpoint{0.809221in}{2.911570in}}{\pgfqpoint{0.843676in}{2.877115in}}%
\pgfpathcurveto{\pgfqpoint{0.878131in}{2.842661in}}{\pgfqpoint{0.924868in}{2.823301in}}{\pgfqpoint{0.973595in}{2.823301in}}%
\pgfpathlineto{\pgfqpoint{0.973595in}{2.823301in}}%
\pgfpathclose%
\pgfusepath{stroke,fill}%
\end{pgfscope}%
\begin{pgfscope}%
\pgfpathrectangle{\pgfqpoint{0.100000in}{0.100000in}}{\pgfqpoint{3.400000in}{3.339972in}}%
\pgfusepath{clip}%
\pgfsetbuttcap%
\pgfsetroundjoin%
\definecolor{currentfill}{rgb}{0.960784,0.521569,0.094118}%
\pgfsetfillcolor{currentfill}%
\pgfsetlinewidth{1.003750pt}%
\definecolor{currentstroke}{rgb}{0.960784,0.521569,0.094118}%
\pgfsetstrokecolor{currentstroke}%
\pgfsetdash{}{0pt}%
\pgfpathmoveto{\pgfqpoint{0.395041in}{2.095963in}}%
\pgfpathcurveto{\pgfqpoint{0.443768in}{2.095963in}}{\pgfqpoint{0.490505in}{2.115322in}}{\pgfqpoint{0.524960in}{2.149777in}}%
\pgfpathcurveto{\pgfqpoint{0.559415in}{2.184231in}}{\pgfqpoint{0.578774in}{2.230969in}}{\pgfqpoint{0.578774in}{2.279695in}}%
\pgfpathcurveto{\pgfqpoint{0.578774in}{2.328422in}}{\pgfqpoint{0.559415in}{2.375159in}}{\pgfqpoint{0.524960in}{2.409614in}}%
\pgfpathcurveto{\pgfqpoint{0.490505in}{2.444069in}}{\pgfqpoint{0.443768in}{2.463428in}}{\pgfqpoint{0.395041in}{2.463428in}}%
\pgfpathcurveto{\pgfqpoint{0.346315in}{2.463428in}}{\pgfqpoint{0.299577in}{2.444069in}}{\pgfqpoint{0.265123in}{2.409614in}}%
\pgfpathcurveto{\pgfqpoint{0.230668in}{2.375159in}}{\pgfqpoint{0.211309in}{2.328422in}}{\pgfqpoint{0.211309in}{2.279695in}}%
\pgfpathcurveto{\pgfqpoint{0.211309in}{2.230969in}}{\pgfqpoint{0.230668in}{2.184231in}}{\pgfqpoint{0.265123in}{2.149777in}}%
\pgfpathcurveto{\pgfqpoint{0.299577in}{2.115322in}}{\pgfqpoint{0.346315in}{2.095963in}}{\pgfqpoint{0.395041in}{2.095963in}}%
\pgfpathlineto{\pgfqpoint{0.395041in}{2.095963in}}%
\pgfpathclose%
\pgfusepath{stroke,fill}%
\end{pgfscope}%
\begin{pgfscope}%
\pgfpathrectangle{\pgfqpoint{0.100000in}{0.100000in}}{\pgfqpoint{3.400000in}{3.339972in}}%
\pgfusepath{clip}%
\pgfsetbuttcap%
\pgfsetroundjoin%
\definecolor{currentfill}{rgb}{0.960784,0.521569,0.094118}%
\pgfsetfillcolor{currentfill}%
\pgfsetlinewidth{1.003750pt}%
\definecolor{currentstroke}{rgb}{0.960784,0.521569,0.094118}%
\pgfsetstrokecolor{currentstroke}%
\pgfsetdash{}{0pt}%
\pgfpathmoveto{\pgfqpoint{1.101336in}{0.473632in}}%
\pgfpathcurveto{\pgfqpoint{1.150062in}{0.473632in}}{\pgfqpoint{1.196799in}{0.492991in}}{\pgfqpoint{1.231254in}{0.527446in}}%
\pgfpathcurveto{\pgfqpoint{1.265709in}{0.561901in}}{\pgfqpoint{1.285068in}{0.608638in}}{\pgfqpoint{1.285068in}{0.657365in}}%
\pgfpathcurveto{\pgfqpoint{1.285068in}{0.706091in}}{\pgfqpoint{1.265709in}{0.752828in}}{\pgfqpoint{1.231254in}{0.787283in}}%
\pgfpathcurveto{\pgfqpoint{1.196799in}{0.821738in}}{\pgfqpoint{1.150062in}{0.841097in}}{\pgfqpoint{1.101336in}{0.841097in}}%
\pgfpathcurveto{\pgfqpoint{1.052609in}{0.841097in}}{\pgfqpoint{1.005872in}{0.821738in}}{\pgfqpoint{0.971417in}{0.787283in}}%
\pgfpathcurveto{\pgfqpoint{0.936962in}{0.752828in}}{\pgfqpoint{0.917603in}{0.706091in}}{\pgfqpoint{0.917603in}{0.657365in}}%
\pgfpathcurveto{\pgfqpoint{0.917603in}{0.608638in}}{\pgfqpoint{0.936962in}{0.561901in}}{\pgfqpoint{0.971417in}{0.527446in}}%
\pgfpathcurveto{\pgfqpoint{1.005872in}{0.492991in}}{\pgfqpoint{1.052609in}{0.473632in}}{\pgfqpoint{1.101336in}{0.473632in}}%
\pgfpathlineto{\pgfqpoint{1.101336in}{0.473632in}}%
\pgfpathclose%
\pgfusepath{stroke,fill}%
\end{pgfscope}%
\begin{pgfscope}%
\pgfpathrectangle{\pgfqpoint{0.100000in}{0.100000in}}{\pgfqpoint{3.400000in}{3.339972in}}%
\pgfusepath{clip}%
\pgfsetbuttcap%
\pgfsetroundjoin%
\definecolor{currentfill}{rgb}{0.960784,0.521569,0.094118}%
\pgfsetfillcolor{currentfill}%
\pgfsetlinewidth{1.003750pt}%
\definecolor{currentstroke}{rgb}{0.960784,0.521569,0.094118}%
\pgfsetstrokecolor{currentstroke}%
\pgfsetdash{}{0pt}%
\pgfpathmoveto{\pgfqpoint{2.378843in}{1.016257in}}%
\pgfpathcurveto{\pgfqpoint{2.427569in}{1.016257in}}{\pgfqpoint{2.474307in}{1.035616in}}{\pgfqpoint{2.508762in}{1.070071in}}%
\pgfpathcurveto{\pgfqpoint{2.543216in}{1.104526in}}{\pgfqpoint{2.562576in}{1.151263in}}{\pgfqpoint{2.562576in}{1.199990in}}%
\pgfpathcurveto{\pgfqpoint{2.562576in}{1.248716in}}{\pgfqpoint{2.543216in}{1.295454in}}{\pgfqpoint{2.508762in}{1.329909in}}%
\pgfpathcurveto{\pgfqpoint{2.474307in}{1.364363in}}{\pgfqpoint{2.427569in}{1.383723in}}{\pgfqpoint{2.378843in}{1.383723in}}%
\pgfpathcurveto{\pgfqpoint{2.330116in}{1.383723in}}{\pgfqpoint{2.283379in}{1.364363in}}{\pgfqpoint{2.248924in}{1.329909in}}%
\pgfpathcurveto{\pgfqpoint{2.214469in}{1.295454in}}{\pgfqpoint{2.195110in}{1.248716in}}{\pgfqpoint{2.195110in}{1.199990in}}%
\pgfpathcurveto{\pgfqpoint{2.195110in}{1.151263in}}{\pgfqpoint{2.214469in}{1.104526in}}{\pgfqpoint{2.248924in}{1.070071in}}%
\pgfpathcurveto{\pgfqpoint{2.283379in}{1.035616in}}{\pgfqpoint{2.330116in}{1.016257in}}{\pgfqpoint{2.378843in}{1.016257in}}%
\pgfpathlineto{\pgfqpoint{2.378843in}{1.016257in}}%
\pgfpathclose%
\pgfusepath{stroke,fill}%
\end{pgfscope}%
\begin{pgfscope}%
\pgfpathrectangle{\pgfqpoint{0.100000in}{0.100000in}}{\pgfqpoint{3.400000in}{3.339972in}}%
\pgfusepath{clip}%
\pgfsetbuttcap%
\pgfsetroundjoin%
\definecolor{currentfill}{rgb}{0.619608,0.619608,0.619608}%
\pgfsetfillcolor{currentfill}%
\pgfsetlinewidth{1.003750pt}%
\definecolor{currentstroke}{rgb}{0.619608,0.619608,0.619608}%
\pgfsetstrokecolor{currentstroke}%
\pgfsetdash{}{0pt}%
\pgfpathmoveto{\pgfqpoint{2.632396in}{0.650692in}}%
\pgfpathlineto{\pgfqpoint{2.999862in}{0.650692in}}%
\pgfpathlineto{\pgfqpoint{2.999862in}{1.018158in}}%
\pgfpathlineto{\pgfqpoint{2.632396in}{1.018158in}}%
\pgfpathlineto{\pgfqpoint{2.632396in}{0.650692in}}%
\pgfpathclose%
\pgfusepath{stroke,fill}%
\end{pgfscope}%
\begin{pgfscope}%
\pgfpathrectangle{\pgfqpoint{0.100000in}{0.100000in}}{\pgfqpoint{3.400000in}{3.339972in}}%
\pgfusepath{clip}%
\pgfsetbuttcap%
\pgfsetroundjoin%
\definecolor{currentfill}{rgb}{0.960784,0.521569,0.094118}%
\pgfsetfillcolor{currentfill}%
\pgfsetlinewidth{1.003750pt}%
\definecolor{currentstroke}{rgb}{0.960784,0.521569,0.094118}%
\pgfsetstrokecolor{currentstroke}%
\pgfsetdash{}{0pt}%
\pgfpathmoveto{\pgfqpoint{2.536698in}{2.497338in}}%
\pgfpathlineto{\pgfqpoint{2.904163in}{2.497338in}}%
\pgfpathlineto{\pgfqpoint{2.904163in}{2.864804in}}%
\pgfpathlineto{\pgfqpoint{2.536698in}{2.864804in}}%
\pgfpathlineto{\pgfqpoint{2.536698in}{2.497338in}}%
\pgfpathclose%
\pgfusepath{stroke,fill}%
\end{pgfscope}%
\begin{pgfscope}%
\definecolor{textcolor}{rgb}{0.000000,0.000000,0.000000}%
\pgfsetstrokecolor{textcolor}%
\pgfsetfillcolor{textcolor}%
\pgftext[x=3.204959in,y=1.790403in,,]{\color{textcolor}\sffamily\fontsize{19.000000}{22.800000}\selectfont \(\displaystyle S_{1}\)}%
\end{pgfscope}%
\begin{pgfscope}%
\definecolor{textcolor}{rgb}{0.000000,0.000000,0.000000}%
\pgfsetstrokecolor{textcolor}%
\pgfsetfillcolor{textcolor}%
\pgftext[x=1.846458in,y=0.832440in,,]{\color{textcolor}\sffamily\fontsize{19.000000}{22.800000}\selectfont \(\displaystyle S_{2}\)}%
\end{pgfscope}%
\begin{pgfscope}%
\definecolor{textcolor}{rgb}{0.000000,0.000000,0.000000}%
\pgfsetstrokecolor{textcolor}%
\pgfsetfillcolor{textcolor}%
\pgftext[x=1.895146in,y=3.150140in,,]{\color{textcolor}\sffamily\fontsize{19.000000}{22.800000}\selectfont \(\displaystyle S_{3}\)}%
\end{pgfscope}%
\begin{pgfscope}%
\definecolor{textcolor}{rgb}{0.000000,0.000000,0.000000}%
\pgfsetstrokecolor{textcolor}%
\pgfsetfillcolor{textcolor}%
\pgftext[x=2.637146in,y=1.991070in,,]{\color{textcolor}\sffamily\fontsize{19.000000}{22.800000}\selectfont \(\displaystyle S_{4}\)}%
\end{pgfscope}%
\begin{pgfscope}%
\definecolor{textcolor}{rgb}{0.000000,0.000000,0.000000}%
\pgfsetstrokecolor{textcolor}%
\pgfsetfillcolor{textcolor}%
\pgftext[x=2.069616in,y=0.389832in,,]{\color{textcolor}\sffamily\fontsize{19.000000}{22.800000}\selectfont \(\displaystyle S_{5}\)}%
\end{pgfscope}%
\begin{pgfscope}%
\definecolor{textcolor}{rgb}{0.000000,0.000000,0.000000}%
\pgfsetstrokecolor{textcolor}%
\pgfsetfillcolor{textcolor}%
\pgftext[x=0.459600in,y=1.349459in,,]{\color{textcolor}\sffamily\fontsize{19.000000}{22.800000}\selectfont \(\displaystyle S_{6}\)}%
\end{pgfscope}%
\begin{pgfscope}%
\definecolor{textcolor}{rgb}{0.000000,0.000000,0.000000}%
\pgfsetstrokecolor{textcolor}%
\pgfsetfillcolor{textcolor}%
\pgftext[x=0.973595in,y=3.007034in,,]{\color{textcolor}\sffamily\fontsize{19.000000}{22.800000}\selectfont \(\displaystyle S_{7}\)}%
\end{pgfscope}%
\begin{pgfscope}%
\definecolor{textcolor}{rgb}{0.000000,0.000000,0.000000}%
\pgfsetstrokecolor{textcolor}%
\pgfsetfillcolor{textcolor}%
\pgftext[x=0.395041in,y=2.279695in,,]{\color{textcolor}\sffamily\fontsize{19.000000}{22.800000}\selectfont \(\displaystyle S_{8}\)}%
\end{pgfscope}%
\begin{pgfscope}%
\definecolor{textcolor}{rgb}{0.000000,0.000000,0.000000}%
\pgfsetstrokecolor{textcolor}%
\pgfsetfillcolor{textcolor}%
\pgftext[x=1.101336in,y=0.657365in,,]{\color{textcolor}\sffamily\fontsize{19.000000}{22.800000}\selectfont \(\displaystyle S_{9}\)}%
\end{pgfscope}%
\begin{pgfscope}%
\definecolor{textcolor}{rgb}{0.000000,0.000000,0.000000}%
\pgfsetstrokecolor{textcolor}%
\pgfsetfillcolor{textcolor}%
\pgftext[x=2.378843in,y=1.199990in,,]{\color{textcolor}\sffamily\fontsize{19.000000}{22.800000}\selectfont \(\displaystyle S_{10}\)}%
\end{pgfscope}%
\begin{pgfscope}%
\definecolor{textcolor}{rgb}{0.000000,0.000000,0.000000}%
\pgfsetstrokecolor{textcolor}%
\pgfsetfillcolor{textcolor}%
\pgftext[x=2.816129in,y=0.834425in,,]{\color{textcolor}\sffamily\fontsize{19.000000}{22.800000}\selectfont \(\displaystyle P_{11}\)}%
\end{pgfscope}%
\begin{pgfscope}%
\definecolor{textcolor}{rgb}{0.000000,0.000000,0.000000}%
\pgfsetstrokecolor{textcolor}%
\pgfsetfillcolor{textcolor}%
\pgftext[x=2.720431in,y=2.681071in,,]{\color{textcolor}\sffamily\fontsize{19.000000}{22.800000}\selectfont \(\displaystyle P_{12}\)}%
\end{pgfscope}%
\end{pgfpicture}%
\makeatother%
\endgroup%

%% file: pgf_example/03_single_plot.tex
\begingroup%
\makeatletter%
\begin{pgfpicture}%
\pgfpathrectangle{\pgfpointorigin}{\pgfqpoint{3.600000in}{3.377118in}}%
\pgfusepath{use as bounding box, clip}%
\begin{pgfscope}%
\pgfsetbuttcap%
\pgfsetmiterjoin%
\definecolor{currentfill}{rgb}{1.000000,1.000000,1.000000}%
\pgfsetfillcolor{currentfill}%
\pgfsetlinewidth{0.000000pt}%
\definecolor{currentstroke}{rgb}{1.000000,1.000000,1.000000}%
\pgfsetstrokecolor{currentstroke}%
\pgfsetdash{}{0pt}%
\pgfpathmoveto{\pgfqpoint{0.000000in}{0.000000in}}%
\pgfpathlineto{\pgfqpoint{3.600000in}{0.000000in}}%
\pgfpathlineto{\pgfqpoint{3.600000in}{3.377118in}}%
\pgfpathlineto{\pgfqpoint{0.000000in}{3.377118in}}%
\pgfpathlineto{\pgfqpoint{0.000000in}{0.000000in}}%
\pgfpathclose%
\pgfusepath{fill}%
\end{pgfscope}%
\begin{pgfscope}%
\pgfpathrectangle{\pgfqpoint{0.100000in}{0.100000in}}{\pgfqpoint{3.400000in}{3.177118in}}%
\pgfusepath{clip}%
\pgfsetbuttcap%
\pgfsetroundjoin%
\pgfsetlinewidth{2.007500pt}%
\definecolor{currentstroke}{rgb}{0.298039,0.470588,0.658824}%
\pgfsetstrokecolor{currentstroke}%
\pgfsetdash{}{0pt}%
\pgfpathmoveto{\pgfqpoint{2.378843in}{1.124601in}}%
\pgfpathlineto{\pgfqpoint{2.816129in}{0.759036in}}%
\pgfusepath{stroke}%
\end{pgfscope}%
\begin{pgfscope}%
\pgfpathrectangle{\pgfqpoint{0.100000in}{0.100000in}}{\pgfqpoint{3.400000in}{3.177118in}}%
\pgfusepath{clip}%
\pgfsetbuttcap%
\pgfsetroundjoin%
\pgfsetlinewidth{2.007500pt}%
\definecolor{currentstroke}{rgb}{0.298039,0.470588,0.658824}%
\pgfsetstrokecolor{currentstroke}%
\pgfsetdash{}{0pt}%
\pgfpathmoveto{\pgfqpoint{2.069616in}{0.314443in}}%
\pgfpathlineto{\pgfqpoint{2.378843in}{1.124601in}}%
\pgfusepath{stroke}%
\end{pgfscope}%
\begin{pgfscope}%
\pgfpathrectangle{\pgfqpoint{0.100000in}{0.100000in}}{\pgfqpoint{3.400000in}{3.177118in}}%
\pgfusepath{clip}%
\pgfsetbuttcap%
\pgfsetroundjoin%
\pgfsetlinewidth{2.007500pt}%
\definecolor{currentstroke}{rgb}{0.298039,0.470588,0.658824}%
\pgfsetstrokecolor{currentstroke}%
\pgfsetdash{}{0pt}%
\pgfpathmoveto{\pgfqpoint{1.846458in}{0.757050in}}%
\pgfpathlineto{\pgfqpoint{2.069616in}{0.314443in}}%
\pgfusepath{stroke}%
\end{pgfscope}%
\begin{pgfscope}%
\pgfpathrectangle{\pgfqpoint{0.100000in}{0.100000in}}{\pgfqpoint{3.400000in}{3.177118in}}%
\pgfusepath{clip}%
\pgfsetbuttcap%
\pgfsetroundjoin%
\pgfsetlinewidth{2.007500pt}%
\definecolor{currentstroke}{rgb}{0.298039,0.470588,0.658824}%
\pgfsetstrokecolor{currentstroke}%
\pgfsetdash{}{0pt}%
\pgfpathmoveto{\pgfqpoint{3.204959in}{1.715014in}}%
\pgfpathlineto{\pgfqpoint{2.816129in}{0.759036in}}%
\pgfusepath{stroke}%
\end{pgfscope}%
\begin{pgfscope}%
\pgfpathrectangle{\pgfqpoint{0.100000in}{0.100000in}}{\pgfqpoint{3.400000in}{3.177118in}}%
\pgfusepath{clip}%
\pgfsetbuttcap%
\pgfsetroundjoin%
\pgfsetlinewidth{2.007500pt}%
\definecolor{currentstroke}{rgb}{0.619608,0.619608,0.619608}%
\pgfsetstrokecolor{currentstroke}%
\pgfsetdash{}{0pt}%
\pgfpathmoveto{\pgfqpoint{0.459600in}{1.274070in}}%
\pgfpathlineto{\pgfqpoint{1.101336in}{0.581975in}}%
\pgfusepath{stroke}%
\end{pgfscope}%
\begin{pgfscope}%
\pgfpathrectangle{\pgfqpoint{0.100000in}{0.100000in}}{\pgfqpoint{3.400000in}{3.177118in}}%
\pgfusepath{clip}%
\pgfsetbuttcap%
\pgfsetroundjoin%
\pgfsetlinewidth{2.007500pt}%
\definecolor{currentstroke}{rgb}{0.960784,0.521569,0.094118}%
\pgfsetstrokecolor{currentstroke}%
\pgfsetdash{}{0pt}%
\pgfpathmoveto{\pgfqpoint{0.973595in}{2.931645in}}%
\pgfpathlineto{\pgfqpoint{0.395041in}{2.204306in}}%
\pgfusepath{stroke}%
\end{pgfscope}%
\begin{pgfscope}%
\pgfpathrectangle{\pgfqpoint{0.100000in}{0.100000in}}{\pgfqpoint{3.400000in}{3.177118in}}%
\pgfusepath{clip}%
\pgfsetbuttcap%
\pgfsetroundjoin%
\pgfsetlinewidth{2.007500pt}%
\definecolor{currentstroke}{rgb}{0.960784,0.521569,0.094118}%
\pgfsetstrokecolor{currentstroke}%
\pgfsetdash{}{0pt}%
\pgfpathmoveto{\pgfqpoint{1.895146in}{3.074750in}}%
\pgfpathlineto{\pgfqpoint{0.973595in}{2.931645in}}%
\pgfusepath{stroke}%
\end{pgfscope}%
\begin{pgfscope}%
\pgfpathrectangle{\pgfqpoint{0.100000in}{0.100000in}}{\pgfqpoint{3.400000in}{3.177118in}}%
\pgfusepath{clip}%
\pgfsetbuttcap%
\pgfsetroundjoin%
\pgfsetlinewidth{2.007500pt}%
\definecolor{currentstroke}{rgb}{0.960784,0.521569,0.094118}%
\pgfsetstrokecolor{currentstroke}%
\pgfsetdash{}{0pt}%
\pgfpathmoveto{\pgfqpoint{1.895146in}{3.074750in}}%
\pgfpathlineto{\pgfqpoint{2.720431in}{2.605682in}}%
\pgfusepath{stroke}%
\end{pgfscope}%
\begin{pgfscope}%
\pgfpathrectangle{\pgfqpoint{0.100000in}{0.100000in}}{\pgfqpoint{3.400000in}{3.177118in}}%
\pgfusepath{clip}%
\pgfsetbuttcap%
\pgfsetroundjoin%
\pgfsetlinewidth{2.007500pt}%
\definecolor{currentstroke}{rgb}{0.960784,0.521569,0.094118}%
\pgfsetstrokecolor{currentstroke}%
\pgfsetdash{}{0pt}%
\pgfpathmoveto{\pgfqpoint{2.637146in}{1.915681in}}%
\pgfpathlineto{\pgfqpoint{2.720431in}{2.605682in}}%
\pgfusepath{stroke}%
\end{pgfscope}%
\begin{pgfscope}%
\pgfpathrectangle{\pgfqpoint{0.100000in}{0.100000in}}{\pgfqpoint{3.400000in}{3.177118in}}%
\pgfusepath{clip}%
\pgfsetbuttcap%
\pgfsetroundjoin%
\pgfsetlinewidth{2.007500pt}%
\definecolor{currentstroke}{rgb}{0.960784,0.521569,0.094118}%
\pgfsetstrokecolor{currentstroke}%
\pgfsetdash{}{0pt}%
\pgfpathmoveto{\pgfqpoint{2.637146in}{1.915681in}}%
\pgfpathlineto{\pgfqpoint{0.395041in}{2.204306in}}%
\pgfusepath{stroke}%
\end{pgfscope}%
\begin{pgfscope}%
\pgfpathrectangle{\pgfqpoint{0.100000in}{0.100000in}}{\pgfqpoint{3.400000in}{3.177118in}}%
\pgfusepath{clip}%
\pgfsetbuttcap%
\pgfsetroundjoin%
\pgfsetlinewidth{2.007500pt}%
\definecolor{currentstroke}{rgb}{0.619608,0.619608,0.619608}%
\pgfsetstrokecolor{currentstroke}%
\pgfsetdash{{7.400000pt}{3.200000pt}}{0.000000pt}%
\pgfpathmoveto{\pgfqpoint{1.846458in}{0.757050in}}%
\pgfpathlineto{\pgfqpoint{1.101336in}{0.581975in}}%
\pgfusepath{stroke}%
\end{pgfscope}%
\begin{pgfscope}%
\pgfpathrectangle{\pgfqpoint{0.100000in}{0.100000in}}{\pgfqpoint{3.400000in}{3.177118in}}%
\pgfusepath{clip}%
\pgfsetbuttcap%
\pgfsetroundjoin%
\pgfsetlinewidth{2.007500pt}%
\definecolor{currentstroke}{rgb}{0.619608,0.619608,0.619608}%
\pgfsetstrokecolor{currentstroke}%
\pgfsetdash{{7.400000pt}{3.200000pt}}{0.000000pt}%
\pgfpathmoveto{\pgfqpoint{0.459600in}{1.274070in}}%
\pgfpathlineto{\pgfqpoint{0.395041in}{2.204306in}}%
\pgfusepath{stroke}%
\end{pgfscope}%
\begin{pgfscope}%
\pgfpathrectangle{\pgfqpoint{0.100000in}{0.100000in}}{\pgfqpoint{3.400000in}{3.177118in}}%
\pgfusepath{clip}%
\pgfsetbuttcap%
\pgfsetroundjoin%
\pgfsetlinewidth{2.007500pt}%
\definecolor{currentstroke}{rgb}{0.619608,0.619608,0.619608}%
\pgfsetstrokecolor{currentstroke}%
\pgfsetdash{{7.400000pt}{3.200000pt}}{0.000000pt}%
\pgfpathmoveto{\pgfqpoint{3.204959in}{1.715014in}}%
\pgfpathlineto{\pgfqpoint{2.637146in}{1.915681in}}%
\pgfusepath{stroke}%
\end{pgfscope}%
\begin{pgfscope}%
\pgfpathrectangle{\pgfqpoint{0.100000in}{0.100000in}}{\pgfqpoint{3.400000in}{3.177118in}}%
\pgfusepath{clip}%
\pgfsetbuttcap%
\pgfsetroundjoin%
\definecolor{currentfill}{rgb}{0.298039,0.470588,0.658824}%
\pgfsetfillcolor{currentfill}%
\pgfsetlinewidth{1.003750pt}%
\definecolor{currentstroke}{rgb}{0.298039,0.470588,0.658824}%
\pgfsetstrokecolor{currentstroke}%
\pgfsetdash{}{0pt}%
\pgfpathmoveto{\pgfqpoint{3.204959in}{1.531281in}}%
\pgfpathcurveto{\pgfqpoint{3.253685in}{1.531281in}}{\pgfqpoint{3.300423in}{1.550641in}}{\pgfqpoint{3.334877in}{1.585095in}}%
\pgfpathcurveto{\pgfqpoint{3.369332in}{1.619550in}}{\pgfqpoint{3.388691in}{1.666288in}}{\pgfqpoint{3.388691in}{1.715014in}}%
\pgfpathcurveto{\pgfqpoint{3.388691in}{1.763741in}}{\pgfqpoint{3.369332in}{1.810478in}}{\pgfqpoint{3.334877in}{1.844933in}}%
\pgfpathcurveto{\pgfqpoint{3.300423in}{1.879388in}}{\pgfqpoint{3.253685in}{1.898747in}}{\pgfqpoint{3.204959in}{1.898747in}}%
\pgfpathcurveto{\pgfqpoint{3.156232in}{1.898747in}}{\pgfqpoint{3.109495in}{1.879388in}}{\pgfqpoint{3.075040in}{1.844933in}}%
\pgfpathcurveto{\pgfqpoint{3.040585in}{1.810478in}}{\pgfqpoint{3.021226in}{1.763741in}}{\pgfqpoint{3.021226in}{1.715014in}}%
\pgfpathcurveto{\pgfqpoint{3.021226in}{1.666288in}}{\pgfqpoint{3.040585in}{1.619550in}}{\pgfqpoint{3.075040in}{1.585095in}}%
\pgfpathcurveto{\pgfqpoint{3.109495in}{1.550641in}}{\pgfqpoint{3.156232in}{1.531281in}}{\pgfqpoint{3.204959in}{1.531281in}}%
\pgfpathlineto{\pgfqpoint{3.204959in}{1.531281in}}%
\pgfpathclose%
\pgfusepath{stroke,fill}%
\end{pgfscope}%
\begin{pgfscope}%
\pgfpathrectangle{\pgfqpoint{0.100000in}{0.100000in}}{\pgfqpoint{3.400000in}{3.177118in}}%
\pgfusepath{clip}%
\pgfsetbuttcap%
\pgfsetroundjoin%
\definecolor{currentfill}{rgb}{0.298039,0.470588,0.658824}%
\pgfsetfillcolor{currentfill}%
\pgfsetlinewidth{1.003750pt}%
\definecolor{currentstroke}{rgb}{0.298039,0.470588,0.658824}%
\pgfsetstrokecolor{currentstroke}%
\pgfsetdash{}{0pt}%
\pgfpathmoveto{\pgfqpoint{1.846458in}{0.573318in}}%
\pgfpathcurveto{\pgfqpoint{1.895185in}{0.573318in}}{\pgfqpoint{1.941922in}{0.592677in}}{\pgfqpoint{1.976377in}{0.627132in}}%
\pgfpathcurveto{\pgfqpoint{2.010832in}{0.661586in}}{\pgfqpoint{2.030191in}{0.708324in}}{\pgfqpoint{2.030191in}{0.757050in}}%
\pgfpathcurveto{\pgfqpoint{2.030191in}{0.805777in}}{\pgfqpoint{2.010832in}{0.852514in}}{\pgfqpoint{1.976377in}{0.886969in}}%
\pgfpathcurveto{\pgfqpoint{1.941922in}{0.921424in}}{\pgfqpoint{1.895185in}{0.940783in}}{\pgfqpoint{1.846458in}{0.940783in}}%
\pgfpathcurveto{\pgfqpoint{1.797732in}{0.940783in}}{\pgfqpoint{1.750994in}{0.921424in}}{\pgfqpoint{1.716539in}{0.886969in}}%
\pgfpathcurveto{\pgfqpoint{1.682085in}{0.852514in}}{\pgfqpoint{1.662725in}{0.805777in}}{\pgfqpoint{1.662725in}{0.757050in}}%
\pgfpathcurveto{\pgfqpoint{1.662725in}{0.708324in}}{\pgfqpoint{1.682085in}{0.661586in}}{\pgfqpoint{1.716539in}{0.627132in}}%
\pgfpathcurveto{\pgfqpoint{1.750994in}{0.592677in}}{\pgfqpoint{1.797732in}{0.573318in}}{\pgfqpoint{1.846458in}{0.573318in}}%
\pgfpathlineto{\pgfqpoint{1.846458in}{0.573318in}}%
\pgfpathclose%
\pgfusepath{stroke,fill}%
\end{pgfscope}%
\begin{pgfscope}%
\pgfpathrectangle{\pgfqpoint{0.100000in}{0.100000in}}{\pgfqpoint{3.400000in}{3.177118in}}%
\pgfusepath{clip}%
\pgfsetbuttcap%
\pgfsetroundjoin%
\definecolor{currentfill}{rgb}{0.960784,0.521569,0.094118}%
\pgfsetfillcolor{currentfill}%
\pgfsetlinewidth{1.003750pt}%
\definecolor{currentstroke}{rgb}{0.960784,0.521569,0.094118}%
\pgfsetstrokecolor{currentstroke}%
\pgfsetdash{}{0pt}%
\pgfpathmoveto{\pgfqpoint{1.895146in}{2.891018in}}%
\pgfpathcurveto{\pgfqpoint{1.943872in}{2.891018in}}{\pgfqpoint{1.990609in}{2.910377in}}{\pgfqpoint{2.025064in}{2.944832in}}%
\pgfpathcurveto{\pgfqpoint{2.059519in}{2.979287in}}{\pgfqpoint{2.078878in}{3.026024in}}{\pgfqpoint{2.078878in}{3.074750in}}%
\pgfpathcurveto{\pgfqpoint{2.078878in}{3.123477in}}{\pgfqpoint{2.059519in}{3.170214in}}{\pgfqpoint{2.025064in}{3.204669in}}%
\pgfpathcurveto{\pgfqpoint{1.990609in}{3.239124in}}{\pgfqpoint{1.943872in}{3.258483in}}{\pgfqpoint{1.895146in}{3.258483in}}%
\pgfpathcurveto{\pgfqpoint{1.846419in}{3.258483in}}{\pgfqpoint{1.799682in}{3.239124in}}{\pgfqpoint{1.765227in}{3.204669in}}%
\pgfpathcurveto{\pgfqpoint{1.730772in}{3.170214in}}{\pgfqpoint{1.711413in}{3.123477in}}{\pgfqpoint{1.711413in}{3.074750in}}%
\pgfpathcurveto{\pgfqpoint{1.711413in}{3.026024in}}{\pgfqpoint{1.730772in}{2.979287in}}{\pgfqpoint{1.765227in}{2.944832in}}%
\pgfpathcurveto{\pgfqpoint{1.799682in}{2.910377in}}{\pgfqpoint{1.846419in}{2.891018in}}{\pgfqpoint{1.895146in}{2.891018in}}%
\pgfpathlineto{\pgfqpoint{1.895146in}{2.891018in}}%
\pgfpathclose%
\pgfusepath{stroke,fill}%
\end{pgfscope}%
\begin{pgfscope}%
\pgfpathrectangle{\pgfqpoint{0.100000in}{0.100000in}}{\pgfqpoint{3.400000in}{3.177118in}}%
\pgfusepath{clip}%
\pgfsetbuttcap%
\pgfsetroundjoin%
\definecolor{currentfill}{rgb}{0.960784,0.521569,0.094118}%
\pgfsetfillcolor{currentfill}%
\pgfsetlinewidth{1.003750pt}%
\definecolor{currentstroke}{rgb}{0.960784,0.521569,0.094118}%
\pgfsetstrokecolor{currentstroke}%
\pgfsetdash{}{0pt}%
\pgfpathmoveto{\pgfqpoint{2.637146in}{1.731948in}}%
\pgfpathcurveto{\pgfqpoint{2.685873in}{1.731948in}}{\pgfqpoint{2.732610in}{1.751307in}}{\pgfqpoint{2.767065in}{1.785762in}}%
\pgfpathcurveto{\pgfqpoint{2.801520in}{1.820217in}}{\pgfqpoint{2.820879in}{1.866954in}}{\pgfqpoint{2.820879in}{1.915681in}}%
\pgfpathcurveto{\pgfqpoint{2.820879in}{1.964407in}}{\pgfqpoint{2.801520in}{2.011145in}}{\pgfqpoint{2.767065in}{2.045600in}}%
\pgfpathcurveto{\pgfqpoint{2.732610in}{2.080054in}}{\pgfqpoint{2.685873in}{2.099414in}}{\pgfqpoint{2.637146in}{2.099414in}}%
\pgfpathcurveto{\pgfqpoint{2.588420in}{2.099414in}}{\pgfqpoint{2.541683in}{2.080054in}}{\pgfqpoint{2.507228in}{2.045600in}}%
\pgfpathcurveto{\pgfqpoint{2.472773in}{2.011145in}}{\pgfqpoint{2.453414in}{1.964407in}}{\pgfqpoint{2.453414in}{1.915681in}}%
\pgfpathcurveto{\pgfqpoint{2.453414in}{1.866954in}}{\pgfqpoint{2.472773in}{1.820217in}}{\pgfqpoint{2.507228in}{1.785762in}}%
\pgfpathcurveto{\pgfqpoint{2.541683in}{1.751307in}}{\pgfqpoint{2.588420in}{1.731948in}}{\pgfqpoint{2.637146in}{1.731948in}}%
\pgfpathlineto{\pgfqpoint{2.637146in}{1.731948in}}%
\pgfpathclose%
\pgfusepath{stroke,fill}%
\end{pgfscope}%
\begin{pgfscope}%
\pgfpathrectangle{\pgfqpoint{0.100000in}{0.100000in}}{\pgfqpoint{3.400000in}{3.177118in}}%
\pgfusepath{clip}%
\pgfsetbuttcap%
\pgfsetroundjoin%
\definecolor{currentfill}{rgb}{0.298039,0.470588,0.658824}%
\pgfsetfillcolor{currentfill}%
\pgfsetlinewidth{1.003750pt}%
\definecolor{currentstroke}{rgb}{0.298039,0.470588,0.658824}%
\pgfsetstrokecolor{currentstroke}%
\pgfsetdash{}{0pt}%
\pgfpathmoveto{\pgfqpoint{2.069616in}{0.130710in}}%
\pgfpathcurveto{\pgfqpoint{2.118342in}{0.130710in}}{\pgfqpoint{2.165080in}{0.150070in}}{\pgfqpoint{2.199535in}{0.184524in}}%
\pgfpathcurveto{\pgfqpoint{2.233989in}{0.218979in}}{\pgfqpoint{2.253349in}{0.265717in}}{\pgfqpoint{2.253349in}{0.314443in}}%
\pgfpathcurveto{\pgfqpoint{2.253349in}{0.363170in}}{\pgfqpoint{2.233989in}{0.409907in}}{\pgfqpoint{2.199535in}{0.444362in}}%
\pgfpathcurveto{\pgfqpoint{2.165080in}{0.478817in}}{\pgfqpoint{2.118342in}{0.498176in}}{\pgfqpoint{2.069616in}{0.498176in}}%
\pgfpathcurveto{\pgfqpoint{2.020889in}{0.498176in}}{\pgfqpoint{1.974152in}{0.478817in}}{\pgfqpoint{1.939697in}{0.444362in}}%
\pgfpathcurveto{\pgfqpoint{1.905242in}{0.409907in}}{\pgfqpoint{1.885883in}{0.363170in}}{\pgfqpoint{1.885883in}{0.314443in}}%
\pgfpathcurveto{\pgfqpoint{1.885883in}{0.265717in}}{\pgfqpoint{1.905242in}{0.218979in}}{\pgfqpoint{1.939697in}{0.184524in}}%
\pgfpathcurveto{\pgfqpoint{1.974152in}{0.150070in}}{\pgfqpoint{2.020889in}{0.130710in}}{\pgfqpoint{2.069616in}{0.130710in}}%
\pgfpathlineto{\pgfqpoint{2.069616in}{0.130710in}}%
\pgfpathclose%
\pgfusepath{stroke,fill}%
\end{pgfscope}%
\begin{pgfscope}%
\pgfpathrectangle{\pgfqpoint{0.100000in}{0.100000in}}{\pgfqpoint{3.400000in}{3.177118in}}%
\pgfusepath{clip}%
\pgfsetbuttcap%
\pgfsetroundjoin%
\definecolor{currentfill}{rgb}{0.619608,0.619608,0.619608}%
\pgfsetfillcolor{currentfill}%
\pgfsetlinewidth{1.003750pt}%
\definecolor{currentstroke}{rgb}{0.619608,0.619608,0.619608}%
\pgfsetstrokecolor{currentstroke}%
\pgfsetdash{}{0pt}%
\pgfpathmoveto{\pgfqpoint{0.459600in}{1.090337in}}%
\pgfpathcurveto{\pgfqpoint{0.508327in}{1.090337in}}{\pgfqpoint{0.555064in}{1.109697in}}{\pgfqpoint{0.589519in}{1.144151in}}%
\pgfpathcurveto{\pgfqpoint{0.623974in}{1.178606in}}{\pgfqpoint{0.643333in}{1.225344in}}{\pgfqpoint{0.643333in}{1.274070in}}%
\pgfpathcurveto{\pgfqpoint{0.643333in}{1.322797in}}{\pgfqpoint{0.623974in}{1.369534in}}{\pgfqpoint{0.589519in}{1.403989in}}%
\pgfpathcurveto{\pgfqpoint{0.555064in}{1.438444in}}{\pgfqpoint{0.508327in}{1.457803in}}{\pgfqpoint{0.459600in}{1.457803in}}%
\pgfpathcurveto{\pgfqpoint{0.410874in}{1.457803in}}{\pgfqpoint{0.364136in}{1.438444in}}{\pgfqpoint{0.329682in}{1.403989in}}%
\pgfpathcurveto{\pgfqpoint{0.295227in}{1.369534in}}{\pgfqpoint{0.275868in}{1.322797in}}{\pgfqpoint{0.275868in}{1.274070in}}%
\pgfpathcurveto{\pgfqpoint{0.275868in}{1.225344in}}{\pgfqpoint{0.295227in}{1.178606in}}{\pgfqpoint{0.329682in}{1.144151in}}%
\pgfpathcurveto{\pgfqpoint{0.364136in}{1.109697in}}{\pgfqpoint{0.410874in}{1.090337in}}{\pgfqpoint{0.459600in}{1.090337in}}%
\pgfpathlineto{\pgfqpoint{0.459600in}{1.090337in}}%
\pgfpathclose%
\pgfusepath{stroke,fill}%
\end{pgfscope}%
\begin{pgfscope}%
\pgfpathrectangle{\pgfqpoint{0.100000in}{0.100000in}}{\pgfqpoint{3.400000in}{3.177118in}}%
\pgfusepath{clip}%
\pgfsetbuttcap%
\pgfsetroundjoin%
\definecolor{currentfill}{rgb}{0.960784,0.521569,0.094118}%
\pgfsetfillcolor{currentfill}%
\pgfsetlinewidth{1.003750pt}%
\definecolor{currentstroke}{rgb}{0.960784,0.521569,0.094118}%
\pgfsetstrokecolor{currentstroke}%
\pgfsetdash{}{0pt}%
\pgfpathmoveto{\pgfqpoint{0.973595in}{2.747912in}}%
\pgfpathcurveto{\pgfqpoint{1.022321in}{2.747912in}}{\pgfqpoint{1.069058in}{2.767271in}}{\pgfqpoint{1.103513in}{2.801726in}}%
\pgfpathcurveto{\pgfqpoint{1.137968in}{2.836181in}}{\pgfqpoint{1.157327in}{2.882918in}}{\pgfqpoint{1.157327in}{2.931645in}}%
\pgfpathcurveto{\pgfqpoint{1.157327in}{2.980371in}}{\pgfqpoint{1.137968in}{3.027109in}}{\pgfqpoint{1.103513in}{3.061563in}}%
\pgfpathcurveto{\pgfqpoint{1.069058in}{3.096018in}}{\pgfqpoint{1.022321in}{3.115378in}}{\pgfqpoint{0.973595in}{3.115378in}}%
\pgfpathcurveto{\pgfqpoint{0.924868in}{3.115378in}}{\pgfqpoint{0.878131in}{3.096018in}}{\pgfqpoint{0.843676in}{3.061563in}}%
\pgfpathcurveto{\pgfqpoint{0.809221in}{3.027109in}}{\pgfqpoint{0.789862in}{2.980371in}}{\pgfqpoint{0.789862in}{2.931645in}}%
\pgfpathcurveto{\pgfqpoint{0.789862in}{2.882918in}}{\pgfqpoint{0.809221in}{2.836181in}}{\pgfqpoint{0.843676in}{2.801726in}}%
\pgfpathcurveto{\pgfqpoint{0.878131in}{2.767271in}}{\pgfqpoint{0.924868in}{2.747912in}}{\pgfqpoint{0.973595in}{2.747912in}}%
\pgfpathlineto{\pgfqpoint{0.973595in}{2.747912in}}%
\pgfpathclose%
\pgfusepath{stroke,fill}%
\end{pgfscope}%
\begin{pgfscope}%
\pgfpathrectangle{\pgfqpoint{0.100000in}{0.100000in}}{\pgfqpoint{3.400000in}{3.177118in}}%
\pgfusepath{clip}%
\pgfsetbuttcap%
\pgfsetroundjoin%
\definecolor{currentfill}{rgb}{0.960784,0.521569,0.094118}%
\pgfsetfillcolor{currentfill}%
\pgfsetlinewidth{1.003750pt}%
\definecolor{currentstroke}{rgb}{0.960784,0.521569,0.094118}%
\pgfsetstrokecolor{currentstroke}%
\pgfsetdash{}{0pt}%
\pgfpathmoveto{\pgfqpoint{0.395041in}{2.020573in}}%
\pgfpathcurveto{\pgfqpoint{0.443768in}{2.020573in}}{\pgfqpoint{0.490505in}{2.039933in}}{\pgfqpoint{0.524960in}{2.074387in}}%
\pgfpathcurveto{\pgfqpoint{0.559415in}{2.108842in}}{\pgfqpoint{0.578774in}{2.155580in}}{\pgfqpoint{0.578774in}{2.204306in}}%
\pgfpathcurveto{\pgfqpoint{0.578774in}{2.253032in}}{\pgfqpoint{0.559415in}{2.299770in}}{\pgfqpoint{0.524960in}{2.334225in}}%
\pgfpathcurveto{\pgfqpoint{0.490505in}{2.368679in}}{\pgfqpoint{0.443768in}{2.388039in}}{\pgfqpoint{0.395041in}{2.388039in}}%
\pgfpathcurveto{\pgfqpoint{0.346315in}{2.388039in}}{\pgfqpoint{0.299577in}{2.368679in}}{\pgfqpoint{0.265123in}{2.334225in}}%
\pgfpathcurveto{\pgfqpoint{0.230668in}{2.299770in}}{\pgfqpoint{0.211309in}{2.253032in}}{\pgfqpoint{0.211309in}{2.204306in}}%
\pgfpathcurveto{\pgfqpoint{0.211309in}{2.155580in}}{\pgfqpoint{0.230668in}{2.108842in}}{\pgfqpoint{0.265123in}{2.074387in}}%
\pgfpathcurveto{\pgfqpoint{0.299577in}{2.039933in}}{\pgfqpoint{0.346315in}{2.020573in}}{\pgfqpoint{0.395041in}{2.020573in}}%
\pgfpathlineto{\pgfqpoint{0.395041in}{2.020573in}}%
\pgfpathclose%
\pgfusepath{stroke,fill}%
\end{pgfscope}%
\begin{pgfscope}%
\pgfpathrectangle{\pgfqpoint{0.100000in}{0.100000in}}{\pgfqpoint{3.400000in}{3.177118in}}%
\pgfusepath{clip}%
\pgfsetbuttcap%
\pgfsetroundjoin%
\definecolor{currentfill}{rgb}{0.619608,0.619608,0.619608}%
\pgfsetfillcolor{currentfill}%
\pgfsetlinewidth{1.003750pt}%
\definecolor{currentstroke}{rgb}{0.619608,0.619608,0.619608}%
\pgfsetstrokecolor{currentstroke}%
\pgfsetdash{}{0pt}%
\pgfpathmoveto{\pgfqpoint{1.101336in}{0.398243in}}%
\pgfpathcurveto{\pgfqpoint{1.150062in}{0.398243in}}{\pgfqpoint{1.196799in}{0.417602in}}{\pgfqpoint{1.231254in}{0.452057in}}%
\pgfpathcurveto{\pgfqpoint{1.265709in}{0.486511in}}{\pgfqpoint{1.285068in}{0.533249in}}{\pgfqpoint{1.285068in}{0.581975in}}%
\pgfpathcurveto{\pgfqpoint{1.285068in}{0.630702in}}{\pgfqpoint{1.265709in}{0.677439in}}{\pgfqpoint{1.231254in}{0.711894in}}%
\pgfpathcurveto{\pgfqpoint{1.196799in}{0.746349in}}{\pgfqpoint{1.150062in}{0.765708in}}{\pgfqpoint{1.101336in}{0.765708in}}%
\pgfpathcurveto{\pgfqpoint{1.052609in}{0.765708in}}{\pgfqpoint{1.005872in}{0.746349in}}{\pgfqpoint{0.971417in}{0.711894in}}%
\pgfpathcurveto{\pgfqpoint{0.936962in}{0.677439in}}{\pgfqpoint{0.917603in}{0.630702in}}{\pgfqpoint{0.917603in}{0.581975in}}%
\pgfpathcurveto{\pgfqpoint{0.917603in}{0.533249in}}{\pgfqpoint{0.936962in}{0.486511in}}{\pgfqpoint{0.971417in}{0.452057in}}%
\pgfpathcurveto{\pgfqpoint{1.005872in}{0.417602in}}{\pgfqpoint{1.052609in}{0.398243in}}{\pgfqpoint{1.101336in}{0.398243in}}%
\pgfpathlineto{\pgfqpoint{1.101336in}{0.398243in}}%
\pgfpathclose%
\pgfusepath{stroke,fill}%
\end{pgfscope}%
\begin{pgfscope}%
\pgfpathrectangle{\pgfqpoint{0.100000in}{0.100000in}}{\pgfqpoint{3.400000in}{3.177118in}}%
\pgfusepath{clip}%
\pgfsetbuttcap%
\pgfsetroundjoin%
\definecolor{currentfill}{rgb}{0.298039,0.470588,0.658824}%
\pgfsetfillcolor{currentfill}%
\pgfsetlinewidth{1.003750pt}%
\definecolor{currentstroke}{rgb}{0.298039,0.470588,0.658824}%
\pgfsetstrokecolor{currentstroke}%
\pgfsetdash{}{0pt}%
\pgfpathmoveto{\pgfqpoint{2.378843in}{0.940868in}}%
\pgfpathcurveto{\pgfqpoint{2.427569in}{0.940868in}}{\pgfqpoint{2.474307in}{0.960227in}}{\pgfqpoint{2.508762in}{0.994682in}}%
\pgfpathcurveto{\pgfqpoint{2.543216in}{1.029137in}}{\pgfqpoint{2.562576in}{1.075874in}}{\pgfqpoint{2.562576in}{1.124601in}}%
\pgfpathcurveto{\pgfqpoint{2.562576in}{1.173327in}}{\pgfqpoint{2.543216in}{1.220065in}}{\pgfqpoint{2.508762in}{1.254519in}}%
\pgfpathcurveto{\pgfqpoint{2.474307in}{1.288974in}}{\pgfqpoint{2.427569in}{1.308333in}}{\pgfqpoint{2.378843in}{1.308333in}}%
\pgfpathcurveto{\pgfqpoint{2.330116in}{1.308333in}}{\pgfqpoint{2.283379in}{1.288974in}}{\pgfqpoint{2.248924in}{1.254519in}}%
\pgfpathcurveto{\pgfqpoint{2.214469in}{1.220065in}}{\pgfqpoint{2.195110in}{1.173327in}}{\pgfqpoint{2.195110in}{1.124601in}}%
\pgfpathcurveto{\pgfqpoint{2.195110in}{1.075874in}}{\pgfqpoint{2.214469in}{1.029137in}}{\pgfqpoint{2.248924in}{0.994682in}}%
\pgfpathcurveto{\pgfqpoint{2.283379in}{0.960227in}}{\pgfqpoint{2.330116in}{0.940868in}}{\pgfqpoint{2.378843in}{0.940868in}}%
\pgfpathlineto{\pgfqpoint{2.378843in}{0.940868in}}%
\pgfpathclose%
\pgfusepath{stroke,fill}%
\end{pgfscope}%
\begin{pgfscope}%
\pgfpathrectangle{\pgfqpoint{0.100000in}{0.100000in}}{\pgfqpoint{3.400000in}{3.177118in}}%
\pgfusepath{clip}%
\pgfsetbuttcap%
\pgfsetroundjoin%
\definecolor{currentfill}{rgb}{0.298039,0.470588,0.658824}%
\pgfsetfillcolor{currentfill}%
\pgfsetlinewidth{1.003750pt}%
\definecolor{currentstroke}{rgb}{0.298039,0.470588,0.658824}%
\pgfsetstrokecolor{currentstroke}%
\pgfsetdash{}{0pt}%
\pgfpathmoveto{\pgfqpoint{2.632396in}{0.575303in}}%
\pgfpathlineto{\pgfqpoint{2.999862in}{0.575303in}}%
\pgfpathlineto{\pgfqpoint{2.999862in}{0.942769in}}%
\pgfpathlineto{\pgfqpoint{2.632396in}{0.942769in}}%
\pgfpathlineto{\pgfqpoint{2.632396in}{0.575303in}}%
\pgfpathclose%
\pgfusepath{stroke,fill}%
\end{pgfscope}%
\begin{pgfscope}%
\pgfpathrectangle{\pgfqpoint{0.100000in}{0.100000in}}{\pgfqpoint{3.400000in}{3.177118in}}%
\pgfusepath{clip}%
\pgfsetbuttcap%
\pgfsetroundjoin%
\definecolor{currentfill}{rgb}{0.960784,0.521569,0.094118}%
\pgfsetfillcolor{currentfill}%
\pgfsetlinewidth{1.003750pt}%
\definecolor{currentstroke}{rgb}{0.960784,0.521569,0.094118}%
\pgfsetstrokecolor{currentstroke}%
\pgfsetdash{}{0pt}%
\pgfpathmoveto{\pgfqpoint{2.536698in}{2.421949in}}%
\pgfpathlineto{\pgfqpoint{2.904163in}{2.421949in}}%
\pgfpathlineto{\pgfqpoint{2.904163in}{2.789415in}}%
\pgfpathlineto{\pgfqpoint{2.536698in}{2.789415in}}%
\pgfpathlineto{\pgfqpoint{2.536698in}{2.421949in}}%
\pgfpathclose%
\pgfusepath{stroke,fill}%
\end{pgfscope}%
\begin{pgfscope}%
\definecolor{textcolor}{rgb}{0.000000,0.000000,0.000000}%
\pgfsetstrokecolor{textcolor}%
\pgfsetfillcolor{textcolor}%
\pgftext[x=3.204959in,y=1.715014in,,]{\color{textcolor}\sffamily\fontsize{19.000000}{22.800000}\selectfont \(\displaystyle S_{1}\)}%
\end{pgfscope}%
\begin{pgfscope}%
\definecolor{textcolor}{rgb}{0.000000,0.000000,0.000000}%
\pgfsetstrokecolor{textcolor}%
\pgfsetfillcolor{textcolor}%
\pgftext[x=1.846458in,y=0.757050in,,]{\color{textcolor}\sffamily\fontsize{19.000000}{22.800000}\selectfont \(\displaystyle S_{2}\)}%
\end{pgfscope}%
\begin{pgfscope}%
\definecolor{textcolor}{rgb}{0.000000,0.000000,0.000000}%
\pgfsetstrokecolor{textcolor}%
\pgfsetfillcolor{textcolor}%
\pgftext[x=1.895146in,y=3.074750in,,]{\color{textcolor}\sffamily\fontsize{19.000000}{22.800000}\selectfont \(\displaystyle S_{3}\)}%
\end{pgfscope}%
\begin{pgfscope}%
\definecolor{textcolor}{rgb}{0.000000,0.000000,0.000000}%
\pgfsetstrokecolor{textcolor}%
\pgfsetfillcolor{textcolor}%
\pgftext[x=2.637146in,y=1.915681in,,]{\color{textcolor}\sffamily\fontsize{19.000000}{22.800000}\selectfont \(\displaystyle S_{4}\)}%
\end{pgfscope}%
\begin{pgfscope}%
\definecolor{textcolor}{rgb}{0.000000,0.000000,0.000000}%
\pgfsetstrokecolor{textcolor}%
\pgfsetfillcolor{textcolor}%
\pgftext[x=2.069616in,y=0.314443in,,]{\color{textcolor}\sffamily\fontsize{19.000000}{22.800000}\selectfont \(\displaystyle S_{5}\)}%
\end{pgfscope}%
\begin{pgfscope}%
\definecolor{textcolor}{rgb}{0.000000,0.000000,0.000000}%
\pgfsetstrokecolor{textcolor}%
\pgfsetfillcolor{textcolor}%
\pgftext[x=0.459600in,y=1.274070in,,]{\color{textcolor}\sffamily\fontsize{19.000000}{22.800000}\selectfont \(\displaystyle S_{6}\)}%
\end{pgfscope}%
\begin{pgfscope}%
\definecolor{textcolor}{rgb}{0.000000,0.000000,0.000000}%
\pgfsetstrokecolor{textcolor}%
\pgfsetfillcolor{textcolor}%
\pgftext[x=0.973595in,y=2.931645in,,]{\color{textcolor}\sffamily\fontsize{19.000000}{22.800000}\selectfont \(\displaystyle S_{7}\)}%
\end{pgfscope}%
\begin{pgfscope}%
\definecolor{textcolor}{rgb}{0.000000,0.000000,0.000000}%
\pgfsetstrokecolor{textcolor}%
\pgfsetfillcolor{textcolor}%
\pgftext[x=0.395041in,y=2.204306in,,]{\color{textcolor}\sffamily\fontsize{19.000000}{22.800000}\selectfont \(\displaystyle S_{8}\)}%
\end{pgfscope}%
\begin{pgfscope}%
\definecolor{textcolor}{rgb}{0.000000,0.000000,0.000000}%
\pgfsetstrokecolor{textcolor}%
\pgfsetfillcolor{textcolor}%
\pgftext[x=1.101336in,y=0.581975in,,]{\color{textcolor}\sffamily\fontsize{19.000000}{22.800000}\selectfont \(\displaystyle S_{9}\)}%
\end{pgfscope}%
\begin{pgfscope}%
\definecolor{textcolor}{rgb}{0.000000,0.000000,0.000000}%
\pgfsetstrokecolor{textcolor}%
\pgfsetfillcolor{textcolor}%
\pgftext[x=2.378843in,y=1.124601in,,]{\color{textcolor}\sffamily\fontsize{19.000000}{22.800000}\selectfont \(\displaystyle S_{10}\)}%
\end{pgfscope}%
\begin{pgfscope}%
\definecolor{textcolor}{rgb}{0.000000,0.000000,0.000000}%
\pgfsetstrokecolor{textcolor}%
\pgfsetfillcolor{textcolor}%
\pgftext[x=2.816129in,y=0.759036in,,]{\color{textcolor}\sffamily\fontsize{19.000000}{22.800000}\selectfont \(\displaystyle P_{11}\)}%
\end{pgfscope}%
\begin{pgfscope}%
\definecolor{textcolor}{rgb}{0.000000,0.000000,0.000000}%
\pgfsetstrokecolor{textcolor}%
\pgfsetfillcolor{textcolor}%
\pgftext[x=2.720431in,y=2.605682in,,]{\color{textcolor}\sffamily\fontsize{19.000000}{22.800000}\selectfont \(\displaystyle P_{12}\)}%
\end{pgfscope}%
\end{pgfpicture}%
\makeatother%
\endgroup%

%% file: pgf_example/04_single_plot.tex
\begingroup%
\makeatletter%
\begin{pgfpicture}%
\pgfpathrectangle{\pgfpointorigin}{\pgfqpoint{3.600000in}{3.555557in}}%
\pgfusepath{use as bounding box, clip}%
\begin{pgfscope}%
\pgfsetbuttcap%
\pgfsetmiterjoin%
\definecolor{currentfill}{rgb}{1.000000,1.000000,1.000000}%
\pgfsetfillcolor{currentfill}%
\pgfsetlinewidth{0.000000pt}%
\definecolor{currentstroke}{rgb}{1.000000,1.000000,1.000000}%
\pgfsetstrokecolor{currentstroke}%
\pgfsetdash{}{0pt}%
\pgfpathmoveto{\pgfqpoint{0.000000in}{0.000000in}}%
\pgfpathlineto{\pgfqpoint{3.600000in}{0.000000in}}%
\pgfpathlineto{\pgfqpoint{3.600000in}{3.555557in}}%
\pgfpathlineto{\pgfqpoint{0.000000in}{3.555557in}}%
\pgfpathlineto{\pgfqpoint{0.000000in}{0.000000in}}%
\pgfpathclose%
\pgfusepath{fill}%
\end{pgfscope}%
\begin{pgfscope}%
\pgfpathrectangle{\pgfqpoint{0.100000in}{0.100000in}}{\pgfqpoint{3.400000in}{3.355557in}}%
\pgfusepath{clip}%
\pgfsetbuttcap%
\pgfsetroundjoin%
\pgfsetlinewidth{2.007500pt}%
\definecolor{currentstroke}{rgb}{0.298039,0.470588,0.658824}%
\pgfsetstrokecolor{currentstroke}%
\pgfsetdash{}{0pt}%
\pgfpathmoveto{\pgfqpoint{2.422510in}{1.161167in}}%
\pgfpathlineto{\pgfqpoint{2.873934in}{0.783784in}}%
\pgfusepath{stroke}%
\end{pgfscope}%
\begin{pgfscope}%
\pgfpathrectangle{\pgfqpoint{0.100000in}{0.100000in}}{\pgfqpoint{3.400000in}{3.355557in}}%
\pgfusepath{clip}%
\pgfsetbuttcap%
\pgfsetroundjoin%
\pgfsetlinewidth{2.007500pt}%
\definecolor{currentstroke}{rgb}{0.298039,0.470588,0.658824}%
\pgfsetstrokecolor{currentstroke}%
\pgfsetdash{}{0pt}%
\pgfpathmoveto{\pgfqpoint{2.103286in}{0.324818in}}%
\pgfpathlineto{\pgfqpoint{2.422510in}{1.161167in}}%
\pgfusepath{stroke}%
\end{pgfscope}%
\begin{pgfscope}%
\pgfpathrectangle{\pgfqpoint{0.100000in}{0.100000in}}{\pgfqpoint{3.400000in}{3.355557in}}%
\pgfusepath{clip}%
\pgfsetbuttcap%
\pgfsetroundjoin%
\pgfsetlinewidth{2.007500pt}%
\definecolor{currentstroke}{rgb}{0.298039,0.470588,0.658824}%
\pgfsetstrokecolor{currentstroke}%
\pgfsetdash{}{0pt}%
\pgfpathmoveto{\pgfqpoint{1.872914in}{0.781734in}}%
\pgfpathlineto{\pgfqpoint{2.103286in}{0.324818in}}%
\pgfusepath{stroke}%
\end{pgfscope}%
\begin{pgfscope}%
\pgfpathrectangle{\pgfqpoint{0.100000in}{0.100000in}}{\pgfqpoint{3.400000in}{3.355557in}}%
\pgfusepath{clip}%
\pgfsetbuttcap%
\pgfsetroundjoin%
\pgfsetlinewidth{2.007500pt}%
\definecolor{currentstroke}{rgb}{0.298039,0.470588,0.658824}%
\pgfsetstrokecolor{currentstroke}%
\pgfsetdash{}{0pt}%
\pgfpathmoveto{\pgfqpoint{3.275334in}{1.770668in}}%
\pgfpathlineto{\pgfqpoint{2.873934in}{0.783784in}}%
\pgfusepath{stroke}%
\end{pgfscope}%
\begin{pgfscope}%
\pgfpathrectangle{\pgfqpoint{0.100000in}{0.100000in}}{\pgfqpoint{3.400000in}{3.355557in}}%
\pgfusepath{clip}%
\pgfsetbuttcap%
\pgfsetroundjoin%
\pgfsetlinewidth{2.007500pt}%
\definecolor{currentstroke}{rgb}{0.960784,0.521569,0.094118}%
\pgfsetstrokecolor{currentstroke}%
\pgfsetdash{}{0pt}%
\pgfpathmoveto{\pgfqpoint{0.441220in}{1.315469in}}%
\pgfpathlineto{\pgfqpoint{1.103702in}{0.600999in}}%
\pgfusepath{stroke}%
\end{pgfscope}%
\begin{pgfscope}%
\pgfpathrectangle{\pgfqpoint{0.100000in}{0.100000in}}{\pgfqpoint{3.400000in}{3.355557in}}%
\pgfusepath{clip}%
\pgfsetbuttcap%
\pgfsetroundjoin%
\pgfsetlinewidth{2.007500pt}%
\definecolor{currentstroke}{rgb}{0.960784,0.521569,0.094118}%
\pgfsetstrokecolor{currentstroke}%
\pgfsetdash{}{0pt}%
\pgfpathmoveto{\pgfqpoint{0.441220in}{1.315469in}}%
\pgfpathlineto{\pgfqpoint{0.374574in}{2.275779in}}%
\pgfusepath{stroke}%
\end{pgfscope}%
\begin{pgfscope}%
\pgfpathrectangle{\pgfqpoint{0.100000in}{0.100000in}}{\pgfqpoint{3.400000in}{3.355557in}}%
\pgfusepath{clip}%
\pgfsetbuttcap%
\pgfsetroundjoin%
\pgfsetlinewidth{2.007500pt}%
\definecolor{currentstroke}{rgb}{0.960784,0.521569,0.094118}%
\pgfsetstrokecolor{currentstroke}%
\pgfsetdash{}{0pt}%
\pgfpathmoveto{\pgfqpoint{0.971831in}{3.026632in}}%
\pgfpathlineto{\pgfqpoint{0.374574in}{2.275779in}}%
\pgfusepath{stroke}%
\end{pgfscope}%
\begin{pgfscope}%
\pgfpathrectangle{\pgfqpoint{0.100000in}{0.100000in}}{\pgfqpoint{3.400000in}{3.355557in}}%
\pgfusepath{clip}%
\pgfsetbuttcap%
\pgfsetroundjoin%
\pgfsetlinewidth{2.007500pt}%
\definecolor{currentstroke}{rgb}{0.960784,0.521569,0.094118}%
\pgfsetstrokecolor{currentstroke}%
\pgfsetdash{}{0pt}%
\pgfpathmoveto{\pgfqpoint{1.923175in}{3.174364in}}%
\pgfpathlineto{\pgfqpoint{0.971831in}{3.026632in}}%
\pgfusepath{stroke}%
\end{pgfscope}%
\begin{pgfscope}%
\pgfpathrectangle{\pgfqpoint{0.100000in}{0.100000in}}{\pgfqpoint{3.400000in}{3.355557in}}%
\pgfusepath{clip}%
\pgfsetbuttcap%
\pgfsetroundjoin%
\pgfsetlinewidth{2.007500pt}%
\definecolor{currentstroke}{rgb}{0.960784,0.521569,0.094118}%
\pgfsetstrokecolor{currentstroke}%
\pgfsetdash{}{0pt}%
\pgfpathmoveto{\pgfqpoint{1.923175in}{3.174364in}}%
\pgfpathlineto{\pgfqpoint{2.775141in}{2.690131in}}%
\pgfusepath{stroke}%
\end{pgfscope}%
\begin{pgfscope}%
\pgfpathrectangle{\pgfqpoint{0.100000in}{0.100000in}}{\pgfqpoint{3.400000in}{3.355557in}}%
\pgfusepath{clip}%
\pgfsetbuttcap%
\pgfsetroundjoin%
\pgfsetlinewidth{2.007500pt}%
\definecolor{currentstroke}{rgb}{0.960784,0.521569,0.094118}%
\pgfsetstrokecolor{currentstroke}%
\pgfsetdash{}{0pt}%
\pgfpathmoveto{\pgfqpoint{2.689164in}{1.977822in}}%
\pgfpathlineto{\pgfqpoint{2.775141in}{2.690131in}}%
\pgfusepath{stroke}%
\end{pgfscope}%
\begin{pgfscope}%
\pgfpathrectangle{\pgfqpoint{0.100000in}{0.100000in}}{\pgfqpoint{3.400000in}{3.355557in}}%
\pgfusepath{clip}%
\pgfsetbuttcap%
\pgfsetroundjoin%
\pgfsetlinewidth{2.007500pt}%
\definecolor{currentstroke}{rgb}{0.619608,0.619608,0.619608}%
\pgfsetstrokecolor{currentstroke}%
\pgfsetdash{{7.400000pt}{3.200000pt}}{0.000000pt}%
\pgfpathmoveto{\pgfqpoint{3.275334in}{1.770668in}}%
\pgfpathlineto{\pgfqpoint{2.689164in}{1.977822in}}%
\pgfusepath{stroke}%
\end{pgfscope}%
\begin{pgfscope}%
\pgfpathrectangle{\pgfqpoint{0.100000in}{0.100000in}}{\pgfqpoint{3.400000in}{3.355557in}}%
\pgfusepath{clip}%
\pgfsetbuttcap%
\pgfsetroundjoin%
\definecolor{currentfill}{rgb}{0.298039,0.470588,0.658824}%
\pgfsetfillcolor{currentfill}%
\pgfsetlinewidth{1.003750pt}%
\definecolor{currentstroke}{rgb}{0.298039,0.470588,0.658824}%
\pgfsetstrokecolor{currentstroke}%
\pgfsetdash{}{0pt}%
\pgfpathmoveto{\pgfqpoint{3.275334in}{1.586936in}}%
\pgfpathcurveto{\pgfqpoint{3.324060in}{1.586936in}}{\pgfqpoint{3.370797in}{1.606295in}}{\pgfqpoint{3.405252in}{1.640750in}}%
\pgfpathcurveto{\pgfqpoint{3.439707in}{1.675204in}}{\pgfqpoint{3.459066in}{1.721942in}}{\pgfqpoint{3.459066in}{1.770668in}}%
\pgfpathcurveto{\pgfqpoint{3.459066in}{1.819395in}}{\pgfqpoint{3.439707in}{1.866132in}}{\pgfqpoint{3.405252in}{1.900587in}}%
\pgfpathcurveto{\pgfqpoint{3.370797in}{1.935042in}}{\pgfqpoint{3.324060in}{1.954401in}}{\pgfqpoint{3.275334in}{1.954401in}}%
\pgfpathcurveto{\pgfqpoint{3.226607in}{1.954401in}}{\pgfqpoint{3.179870in}{1.935042in}}{\pgfqpoint{3.145415in}{1.900587in}}%
\pgfpathcurveto{\pgfqpoint{3.110960in}{1.866132in}}{\pgfqpoint{3.091601in}{1.819395in}}{\pgfqpoint{3.091601in}{1.770668in}}%
\pgfpathcurveto{\pgfqpoint{3.091601in}{1.721942in}}{\pgfqpoint{3.110960in}{1.675204in}}{\pgfqpoint{3.145415in}{1.640750in}}%
\pgfpathcurveto{\pgfqpoint{3.179870in}{1.606295in}}{\pgfqpoint{3.226607in}{1.586936in}}{\pgfqpoint{3.275334in}{1.586936in}}%
\pgfpathlineto{\pgfqpoint{3.275334in}{1.586936in}}%
\pgfpathclose%
\pgfusepath{stroke,fill}%
\end{pgfscope}%
\begin{pgfscope}%
\pgfpathrectangle{\pgfqpoint{0.100000in}{0.100000in}}{\pgfqpoint{3.400000in}{3.355557in}}%
\pgfusepath{clip}%
\pgfsetbuttcap%
\pgfsetroundjoin%
\definecolor{currentfill}{rgb}{0.298039,0.470588,0.658824}%
\pgfsetfillcolor{currentfill}%
\pgfsetlinewidth{1.003750pt}%
\definecolor{currentstroke}{rgb}{0.298039,0.470588,0.658824}%
\pgfsetstrokecolor{currentstroke}%
\pgfsetdash{}{0pt}%
\pgfpathmoveto{\pgfqpoint{1.872914in}{0.598002in}}%
\pgfpathcurveto{\pgfqpoint{1.921640in}{0.598002in}}{\pgfqpoint{1.968378in}{0.617361in}}{\pgfqpoint{2.002832in}{0.651816in}}%
\pgfpathcurveto{\pgfqpoint{2.037287in}{0.686270in}}{\pgfqpoint{2.056646in}{0.733008in}}{\pgfqpoint{2.056646in}{0.781734in}}%
\pgfpathcurveto{\pgfqpoint{2.056646in}{0.830461in}}{\pgfqpoint{2.037287in}{0.877198in}}{\pgfqpoint{2.002832in}{0.911653in}}%
\pgfpathcurveto{\pgfqpoint{1.968378in}{0.946108in}}{\pgfqpoint{1.921640in}{0.965467in}}{\pgfqpoint{1.872914in}{0.965467in}}%
\pgfpathcurveto{\pgfqpoint{1.824187in}{0.965467in}}{\pgfqpoint{1.777450in}{0.946108in}}{\pgfqpoint{1.742995in}{0.911653in}}%
\pgfpathcurveto{\pgfqpoint{1.708540in}{0.877198in}}{\pgfqpoint{1.689181in}{0.830461in}}{\pgfqpoint{1.689181in}{0.781734in}}%
\pgfpathcurveto{\pgfqpoint{1.689181in}{0.733008in}}{\pgfqpoint{1.708540in}{0.686270in}}{\pgfqpoint{1.742995in}{0.651816in}}%
\pgfpathcurveto{\pgfqpoint{1.777450in}{0.617361in}}{\pgfqpoint{1.824187in}{0.598002in}}{\pgfqpoint{1.872914in}{0.598002in}}%
\pgfpathlineto{\pgfqpoint{1.872914in}{0.598002in}}%
\pgfpathclose%
\pgfusepath{stroke,fill}%
\end{pgfscope}%
\begin{pgfscope}%
\pgfpathrectangle{\pgfqpoint{0.100000in}{0.100000in}}{\pgfqpoint{3.400000in}{3.355557in}}%
\pgfusepath{clip}%
\pgfsetbuttcap%
\pgfsetroundjoin%
\definecolor{currentfill}{rgb}{0.960784,0.521569,0.094118}%
\pgfsetfillcolor{currentfill}%
\pgfsetlinewidth{1.003750pt}%
\definecolor{currentstroke}{rgb}{0.960784,0.521569,0.094118}%
\pgfsetstrokecolor{currentstroke}%
\pgfsetdash{}{0pt}%
\pgfpathmoveto{\pgfqpoint{1.923175in}{2.990631in}}%
\pgfpathcurveto{\pgfqpoint{1.971902in}{2.990631in}}{\pgfqpoint{2.018639in}{3.009990in}}{\pgfqpoint{2.053094in}{3.044445in}}%
\pgfpathcurveto{\pgfqpoint{2.087549in}{3.078900in}}{\pgfqpoint{2.106908in}{3.125637in}}{\pgfqpoint{2.106908in}{3.174364in}}%
\pgfpathcurveto{\pgfqpoint{2.106908in}{3.223090in}}{\pgfqpoint{2.087549in}{3.269828in}}{\pgfqpoint{2.053094in}{3.304282in}}%
\pgfpathcurveto{\pgfqpoint{2.018639in}{3.338737in}}{\pgfqpoint{1.971902in}{3.358096in}}{\pgfqpoint{1.923175in}{3.358096in}}%
\pgfpathcurveto{\pgfqpoint{1.874449in}{3.358096in}}{\pgfqpoint{1.827711in}{3.338737in}}{\pgfqpoint{1.793257in}{3.304282in}}%
\pgfpathcurveto{\pgfqpoint{1.758802in}{3.269828in}}{\pgfqpoint{1.739442in}{3.223090in}}{\pgfqpoint{1.739442in}{3.174364in}}%
\pgfpathcurveto{\pgfqpoint{1.739442in}{3.125637in}}{\pgfqpoint{1.758802in}{3.078900in}}{\pgfqpoint{1.793257in}{3.044445in}}%
\pgfpathcurveto{\pgfqpoint{1.827711in}{3.009990in}}{\pgfqpoint{1.874449in}{2.990631in}}{\pgfqpoint{1.923175in}{2.990631in}}%
\pgfpathlineto{\pgfqpoint{1.923175in}{2.990631in}}%
\pgfpathclose%
\pgfusepath{stroke,fill}%
\end{pgfscope}%
\begin{pgfscope}%
\pgfpathrectangle{\pgfqpoint{0.100000in}{0.100000in}}{\pgfqpoint{3.400000in}{3.355557in}}%
\pgfusepath{clip}%
\pgfsetbuttcap%
\pgfsetroundjoin%
\definecolor{currentfill}{rgb}{0.960784,0.521569,0.094118}%
\pgfsetfillcolor{currentfill}%
\pgfsetlinewidth{1.003750pt}%
\definecolor{currentstroke}{rgb}{0.960784,0.521569,0.094118}%
\pgfsetstrokecolor{currentstroke}%
\pgfsetdash{}{0pt}%
\pgfpathmoveto{\pgfqpoint{2.689164in}{1.794090in}}%
\pgfpathcurveto{\pgfqpoint{2.737891in}{1.794090in}}{\pgfqpoint{2.784628in}{1.813449in}}{\pgfqpoint{2.819083in}{1.847904in}}%
\pgfpathcurveto{\pgfqpoint{2.853538in}{1.882359in}}{\pgfqpoint{2.872897in}{1.929096in}}{\pgfqpoint{2.872897in}{1.977822in}}%
\pgfpathcurveto{\pgfqpoint{2.872897in}{2.026549in}}{\pgfqpoint{2.853538in}{2.073286in}}{\pgfqpoint{2.819083in}{2.107741in}}%
\pgfpathcurveto{\pgfqpoint{2.784628in}{2.142196in}}{\pgfqpoint{2.737891in}{2.161555in}}{\pgfqpoint{2.689164in}{2.161555in}}%
\pgfpathcurveto{\pgfqpoint{2.640438in}{2.161555in}}{\pgfqpoint{2.593701in}{2.142196in}}{\pgfqpoint{2.559246in}{2.107741in}}%
\pgfpathcurveto{\pgfqpoint{2.524791in}{2.073286in}}{\pgfqpoint{2.505432in}{2.026549in}}{\pgfqpoint{2.505432in}{1.977822in}}%
\pgfpathcurveto{\pgfqpoint{2.505432in}{1.929096in}}{\pgfqpoint{2.524791in}{1.882359in}}{\pgfqpoint{2.559246in}{1.847904in}}%
\pgfpathcurveto{\pgfqpoint{2.593701in}{1.813449in}}{\pgfqpoint{2.640438in}{1.794090in}}{\pgfqpoint{2.689164in}{1.794090in}}%
\pgfpathlineto{\pgfqpoint{2.689164in}{1.794090in}}%
\pgfpathclose%
\pgfusepath{stroke,fill}%
\end{pgfscope}%
\begin{pgfscope}%
\pgfpathrectangle{\pgfqpoint{0.100000in}{0.100000in}}{\pgfqpoint{3.400000in}{3.355557in}}%
\pgfusepath{clip}%
\pgfsetbuttcap%
\pgfsetroundjoin%
\definecolor{currentfill}{rgb}{0.298039,0.470588,0.658824}%
\pgfsetfillcolor{currentfill}%
\pgfsetlinewidth{1.003750pt}%
\definecolor{currentstroke}{rgb}{0.298039,0.470588,0.658824}%
\pgfsetstrokecolor{currentstroke}%
\pgfsetdash{}{0pt}%
\pgfpathmoveto{\pgfqpoint{2.103286in}{0.141085in}}%
\pgfpathcurveto{\pgfqpoint{2.152013in}{0.141085in}}{\pgfqpoint{2.198750in}{0.160444in}}{\pgfqpoint{2.233205in}{0.194899in}}%
\pgfpathcurveto{\pgfqpoint{2.267660in}{0.229354in}}{\pgfqpoint{2.287019in}{0.276091in}}{\pgfqpoint{2.287019in}{0.324818in}}%
\pgfpathcurveto{\pgfqpoint{2.287019in}{0.373544in}}{\pgfqpoint{2.267660in}{0.420282in}}{\pgfqpoint{2.233205in}{0.454737in}}%
\pgfpathcurveto{\pgfqpoint{2.198750in}{0.489191in}}{\pgfqpoint{2.152013in}{0.508551in}}{\pgfqpoint{2.103286in}{0.508551in}}%
\pgfpathcurveto{\pgfqpoint{2.054560in}{0.508551in}}{\pgfqpoint{2.007822in}{0.489191in}}{\pgfqpoint{1.973367in}{0.454737in}}%
\pgfpathcurveto{\pgfqpoint{1.938913in}{0.420282in}}{\pgfqpoint{1.919553in}{0.373544in}}{\pgfqpoint{1.919553in}{0.324818in}}%
\pgfpathcurveto{\pgfqpoint{1.919553in}{0.276091in}}{\pgfqpoint{1.938913in}{0.229354in}}{\pgfqpoint{1.973367in}{0.194899in}}%
\pgfpathcurveto{\pgfqpoint{2.007822in}{0.160444in}}{\pgfqpoint{2.054560in}{0.141085in}}{\pgfqpoint{2.103286in}{0.141085in}}%
\pgfpathlineto{\pgfqpoint{2.103286in}{0.141085in}}%
\pgfpathclose%
\pgfusepath{stroke,fill}%
\end{pgfscope}%
\begin{pgfscope}%
\pgfpathrectangle{\pgfqpoint{0.100000in}{0.100000in}}{\pgfqpoint{3.400000in}{3.355557in}}%
\pgfusepath{clip}%
\pgfsetbuttcap%
\pgfsetroundjoin%
\definecolor{currentfill}{rgb}{0.960784,0.521569,0.094118}%
\pgfsetfillcolor{currentfill}%
\pgfsetlinewidth{1.003750pt}%
\definecolor{currentstroke}{rgb}{0.960784,0.521569,0.094118}%
\pgfsetstrokecolor{currentstroke}%
\pgfsetdash{}{0pt}%
\pgfpathmoveto{\pgfqpoint{0.441220in}{1.131736in}}%
\pgfpathcurveto{\pgfqpoint{0.489946in}{1.131736in}}{\pgfqpoint{0.536684in}{1.151095in}}{\pgfqpoint{0.571139in}{1.185550in}}%
\pgfpathcurveto{\pgfqpoint{0.605593in}{1.220005in}}{\pgfqpoint{0.624953in}{1.266742in}}{\pgfqpoint{0.624953in}{1.315469in}}%
\pgfpathcurveto{\pgfqpoint{0.624953in}{1.364195in}}{\pgfqpoint{0.605593in}{1.410933in}}{\pgfqpoint{0.571139in}{1.445388in}}%
\pgfpathcurveto{\pgfqpoint{0.536684in}{1.479842in}}{\pgfqpoint{0.489946in}{1.499202in}}{\pgfqpoint{0.441220in}{1.499202in}}%
\pgfpathcurveto{\pgfqpoint{0.392493in}{1.499202in}}{\pgfqpoint{0.345756in}{1.479842in}}{\pgfqpoint{0.311301in}{1.445388in}}%
\pgfpathcurveto{\pgfqpoint{0.276846in}{1.410933in}}{\pgfqpoint{0.257487in}{1.364195in}}{\pgfqpoint{0.257487in}{1.315469in}}%
\pgfpathcurveto{\pgfqpoint{0.257487in}{1.266742in}}{\pgfqpoint{0.276846in}{1.220005in}}{\pgfqpoint{0.311301in}{1.185550in}}%
\pgfpathcurveto{\pgfqpoint{0.345756in}{1.151095in}}{\pgfqpoint{0.392493in}{1.131736in}}{\pgfqpoint{0.441220in}{1.131736in}}%
\pgfpathlineto{\pgfqpoint{0.441220in}{1.131736in}}%
\pgfpathclose%
\pgfusepath{stroke,fill}%
\end{pgfscope}%
\begin{pgfscope}%
\pgfpathrectangle{\pgfqpoint{0.100000in}{0.100000in}}{\pgfqpoint{3.400000in}{3.355557in}}%
\pgfusepath{clip}%
\pgfsetbuttcap%
\pgfsetroundjoin%
\definecolor{currentfill}{rgb}{0.960784,0.521569,0.094118}%
\pgfsetfillcolor{currentfill}%
\pgfsetlinewidth{1.003750pt}%
\definecolor{currentstroke}{rgb}{0.960784,0.521569,0.094118}%
\pgfsetstrokecolor{currentstroke}%
\pgfsetdash{}{0pt}%
\pgfpathmoveto{\pgfqpoint{0.971831in}{2.842899in}}%
\pgfpathcurveto{\pgfqpoint{1.020558in}{2.842899in}}{\pgfqpoint{1.067295in}{2.862258in}}{\pgfqpoint{1.101750in}{2.896713in}}%
\pgfpathcurveto{\pgfqpoint{1.136205in}{2.931168in}}{\pgfqpoint{1.155564in}{2.977905in}}{\pgfqpoint{1.155564in}{3.026632in}}%
\pgfpathcurveto{\pgfqpoint{1.155564in}{3.075358in}}{\pgfqpoint{1.136205in}{3.122095in}}{\pgfqpoint{1.101750in}{3.156550in}}%
\pgfpathcurveto{\pgfqpoint{1.067295in}{3.191005in}}{\pgfqpoint{1.020558in}{3.210364in}}{\pgfqpoint{0.971831in}{3.210364in}}%
\pgfpathcurveto{\pgfqpoint{0.923105in}{3.210364in}}{\pgfqpoint{0.876367in}{3.191005in}}{\pgfqpoint{0.841913in}{3.156550in}}%
\pgfpathcurveto{\pgfqpoint{0.807458in}{3.122095in}}{\pgfqpoint{0.788098in}{3.075358in}}{\pgfqpoint{0.788098in}{3.026632in}}%
\pgfpathcurveto{\pgfqpoint{0.788098in}{2.977905in}}{\pgfqpoint{0.807458in}{2.931168in}}{\pgfqpoint{0.841913in}{2.896713in}}%
\pgfpathcurveto{\pgfqpoint{0.876367in}{2.862258in}}{\pgfqpoint{0.923105in}{2.842899in}}{\pgfqpoint{0.971831in}{2.842899in}}%
\pgfpathlineto{\pgfqpoint{0.971831in}{2.842899in}}%
\pgfpathclose%
\pgfusepath{stroke,fill}%
\end{pgfscope}%
\begin{pgfscope}%
\pgfpathrectangle{\pgfqpoint{0.100000in}{0.100000in}}{\pgfqpoint{3.400000in}{3.355557in}}%
\pgfusepath{clip}%
\pgfsetbuttcap%
\pgfsetroundjoin%
\definecolor{currentfill}{rgb}{0.960784,0.521569,0.094118}%
\pgfsetfillcolor{currentfill}%
\pgfsetlinewidth{1.003750pt}%
\definecolor{currentstroke}{rgb}{0.960784,0.521569,0.094118}%
\pgfsetstrokecolor{currentstroke}%
\pgfsetdash{}{0pt}%
\pgfpathmoveto{\pgfqpoint{0.374574in}{2.092046in}}%
\pgfpathcurveto{\pgfqpoint{0.423300in}{2.092046in}}{\pgfqpoint{0.470038in}{2.111405in}}{\pgfqpoint{0.504492in}{2.145860in}}%
\pgfpathcurveto{\pgfqpoint{0.538947in}{2.180315in}}{\pgfqpoint{0.558307in}{2.227052in}}{\pgfqpoint{0.558307in}{2.275779in}}%
\pgfpathcurveto{\pgfqpoint{0.558307in}{2.324505in}}{\pgfqpoint{0.538947in}{2.371242in}}{\pgfqpoint{0.504492in}{2.405697in}}%
\pgfpathcurveto{\pgfqpoint{0.470038in}{2.440152in}}{\pgfqpoint{0.423300in}{2.459511in}}{\pgfqpoint{0.374574in}{2.459511in}}%
\pgfpathcurveto{\pgfqpoint{0.325847in}{2.459511in}}{\pgfqpoint{0.279110in}{2.440152in}}{\pgfqpoint{0.244655in}{2.405697in}}%
\pgfpathcurveto{\pgfqpoint{0.210200in}{2.371242in}}{\pgfqpoint{0.190841in}{2.324505in}}{\pgfqpoint{0.190841in}{2.275779in}}%
\pgfpathcurveto{\pgfqpoint{0.190841in}{2.227052in}}{\pgfqpoint{0.210200in}{2.180315in}}{\pgfqpoint{0.244655in}{2.145860in}}%
\pgfpathcurveto{\pgfqpoint{0.279110in}{2.111405in}}{\pgfqpoint{0.325847in}{2.092046in}}{\pgfqpoint{0.374574in}{2.092046in}}%
\pgfpathlineto{\pgfqpoint{0.374574in}{2.092046in}}%
\pgfpathclose%
\pgfusepath{stroke,fill}%
\end{pgfscope}%
\begin{pgfscope}%
\pgfpathrectangle{\pgfqpoint{0.100000in}{0.100000in}}{\pgfqpoint{3.400000in}{3.355557in}}%
\pgfusepath{clip}%
\pgfsetbuttcap%
\pgfsetroundjoin%
\definecolor{currentfill}{rgb}{0.960784,0.521569,0.094118}%
\pgfsetfillcolor{currentfill}%
\pgfsetlinewidth{1.003750pt}%
\definecolor{currentstroke}{rgb}{0.960784,0.521569,0.094118}%
\pgfsetstrokecolor{currentstroke}%
\pgfsetdash{}{0pt}%
\pgfpathmoveto{\pgfqpoint{1.103702in}{0.417266in}}%
\pgfpathcurveto{\pgfqpoint{1.152429in}{0.417266in}}{\pgfqpoint{1.199166in}{0.436626in}}{\pgfqpoint{1.233621in}{0.471081in}}%
\pgfpathcurveto{\pgfqpoint{1.268076in}{0.505535in}}{\pgfqpoint{1.287435in}{0.552273in}}{\pgfqpoint{1.287435in}{0.600999in}}%
\pgfpathcurveto{\pgfqpoint{1.287435in}{0.649726in}}{\pgfqpoint{1.268076in}{0.696463in}}{\pgfqpoint{1.233621in}{0.730918in}}%
\pgfpathcurveto{\pgfqpoint{1.199166in}{0.765373in}}{\pgfqpoint{1.152429in}{0.784732in}}{\pgfqpoint{1.103702in}{0.784732in}}%
\pgfpathcurveto{\pgfqpoint{1.054976in}{0.784732in}}{\pgfqpoint{1.008238in}{0.765373in}}{\pgfqpoint{0.973783in}{0.730918in}}%
\pgfpathcurveto{\pgfqpoint{0.939329in}{0.696463in}}{\pgfqpoint{0.919969in}{0.649726in}}{\pgfqpoint{0.919969in}{0.600999in}}%
\pgfpathcurveto{\pgfqpoint{0.919969in}{0.552273in}}{\pgfqpoint{0.939329in}{0.505535in}}{\pgfqpoint{0.973783in}{0.471081in}}%
\pgfpathcurveto{\pgfqpoint{1.008238in}{0.436626in}}{\pgfqpoint{1.054976in}{0.417266in}}{\pgfqpoint{1.103702in}{0.417266in}}%
\pgfpathlineto{\pgfqpoint{1.103702in}{0.417266in}}%
\pgfpathclose%
\pgfusepath{stroke,fill}%
\end{pgfscope}%
\begin{pgfscope}%
\pgfpathrectangle{\pgfqpoint{0.100000in}{0.100000in}}{\pgfqpoint{3.400000in}{3.355557in}}%
\pgfusepath{clip}%
\pgfsetbuttcap%
\pgfsetroundjoin%
\definecolor{currentfill}{rgb}{0.298039,0.470588,0.658824}%
\pgfsetfillcolor{currentfill}%
\pgfsetlinewidth{1.003750pt}%
\definecolor{currentstroke}{rgb}{0.298039,0.470588,0.658824}%
\pgfsetstrokecolor{currentstroke}%
\pgfsetdash{}{0pt}%
\pgfpathmoveto{\pgfqpoint{2.422510in}{0.977435in}}%
\pgfpathcurveto{\pgfqpoint{2.471237in}{0.977435in}}{\pgfqpoint{2.517974in}{0.996794in}}{\pgfqpoint{2.552429in}{1.031249in}}%
\pgfpathcurveto{\pgfqpoint{2.586884in}{1.065703in}}{\pgfqpoint{2.606243in}{1.112441in}}{\pgfqpoint{2.606243in}{1.161167in}}%
\pgfpathcurveto{\pgfqpoint{2.606243in}{1.209894in}}{\pgfqpoint{2.586884in}{1.256631in}}{\pgfqpoint{2.552429in}{1.291086in}}%
\pgfpathcurveto{\pgfqpoint{2.517974in}{1.325541in}}{\pgfqpoint{2.471237in}{1.344900in}}{\pgfqpoint{2.422510in}{1.344900in}}%
\pgfpathcurveto{\pgfqpoint{2.373784in}{1.344900in}}{\pgfqpoint{2.327046in}{1.325541in}}{\pgfqpoint{2.292591in}{1.291086in}}%
\pgfpathcurveto{\pgfqpoint{2.258137in}{1.256631in}}{\pgfqpoint{2.238777in}{1.209894in}}{\pgfqpoint{2.238777in}{1.161167in}}%
\pgfpathcurveto{\pgfqpoint{2.238777in}{1.112441in}}{\pgfqpoint{2.258137in}{1.065703in}}{\pgfqpoint{2.292591in}{1.031249in}}%
\pgfpathcurveto{\pgfqpoint{2.327046in}{0.996794in}}{\pgfqpoint{2.373784in}{0.977435in}}{\pgfqpoint{2.422510in}{0.977435in}}%
\pgfpathlineto{\pgfqpoint{2.422510in}{0.977435in}}%
\pgfpathclose%
\pgfusepath{stroke,fill}%
\end{pgfscope}%
\begin{pgfscope}%
\pgfpathrectangle{\pgfqpoint{0.100000in}{0.100000in}}{\pgfqpoint{3.400000in}{3.355557in}}%
\pgfusepath{clip}%
\pgfsetbuttcap%
\pgfsetroundjoin%
\definecolor{currentfill}{rgb}{0.298039,0.470588,0.658824}%
\pgfsetfillcolor{currentfill}%
\pgfsetlinewidth{1.003750pt}%
\definecolor{currentstroke}{rgb}{0.298039,0.470588,0.658824}%
\pgfsetstrokecolor{currentstroke}%
\pgfsetdash{}{0pt}%
\pgfpathmoveto{\pgfqpoint{2.690201in}{0.600051in}}%
\pgfpathlineto{\pgfqpoint{3.057666in}{0.600051in}}%
\pgfpathlineto{\pgfqpoint{3.057666in}{0.967517in}}%
\pgfpathlineto{\pgfqpoint{2.690201in}{0.967517in}}%
\pgfpathlineto{\pgfqpoint{2.690201in}{0.600051in}}%
\pgfpathclose%
\pgfusepath{stroke,fill}%
\end{pgfscope}%
\begin{pgfscope}%
\pgfpathrectangle{\pgfqpoint{0.100000in}{0.100000in}}{\pgfqpoint{3.400000in}{3.355557in}}%
\pgfusepath{clip}%
\pgfsetbuttcap%
\pgfsetroundjoin%
\definecolor{currentfill}{rgb}{0.960784,0.521569,0.094118}%
\pgfsetfillcolor{currentfill}%
\pgfsetlinewidth{1.003750pt}%
\definecolor{currentstroke}{rgb}{0.960784,0.521569,0.094118}%
\pgfsetstrokecolor{currentstroke}%
\pgfsetdash{}{0pt}%
\pgfpathmoveto{\pgfqpoint{2.591408in}{2.506398in}}%
\pgfpathlineto{\pgfqpoint{2.958874in}{2.506398in}}%
\pgfpathlineto{\pgfqpoint{2.958874in}{2.873863in}}%
\pgfpathlineto{\pgfqpoint{2.591408in}{2.873863in}}%
\pgfpathlineto{\pgfqpoint{2.591408in}{2.506398in}}%
\pgfpathclose%
\pgfusepath{stroke,fill}%
\end{pgfscope}%
\begin{pgfscope}%
\definecolor{textcolor}{rgb}{0.000000,0.000000,0.000000}%
\pgfsetstrokecolor{textcolor}%
\pgfsetfillcolor{textcolor}%
\pgftext[x=3.275334in,y=1.770668in,,]{\color{textcolor}\sffamily\fontsize{19.000000}{22.800000}\selectfont \(\displaystyle S_{1}\)}%
\end{pgfscope}%
\begin{pgfscope}%
\definecolor{textcolor}{rgb}{0.000000,0.000000,0.000000}%
\pgfsetstrokecolor{textcolor}%
\pgfsetfillcolor{textcolor}%
\pgftext[x=1.872914in,y=0.781734in,,]{\color{textcolor}\sffamily\fontsize{19.000000}{22.800000}\selectfont \(\displaystyle S_{2}\)}%
\end{pgfscope}%
\begin{pgfscope}%
\definecolor{textcolor}{rgb}{0.000000,0.000000,0.000000}%
\pgfsetstrokecolor{textcolor}%
\pgfsetfillcolor{textcolor}%
\pgftext[x=1.923175in,y=3.174364in,,]{\color{textcolor}\sffamily\fontsize{19.000000}{22.800000}\selectfont \(\displaystyle S_{3}\)}%
\end{pgfscope}%
\begin{pgfscope}%
\definecolor{textcolor}{rgb}{0.000000,0.000000,0.000000}%
\pgfsetstrokecolor{textcolor}%
\pgfsetfillcolor{textcolor}%
\pgftext[x=2.689164in,y=1.977822in,,]{\color{textcolor}\sffamily\fontsize{19.000000}{22.800000}\selectfont \(\displaystyle S_{4}\)}%
\end{pgfscope}%
\begin{pgfscope}%
\definecolor{textcolor}{rgb}{0.000000,0.000000,0.000000}%
\pgfsetstrokecolor{textcolor}%
\pgfsetfillcolor{textcolor}%
\pgftext[x=2.103286in,y=0.324818in,,]{\color{textcolor}\sffamily\fontsize{19.000000}{22.800000}\selectfont \(\displaystyle S_{5}\)}%
\end{pgfscope}%
\begin{pgfscope}%
\definecolor{textcolor}{rgb}{0.000000,0.000000,0.000000}%
\pgfsetstrokecolor{textcolor}%
\pgfsetfillcolor{textcolor}%
\pgftext[x=0.441220in,y=1.315469in,,]{\color{textcolor}\sffamily\fontsize{19.000000}{22.800000}\selectfont \(\displaystyle S_{6}\)}%
\end{pgfscope}%
\begin{pgfscope}%
\definecolor{textcolor}{rgb}{0.000000,0.000000,0.000000}%
\pgfsetstrokecolor{textcolor}%
\pgfsetfillcolor{textcolor}%
\pgftext[x=0.971831in,y=3.026632in,,]{\color{textcolor}\sffamily\fontsize{19.000000}{22.800000}\selectfont \(\displaystyle S_{7}\)}%
\end{pgfscope}%
\begin{pgfscope}%
\definecolor{textcolor}{rgb}{0.000000,0.000000,0.000000}%
\pgfsetstrokecolor{textcolor}%
\pgfsetfillcolor{textcolor}%
\pgftext[x=0.374574in,y=2.275779in,,]{\color{textcolor}\sffamily\fontsize{19.000000}{22.800000}\selectfont \(\displaystyle S_{8}\)}%
\end{pgfscope}%
\begin{pgfscope}%
\definecolor{textcolor}{rgb}{0.000000,0.000000,0.000000}%
\pgfsetstrokecolor{textcolor}%
\pgfsetfillcolor{textcolor}%
\pgftext[x=1.103702in,y=0.600999in,,]{\color{textcolor}\sffamily\fontsize{19.000000}{22.800000}\selectfont \(\displaystyle S_{9}\)}%
\end{pgfscope}%
\begin{pgfscope}%
\definecolor{textcolor}{rgb}{0.000000,0.000000,0.000000}%
\pgfsetstrokecolor{textcolor}%
\pgfsetfillcolor{textcolor}%
\pgftext[x=2.422510in,y=1.161167in,,]{\color{textcolor}\sffamily\fontsize{19.000000}{22.800000}\selectfont \(\displaystyle S_{10}\)}%
\end{pgfscope}%
\begin{pgfscope}%
\definecolor{textcolor}{rgb}{0.000000,0.000000,0.000000}%
\pgfsetstrokecolor{textcolor}%
\pgfsetfillcolor{textcolor}%
\pgftext[x=2.873934in,y=0.783784in,,]{\color{textcolor}\sffamily\fontsize{19.000000}{22.800000}\selectfont \(\displaystyle P_{11}\)}%
\end{pgfscope}%
\begin{pgfscope}%
\definecolor{textcolor}{rgb}{0.000000,0.000000,0.000000}%
\pgfsetstrokecolor{textcolor}%
\pgfsetfillcolor{textcolor}%
\pgftext[x=2.775141in,y=2.690131in,,]{\color{textcolor}\sffamily\fontsize{19.000000}{22.800000}\selectfont \(\displaystyle P_{12}\)}%
\end{pgfscope}%
\end{pgfpicture}%
\makeatother%
\endgroup%

%% file: pgf_example/05_single_plot.tex
\begingroup%
\makeatletter%
\begin{pgfpicture}%
\pgfpathrectangle{\pgfpointorigin}{\pgfqpoint{3.600000in}{3.450472in}}%
\pgfusepath{use as bounding box, clip}%
\begin{pgfscope}%
\pgfsetbuttcap%
\pgfsetmiterjoin%
\definecolor{currentfill}{rgb}{1.000000,1.000000,1.000000}%
\pgfsetfillcolor{currentfill}%
\pgfsetlinewidth{0.000000pt}%
\definecolor{currentstroke}{rgb}{1.000000,1.000000,1.000000}%
\pgfsetstrokecolor{currentstroke}%
\pgfsetdash{}{0pt}%
\pgfpathmoveto{\pgfqpoint{0.000000in}{0.000000in}}%
\pgfpathlineto{\pgfqpoint{3.600000in}{0.000000in}}%
\pgfpathlineto{\pgfqpoint{3.600000in}{3.450472in}}%
\pgfpathlineto{\pgfqpoint{0.000000in}{3.450472in}}%
\pgfpathlineto{\pgfqpoint{0.000000in}{0.000000in}}%
\pgfpathclose%
\pgfusepath{fill}%
\end{pgfscope}%
\begin{pgfscope}%
\pgfpathrectangle{\pgfqpoint{0.100000in}{0.100000in}}{\pgfqpoint{3.400000in}{3.250472in}}%
\pgfusepath{clip}%
\pgfsetbuttcap%
\pgfsetroundjoin%
\pgfsetlinewidth{2.007500pt}%
\definecolor{currentstroke}{rgb}{0.619608,0.619608,0.619608}%
\pgfsetstrokecolor{currentstroke}%
\pgfsetdash{}{0pt}%
\pgfpathmoveto{\pgfqpoint{2.378843in}{1.127935in}}%
\pgfpathlineto{\pgfqpoint{2.816129in}{0.762370in}}%
\pgfusepath{stroke}%
\end{pgfscope}%
\begin{pgfscope}%
\pgfpathrectangle{\pgfqpoint{0.100000in}{0.100000in}}{\pgfqpoint{3.400000in}{3.250472in}}%
\pgfusepath{clip}%
\pgfsetbuttcap%
\pgfsetroundjoin%
\pgfsetlinewidth{2.007500pt}%
\definecolor{currentstroke}{rgb}{0.619608,0.619608,0.619608}%
\pgfsetstrokecolor{currentstroke}%
\pgfsetdash{}{0pt}%
\pgfpathmoveto{\pgfqpoint{2.069616in}{0.317777in}}%
\pgfpathlineto{\pgfqpoint{2.378843in}{1.127935in}}%
\pgfusepath{stroke}%
\end{pgfscope}%
\begin{pgfscope}%
\pgfpathrectangle{\pgfqpoint{0.100000in}{0.100000in}}{\pgfqpoint{3.400000in}{3.250472in}}%
\pgfusepath{clip}%
\pgfsetbuttcap%
\pgfsetroundjoin%
\pgfsetlinewidth{2.007500pt}%
\definecolor{currentstroke}{rgb}{0.619608,0.619608,0.619608}%
\pgfsetstrokecolor{currentstroke}%
\pgfsetdash{}{0pt}%
\pgfpathmoveto{\pgfqpoint{1.846458in}{0.760385in}}%
\pgfpathlineto{\pgfqpoint{2.069616in}{0.317777in}}%
\pgfusepath{stroke}%
\end{pgfscope}%
\begin{pgfscope}%
\pgfpathrectangle{\pgfqpoint{0.100000in}{0.100000in}}{\pgfqpoint{3.400000in}{3.250472in}}%
\pgfusepath{clip}%
\pgfsetbuttcap%
\pgfsetroundjoin%
\pgfsetlinewidth{2.007500pt}%
\definecolor{currentstroke}{rgb}{0.619608,0.619608,0.619608}%
\pgfsetstrokecolor{currentstroke}%
\pgfsetdash{}{0pt}%
\pgfpathmoveto{\pgfqpoint{3.204959in}{1.718348in}}%
\pgfpathlineto{\pgfqpoint{2.816129in}{0.762370in}}%
\pgfusepath{stroke}%
\end{pgfscope}%
\begin{pgfscope}%
\pgfpathrectangle{\pgfqpoint{0.100000in}{0.100000in}}{\pgfqpoint{3.400000in}{3.250472in}}%
\pgfusepath{clip}%
\pgfsetbuttcap%
\pgfsetroundjoin%
\pgfsetlinewidth{2.007500pt}%
\definecolor{currentstroke}{rgb}{0.960784,0.521569,0.094118}%
\pgfsetstrokecolor{currentstroke}%
\pgfsetdash{}{0pt}%
\pgfpathmoveto{\pgfqpoint{0.459600in}{1.277404in}}%
\pgfpathlineto{\pgfqpoint{1.101336in}{0.585310in}}%
\pgfusepath{stroke}%
\end{pgfscope}%
\begin{pgfscope}%
\pgfpathrectangle{\pgfqpoint{0.100000in}{0.100000in}}{\pgfqpoint{3.400000in}{3.250472in}}%
\pgfusepath{clip}%
\pgfsetbuttcap%
\pgfsetroundjoin%
\pgfsetlinewidth{2.007500pt}%
\definecolor{currentstroke}{rgb}{0.960784,0.521569,0.094118}%
\pgfsetstrokecolor{currentstroke}%
\pgfsetdash{}{0pt}%
\pgfpathmoveto{\pgfqpoint{0.459600in}{1.277404in}}%
\pgfpathlineto{\pgfqpoint{0.395041in}{2.207640in}}%
\pgfusepath{stroke}%
\end{pgfscope}%
\begin{pgfscope}%
\pgfpathrectangle{\pgfqpoint{0.100000in}{0.100000in}}{\pgfqpoint{3.400000in}{3.250472in}}%
\pgfusepath{clip}%
\pgfsetbuttcap%
\pgfsetroundjoin%
\pgfsetlinewidth{2.007500pt}%
\definecolor{currentstroke}{rgb}{0.960784,0.521569,0.094118}%
\pgfsetstrokecolor{currentstroke}%
\pgfsetdash{}{0pt}%
\pgfpathmoveto{\pgfqpoint{0.973595in}{2.934979in}}%
\pgfpathlineto{\pgfqpoint{0.395041in}{2.207640in}}%
\pgfusepath{stroke}%
\end{pgfscope}%
\begin{pgfscope}%
\pgfpathrectangle{\pgfqpoint{0.100000in}{0.100000in}}{\pgfqpoint{3.400000in}{3.250472in}}%
\pgfusepath{clip}%
\pgfsetbuttcap%
\pgfsetroundjoin%
\pgfsetlinewidth{2.007500pt}%
\definecolor{currentstroke}{rgb}{0.960784,0.521569,0.094118}%
\pgfsetstrokecolor{currentstroke}%
\pgfsetdash{}{0pt}%
\pgfpathmoveto{\pgfqpoint{1.895146in}{3.078085in}}%
\pgfpathlineto{\pgfqpoint{0.973595in}{2.934979in}}%
\pgfusepath{stroke}%
\end{pgfscope}%
\begin{pgfscope}%
\pgfpathrectangle{\pgfqpoint{0.100000in}{0.100000in}}{\pgfqpoint{3.400000in}{3.250472in}}%
\pgfusepath{clip}%
\pgfsetbuttcap%
\pgfsetroundjoin%
\pgfsetlinewidth{2.007500pt}%
\definecolor{currentstroke}{rgb}{0.960784,0.521569,0.094118}%
\pgfsetstrokecolor{currentstroke}%
\pgfsetdash{}{0pt}%
\pgfpathmoveto{\pgfqpoint{1.895146in}{3.078085in}}%
\pgfpathlineto{\pgfqpoint{2.720431in}{2.609016in}}%
\pgfusepath{stroke}%
\end{pgfscope}%
\begin{pgfscope}%
\pgfpathrectangle{\pgfqpoint{0.100000in}{0.100000in}}{\pgfqpoint{3.400000in}{3.250472in}}%
\pgfusepath{clip}%
\pgfsetbuttcap%
\pgfsetroundjoin%
\pgfsetlinewidth{2.007500pt}%
\definecolor{currentstroke}{rgb}{0.960784,0.521569,0.094118}%
\pgfsetstrokecolor{currentstroke}%
\pgfsetdash{}{0pt}%
\pgfpathmoveto{\pgfqpoint{3.204959in}{1.718348in}}%
\pgfpathlineto{\pgfqpoint{2.637146in}{1.919015in}}%
\pgfusepath{stroke}%
\end{pgfscope}%
\begin{pgfscope}%
\pgfpathrectangle{\pgfqpoint{0.100000in}{0.100000in}}{\pgfqpoint{3.400000in}{3.250472in}}%
\pgfusepath{clip}%
\pgfsetbuttcap%
\pgfsetroundjoin%
\pgfsetlinewidth{2.007500pt}%
\definecolor{currentstroke}{rgb}{0.960784,0.521569,0.094118}%
\pgfsetstrokecolor{currentstroke}%
\pgfsetdash{}{0pt}%
\pgfpathmoveto{\pgfqpoint{2.637146in}{1.919015in}}%
\pgfpathlineto{\pgfqpoint{2.720431in}{2.609016in}}%
\pgfusepath{stroke}%
\end{pgfscope}%
\begin{pgfscope}%
\pgfpathrectangle{\pgfqpoint{0.100000in}{0.100000in}}{\pgfqpoint{3.400000in}{3.250472in}}%
\pgfusepath{clip}%
\pgfsetbuttcap%
\pgfsetroundjoin%
\definecolor{currentfill}{rgb}{0.960784,0.521569,0.094118}%
\pgfsetfillcolor{currentfill}%
\pgfsetlinewidth{1.003750pt}%
\definecolor{currentstroke}{rgb}{0.960784,0.521569,0.094118}%
\pgfsetstrokecolor{currentstroke}%
\pgfsetdash{}{0pt}%
\pgfpathmoveto{\pgfqpoint{3.204959in}{1.534616in}}%
\pgfpathcurveto{\pgfqpoint{3.253685in}{1.534616in}}{\pgfqpoint{3.300423in}{1.553975in}}{\pgfqpoint{3.334877in}{1.588430in}}%
\pgfpathcurveto{\pgfqpoint{3.369332in}{1.622884in}}{\pgfqpoint{3.388691in}{1.669622in}}{\pgfqpoint{3.388691in}{1.718348in}}%
\pgfpathcurveto{\pgfqpoint{3.388691in}{1.767075in}}{\pgfqpoint{3.369332in}{1.813812in}}{\pgfqpoint{3.334877in}{1.848267in}}%
\pgfpathcurveto{\pgfqpoint{3.300423in}{1.882722in}}{\pgfqpoint{3.253685in}{1.902081in}}{\pgfqpoint{3.204959in}{1.902081in}}%
\pgfpathcurveto{\pgfqpoint{3.156232in}{1.902081in}}{\pgfqpoint{3.109495in}{1.882722in}}{\pgfqpoint{3.075040in}{1.848267in}}%
\pgfpathcurveto{\pgfqpoint{3.040585in}{1.813812in}}{\pgfqpoint{3.021226in}{1.767075in}}{\pgfqpoint{3.021226in}{1.718348in}}%
\pgfpathcurveto{\pgfqpoint{3.021226in}{1.669622in}}{\pgfqpoint{3.040585in}{1.622884in}}{\pgfqpoint{3.075040in}{1.588430in}}%
\pgfpathcurveto{\pgfqpoint{3.109495in}{1.553975in}}{\pgfqpoint{3.156232in}{1.534616in}}{\pgfqpoint{3.204959in}{1.534616in}}%
\pgfpathlineto{\pgfqpoint{3.204959in}{1.534616in}}%
\pgfpathclose%
\pgfusepath{stroke,fill}%
\end{pgfscope}%
\begin{pgfscope}%
\pgfpathrectangle{\pgfqpoint{0.100000in}{0.100000in}}{\pgfqpoint{3.400000in}{3.250472in}}%
\pgfusepath{clip}%
\pgfsetbuttcap%
\pgfsetroundjoin%
\definecolor{currentfill}{rgb}{0.619608,0.619608,0.619608}%
\pgfsetfillcolor{currentfill}%
\pgfsetlinewidth{1.003750pt}%
\definecolor{currentstroke}{rgb}{0.619608,0.619608,0.619608}%
\pgfsetstrokecolor{currentstroke}%
\pgfsetdash{}{0pt}%
\pgfpathmoveto{\pgfqpoint{1.846458in}{0.576652in}}%
\pgfpathcurveto{\pgfqpoint{1.895185in}{0.576652in}}{\pgfqpoint{1.941922in}{0.596011in}}{\pgfqpoint{1.976377in}{0.630466in}}%
\pgfpathcurveto{\pgfqpoint{2.010832in}{0.664921in}}{\pgfqpoint{2.030191in}{0.711658in}}{\pgfqpoint{2.030191in}{0.760385in}}%
\pgfpathcurveto{\pgfqpoint{2.030191in}{0.809111in}}{\pgfqpoint{2.010832in}{0.855848in}}{\pgfqpoint{1.976377in}{0.890303in}}%
\pgfpathcurveto{\pgfqpoint{1.941922in}{0.924758in}}{\pgfqpoint{1.895185in}{0.944117in}}{\pgfqpoint{1.846458in}{0.944117in}}%
\pgfpathcurveto{\pgfqpoint{1.797732in}{0.944117in}}{\pgfqpoint{1.750994in}{0.924758in}}{\pgfqpoint{1.716539in}{0.890303in}}%
\pgfpathcurveto{\pgfqpoint{1.682085in}{0.855848in}}{\pgfqpoint{1.662725in}{0.809111in}}{\pgfqpoint{1.662725in}{0.760385in}}%
\pgfpathcurveto{\pgfqpoint{1.662725in}{0.711658in}}{\pgfqpoint{1.682085in}{0.664921in}}{\pgfqpoint{1.716539in}{0.630466in}}%
\pgfpathcurveto{\pgfqpoint{1.750994in}{0.596011in}}{\pgfqpoint{1.797732in}{0.576652in}}{\pgfqpoint{1.846458in}{0.576652in}}%
\pgfpathlineto{\pgfqpoint{1.846458in}{0.576652in}}%
\pgfpathclose%
\pgfusepath{stroke,fill}%
\end{pgfscope}%
\begin{pgfscope}%
\pgfpathrectangle{\pgfqpoint{0.100000in}{0.100000in}}{\pgfqpoint{3.400000in}{3.250472in}}%
\pgfusepath{clip}%
\pgfsetbuttcap%
\pgfsetroundjoin%
\definecolor{currentfill}{rgb}{0.960784,0.521569,0.094118}%
\pgfsetfillcolor{currentfill}%
\pgfsetlinewidth{1.003750pt}%
\definecolor{currentstroke}{rgb}{0.960784,0.521569,0.094118}%
\pgfsetstrokecolor{currentstroke}%
\pgfsetdash{}{0pt}%
\pgfpathmoveto{\pgfqpoint{1.895146in}{2.894352in}}%
\pgfpathcurveto{\pgfqpoint{1.943872in}{2.894352in}}{\pgfqpoint{1.990609in}{2.913711in}}{\pgfqpoint{2.025064in}{2.948166in}}%
\pgfpathcurveto{\pgfqpoint{2.059519in}{2.982621in}}{\pgfqpoint{2.078878in}{3.029358in}}{\pgfqpoint{2.078878in}{3.078085in}}%
\pgfpathcurveto{\pgfqpoint{2.078878in}{3.126811in}}{\pgfqpoint{2.059519in}{3.173548in}}{\pgfqpoint{2.025064in}{3.208003in}}%
\pgfpathcurveto{\pgfqpoint{1.990609in}{3.242458in}}{\pgfqpoint{1.943872in}{3.261817in}}{\pgfqpoint{1.895146in}{3.261817in}}%
\pgfpathcurveto{\pgfqpoint{1.846419in}{3.261817in}}{\pgfqpoint{1.799682in}{3.242458in}}{\pgfqpoint{1.765227in}{3.208003in}}%
\pgfpathcurveto{\pgfqpoint{1.730772in}{3.173548in}}{\pgfqpoint{1.711413in}{3.126811in}}{\pgfqpoint{1.711413in}{3.078085in}}%
\pgfpathcurveto{\pgfqpoint{1.711413in}{3.029358in}}{\pgfqpoint{1.730772in}{2.982621in}}{\pgfqpoint{1.765227in}{2.948166in}}%
\pgfpathcurveto{\pgfqpoint{1.799682in}{2.913711in}}{\pgfqpoint{1.846419in}{2.894352in}}{\pgfqpoint{1.895146in}{2.894352in}}%
\pgfpathlineto{\pgfqpoint{1.895146in}{2.894352in}}%
\pgfpathclose%
\pgfusepath{stroke,fill}%
\end{pgfscope}%
\begin{pgfscope}%
\pgfpathrectangle{\pgfqpoint{0.100000in}{0.100000in}}{\pgfqpoint{3.400000in}{3.250472in}}%
\pgfusepath{clip}%
\pgfsetbuttcap%
\pgfsetroundjoin%
\definecolor{currentfill}{rgb}{0.960784,0.521569,0.094118}%
\pgfsetfillcolor{currentfill}%
\pgfsetlinewidth{1.003750pt}%
\definecolor{currentstroke}{rgb}{0.960784,0.521569,0.094118}%
\pgfsetstrokecolor{currentstroke}%
\pgfsetdash{}{0pt}%
\pgfpathmoveto{\pgfqpoint{2.637146in}{1.735282in}}%
\pgfpathcurveto{\pgfqpoint{2.685873in}{1.735282in}}{\pgfqpoint{2.732610in}{1.754642in}}{\pgfqpoint{2.767065in}{1.789096in}}%
\pgfpathcurveto{\pgfqpoint{2.801520in}{1.823551in}}{\pgfqpoint{2.820879in}{1.870289in}}{\pgfqpoint{2.820879in}{1.919015in}}%
\pgfpathcurveto{\pgfqpoint{2.820879in}{1.967742in}}{\pgfqpoint{2.801520in}{2.014479in}}{\pgfqpoint{2.767065in}{2.048934in}}%
\pgfpathcurveto{\pgfqpoint{2.732610in}{2.083389in}}{\pgfqpoint{2.685873in}{2.102748in}}{\pgfqpoint{2.637146in}{2.102748in}}%
\pgfpathcurveto{\pgfqpoint{2.588420in}{2.102748in}}{\pgfqpoint{2.541683in}{2.083389in}}{\pgfqpoint{2.507228in}{2.048934in}}%
\pgfpathcurveto{\pgfqpoint{2.472773in}{2.014479in}}{\pgfqpoint{2.453414in}{1.967742in}}{\pgfqpoint{2.453414in}{1.919015in}}%
\pgfpathcurveto{\pgfqpoint{2.453414in}{1.870289in}}{\pgfqpoint{2.472773in}{1.823551in}}{\pgfqpoint{2.507228in}{1.789096in}}%
\pgfpathcurveto{\pgfqpoint{2.541683in}{1.754642in}}{\pgfqpoint{2.588420in}{1.735282in}}{\pgfqpoint{2.637146in}{1.735282in}}%
\pgfpathlineto{\pgfqpoint{2.637146in}{1.735282in}}%
\pgfpathclose%
\pgfusepath{stroke,fill}%
\end{pgfscope}%
\begin{pgfscope}%
\pgfpathrectangle{\pgfqpoint{0.100000in}{0.100000in}}{\pgfqpoint{3.400000in}{3.250472in}}%
\pgfusepath{clip}%
\pgfsetbuttcap%
\pgfsetroundjoin%
\definecolor{currentfill}{rgb}{0.619608,0.619608,0.619608}%
\pgfsetfillcolor{currentfill}%
\pgfsetlinewidth{1.003750pt}%
\definecolor{currentstroke}{rgb}{0.619608,0.619608,0.619608}%
\pgfsetstrokecolor{currentstroke}%
\pgfsetdash{}{0pt}%
\pgfpathmoveto{\pgfqpoint{2.069616in}{0.134045in}}%
\pgfpathcurveto{\pgfqpoint{2.118342in}{0.134045in}}{\pgfqpoint{2.165080in}{0.153404in}}{\pgfqpoint{2.199535in}{0.187859in}}%
\pgfpathcurveto{\pgfqpoint{2.233989in}{0.222313in}}{\pgfqpoint{2.253349in}{0.269051in}}{\pgfqpoint{2.253349in}{0.317777in}}%
\pgfpathcurveto{\pgfqpoint{2.253349in}{0.366504in}}{\pgfqpoint{2.233989in}{0.413241in}}{\pgfqpoint{2.199535in}{0.447696in}}%
\pgfpathcurveto{\pgfqpoint{2.165080in}{0.482151in}}{\pgfqpoint{2.118342in}{0.501510in}}{\pgfqpoint{2.069616in}{0.501510in}}%
\pgfpathcurveto{\pgfqpoint{2.020889in}{0.501510in}}{\pgfqpoint{1.974152in}{0.482151in}}{\pgfqpoint{1.939697in}{0.447696in}}%
\pgfpathcurveto{\pgfqpoint{1.905242in}{0.413241in}}{\pgfqpoint{1.885883in}{0.366504in}}{\pgfqpoint{1.885883in}{0.317777in}}%
\pgfpathcurveto{\pgfqpoint{1.885883in}{0.269051in}}{\pgfqpoint{1.905242in}{0.222313in}}{\pgfqpoint{1.939697in}{0.187859in}}%
\pgfpathcurveto{\pgfqpoint{1.974152in}{0.153404in}}{\pgfqpoint{2.020889in}{0.134045in}}{\pgfqpoint{2.069616in}{0.134045in}}%
\pgfpathlineto{\pgfqpoint{2.069616in}{0.134045in}}%
\pgfpathclose%
\pgfusepath{stroke,fill}%
\end{pgfscope}%
\begin{pgfscope}%
\pgfpathrectangle{\pgfqpoint{0.100000in}{0.100000in}}{\pgfqpoint{3.400000in}{3.250472in}}%
\pgfusepath{clip}%
\pgfsetbuttcap%
\pgfsetroundjoin%
\definecolor{currentfill}{rgb}{0.960784,0.521569,0.094118}%
\pgfsetfillcolor{currentfill}%
\pgfsetlinewidth{1.003750pt}%
\definecolor{currentstroke}{rgb}{0.960784,0.521569,0.094118}%
\pgfsetstrokecolor{currentstroke}%
\pgfsetdash{}{0pt}%
\pgfpathmoveto{\pgfqpoint{0.459600in}{1.093672in}}%
\pgfpathcurveto{\pgfqpoint{0.508327in}{1.093672in}}{\pgfqpoint{0.555064in}{1.113031in}}{\pgfqpoint{0.589519in}{1.147486in}}%
\pgfpathcurveto{\pgfqpoint{0.623974in}{1.181941in}}{\pgfqpoint{0.643333in}{1.228678in}}{\pgfqpoint{0.643333in}{1.277404in}}%
\pgfpathcurveto{\pgfqpoint{0.643333in}{1.326131in}}{\pgfqpoint{0.623974in}{1.372868in}}{\pgfqpoint{0.589519in}{1.407323in}}%
\pgfpathcurveto{\pgfqpoint{0.555064in}{1.441778in}}{\pgfqpoint{0.508327in}{1.461137in}}{\pgfqpoint{0.459600in}{1.461137in}}%
\pgfpathcurveto{\pgfqpoint{0.410874in}{1.461137in}}{\pgfqpoint{0.364136in}{1.441778in}}{\pgfqpoint{0.329682in}{1.407323in}}%
\pgfpathcurveto{\pgfqpoint{0.295227in}{1.372868in}}{\pgfqpoint{0.275868in}{1.326131in}}{\pgfqpoint{0.275868in}{1.277404in}}%
\pgfpathcurveto{\pgfqpoint{0.275868in}{1.228678in}}{\pgfqpoint{0.295227in}{1.181941in}}{\pgfqpoint{0.329682in}{1.147486in}}%
\pgfpathcurveto{\pgfqpoint{0.364136in}{1.113031in}}{\pgfqpoint{0.410874in}{1.093672in}}{\pgfqpoint{0.459600in}{1.093672in}}%
\pgfpathlineto{\pgfqpoint{0.459600in}{1.093672in}}%
\pgfpathclose%
\pgfusepath{stroke,fill}%
\end{pgfscope}%
\begin{pgfscope}%
\pgfpathrectangle{\pgfqpoint{0.100000in}{0.100000in}}{\pgfqpoint{3.400000in}{3.250472in}}%
\pgfusepath{clip}%
\pgfsetbuttcap%
\pgfsetroundjoin%
\definecolor{currentfill}{rgb}{0.960784,0.521569,0.094118}%
\pgfsetfillcolor{currentfill}%
\pgfsetlinewidth{1.003750pt}%
\definecolor{currentstroke}{rgb}{0.960784,0.521569,0.094118}%
\pgfsetstrokecolor{currentstroke}%
\pgfsetdash{}{0pt}%
\pgfpathmoveto{\pgfqpoint{0.973595in}{2.751246in}}%
\pgfpathcurveto{\pgfqpoint{1.022321in}{2.751246in}}{\pgfqpoint{1.069058in}{2.770606in}}{\pgfqpoint{1.103513in}{2.805060in}}%
\pgfpathcurveto{\pgfqpoint{1.137968in}{2.839515in}}{\pgfqpoint{1.157327in}{2.886253in}}{\pgfqpoint{1.157327in}{2.934979in}}%
\pgfpathcurveto{\pgfqpoint{1.157327in}{2.983706in}}{\pgfqpoint{1.137968in}{3.030443in}}{\pgfqpoint{1.103513in}{3.064898in}}%
\pgfpathcurveto{\pgfqpoint{1.069058in}{3.099353in}}{\pgfqpoint{1.022321in}{3.118712in}}{\pgfqpoint{0.973595in}{3.118712in}}%
\pgfpathcurveto{\pgfqpoint{0.924868in}{3.118712in}}{\pgfqpoint{0.878131in}{3.099353in}}{\pgfqpoint{0.843676in}{3.064898in}}%
\pgfpathcurveto{\pgfqpoint{0.809221in}{3.030443in}}{\pgfqpoint{0.789862in}{2.983706in}}{\pgfqpoint{0.789862in}{2.934979in}}%
\pgfpathcurveto{\pgfqpoint{0.789862in}{2.886253in}}{\pgfqpoint{0.809221in}{2.839515in}}{\pgfqpoint{0.843676in}{2.805060in}}%
\pgfpathcurveto{\pgfqpoint{0.878131in}{2.770606in}}{\pgfqpoint{0.924868in}{2.751246in}}{\pgfqpoint{0.973595in}{2.751246in}}%
\pgfpathlineto{\pgfqpoint{0.973595in}{2.751246in}}%
\pgfpathclose%
\pgfusepath{stroke,fill}%
\end{pgfscope}%
\begin{pgfscope}%
\pgfpathrectangle{\pgfqpoint{0.100000in}{0.100000in}}{\pgfqpoint{3.400000in}{3.250472in}}%
\pgfusepath{clip}%
\pgfsetbuttcap%
\pgfsetroundjoin%
\definecolor{currentfill}{rgb}{0.960784,0.521569,0.094118}%
\pgfsetfillcolor{currentfill}%
\pgfsetlinewidth{1.003750pt}%
\definecolor{currentstroke}{rgb}{0.960784,0.521569,0.094118}%
\pgfsetstrokecolor{currentstroke}%
\pgfsetdash{}{0pt}%
\pgfpathmoveto{\pgfqpoint{0.395041in}{2.023908in}}%
\pgfpathcurveto{\pgfqpoint{0.443768in}{2.023908in}}{\pgfqpoint{0.490505in}{2.043267in}}{\pgfqpoint{0.524960in}{2.077722in}}%
\pgfpathcurveto{\pgfqpoint{0.559415in}{2.112176in}}{\pgfqpoint{0.578774in}{2.158914in}}{\pgfqpoint{0.578774in}{2.207640in}}%
\pgfpathcurveto{\pgfqpoint{0.578774in}{2.256367in}}{\pgfqpoint{0.559415in}{2.303104in}}{\pgfqpoint{0.524960in}{2.337559in}}%
\pgfpathcurveto{\pgfqpoint{0.490505in}{2.372014in}}{\pgfqpoint{0.443768in}{2.391373in}}{\pgfqpoint{0.395041in}{2.391373in}}%
\pgfpathcurveto{\pgfqpoint{0.346315in}{2.391373in}}{\pgfqpoint{0.299577in}{2.372014in}}{\pgfqpoint{0.265123in}{2.337559in}}%
\pgfpathcurveto{\pgfqpoint{0.230668in}{2.303104in}}{\pgfqpoint{0.211309in}{2.256367in}}{\pgfqpoint{0.211309in}{2.207640in}}%
\pgfpathcurveto{\pgfqpoint{0.211309in}{2.158914in}}{\pgfqpoint{0.230668in}{2.112176in}}{\pgfqpoint{0.265123in}{2.077722in}}%
\pgfpathcurveto{\pgfqpoint{0.299577in}{2.043267in}}{\pgfqpoint{0.346315in}{2.023908in}}{\pgfqpoint{0.395041in}{2.023908in}}%
\pgfpathlineto{\pgfqpoint{0.395041in}{2.023908in}}%
\pgfpathclose%
\pgfusepath{stroke,fill}%
\end{pgfscope}%
\begin{pgfscope}%
\pgfpathrectangle{\pgfqpoint{0.100000in}{0.100000in}}{\pgfqpoint{3.400000in}{3.250472in}}%
\pgfusepath{clip}%
\pgfsetbuttcap%
\pgfsetroundjoin%
\definecolor{currentfill}{rgb}{0.960784,0.521569,0.094118}%
\pgfsetfillcolor{currentfill}%
\pgfsetlinewidth{1.003750pt}%
\definecolor{currentstroke}{rgb}{0.960784,0.521569,0.094118}%
\pgfsetstrokecolor{currentstroke}%
\pgfsetdash{}{0pt}%
\pgfpathmoveto{\pgfqpoint{1.101336in}{0.401577in}}%
\pgfpathcurveto{\pgfqpoint{1.150062in}{0.401577in}}{\pgfqpoint{1.196799in}{0.420936in}}{\pgfqpoint{1.231254in}{0.455391in}}%
\pgfpathcurveto{\pgfqpoint{1.265709in}{0.489846in}}{\pgfqpoint{1.285068in}{0.536583in}}{\pgfqpoint{1.285068in}{0.585310in}}%
\pgfpathcurveto{\pgfqpoint{1.285068in}{0.634036in}}{\pgfqpoint{1.265709in}{0.680773in}}{\pgfqpoint{1.231254in}{0.715228in}}%
\pgfpathcurveto{\pgfqpoint{1.196799in}{0.749683in}}{\pgfqpoint{1.150062in}{0.769042in}}{\pgfqpoint{1.101336in}{0.769042in}}%
\pgfpathcurveto{\pgfqpoint{1.052609in}{0.769042in}}{\pgfqpoint{1.005872in}{0.749683in}}{\pgfqpoint{0.971417in}{0.715228in}}%
\pgfpathcurveto{\pgfqpoint{0.936962in}{0.680773in}}{\pgfqpoint{0.917603in}{0.634036in}}{\pgfqpoint{0.917603in}{0.585310in}}%
\pgfpathcurveto{\pgfqpoint{0.917603in}{0.536583in}}{\pgfqpoint{0.936962in}{0.489846in}}{\pgfqpoint{0.971417in}{0.455391in}}%
\pgfpathcurveto{\pgfqpoint{1.005872in}{0.420936in}}{\pgfqpoint{1.052609in}{0.401577in}}{\pgfqpoint{1.101336in}{0.401577in}}%
\pgfpathlineto{\pgfqpoint{1.101336in}{0.401577in}}%
\pgfpathclose%
\pgfusepath{stroke,fill}%
\end{pgfscope}%
\begin{pgfscope}%
\pgfpathrectangle{\pgfqpoint{0.100000in}{0.100000in}}{\pgfqpoint{3.400000in}{3.250472in}}%
\pgfusepath{clip}%
\pgfsetbuttcap%
\pgfsetroundjoin%
\definecolor{currentfill}{rgb}{0.619608,0.619608,0.619608}%
\pgfsetfillcolor{currentfill}%
\pgfsetlinewidth{1.003750pt}%
\definecolor{currentstroke}{rgb}{0.619608,0.619608,0.619608}%
\pgfsetstrokecolor{currentstroke}%
\pgfsetdash{}{0pt}%
\pgfpathmoveto{\pgfqpoint{2.378843in}{0.944202in}}%
\pgfpathcurveto{\pgfqpoint{2.427569in}{0.944202in}}{\pgfqpoint{2.474307in}{0.963561in}}{\pgfqpoint{2.508762in}{0.998016in}}%
\pgfpathcurveto{\pgfqpoint{2.543216in}{1.032471in}}{\pgfqpoint{2.562576in}{1.079208in}}{\pgfqpoint{2.562576in}{1.127935in}}%
\pgfpathcurveto{\pgfqpoint{2.562576in}{1.176661in}}{\pgfqpoint{2.543216in}{1.223399in}}{\pgfqpoint{2.508762in}{1.257854in}}%
\pgfpathcurveto{\pgfqpoint{2.474307in}{1.292308in}}{\pgfqpoint{2.427569in}{1.311668in}}{\pgfqpoint{2.378843in}{1.311668in}}%
\pgfpathcurveto{\pgfqpoint{2.330116in}{1.311668in}}{\pgfqpoint{2.283379in}{1.292308in}}{\pgfqpoint{2.248924in}{1.257854in}}%
\pgfpathcurveto{\pgfqpoint{2.214469in}{1.223399in}}{\pgfqpoint{2.195110in}{1.176661in}}{\pgfqpoint{2.195110in}{1.127935in}}%
\pgfpathcurveto{\pgfqpoint{2.195110in}{1.079208in}}{\pgfqpoint{2.214469in}{1.032471in}}{\pgfqpoint{2.248924in}{0.998016in}}%
\pgfpathcurveto{\pgfqpoint{2.283379in}{0.963561in}}{\pgfqpoint{2.330116in}{0.944202in}}{\pgfqpoint{2.378843in}{0.944202in}}%
\pgfpathlineto{\pgfqpoint{2.378843in}{0.944202in}}%
\pgfpathclose%
\pgfusepath{stroke,fill}%
\end{pgfscope}%
\begin{pgfscope}%
\pgfpathrectangle{\pgfqpoint{0.100000in}{0.100000in}}{\pgfqpoint{3.400000in}{3.250472in}}%
\pgfusepath{clip}%
\pgfsetbuttcap%
\pgfsetroundjoin%
\definecolor{currentfill}{rgb}{0.619608,0.619608,0.619608}%
\pgfsetfillcolor{currentfill}%
\pgfsetlinewidth{1.003750pt}%
\definecolor{currentstroke}{rgb}{0.619608,0.619608,0.619608}%
\pgfsetstrokecolor{currentstroke}%
\pgfsetdash{}{0pt}%
\pgfpathmoveto{\pgfqpoint{2.632396in}{0.578638in}}%
\pgfpathlineto{\pgfqpoint{2.999862in}{0.578638in}}%
\pgfpathlineto{\pgfqpoint{2.999862in}{0.946103in}}%
\pgfpathlineto{\pgfqpoint{2.632396in}{0.946103in}}%
\pgfpathlineto{\pgfqpoint{2.632396in}{0.578638in}}%
\pgfpathclose%
\pgfusepath{stroke,fill}%
\end{pgfscope}%
\begin{pgfscope}%
\pgfpathrectangle{\pgfqpoint{0.100000in}{0.100000in}}{\pgfqpoint{3.400000in}{3.250472in}}%
\pgfusepath{clip}%
\pgfsetbuttcap%
\pgfsetroundjoin%
\definecolor{currentfill}{rgb}{0.960784,0.521569,0.094118}%
\pgfsetfillcolor{currentfill}%
\pgfsetlinewidth{1.003750pt}%
\definecolor{currentstroke}{rgb}{0.960784,0.521569,0.094118}%
\pgfsetstrokecolor{currentstroke}%
\pgfsetdash{}{0pt}%
\pgfpathmoveto{\pgfqpoint{2.536698in}{2.425283in}}%
\pgfpathlineto{\pgfqpoint{2.904163in}{2.425283in}}%
\pgfpathlineto{\pgfqpoint{2.904163in}{2.792749in}}%
\pgfpathlineto{\pgfqpoint{2.536698in}{2.792749in}}%
\pgfpathlineto{\pgfqpoint{2.536698in}{2.425283in}}%
\pgfpathclose%
\pgfusepath{stroke,fill}%
\end{pgfscope}%
\begin{pgfscope}%
\definecolor{textcolor}{rgb}{0.000000,0.000000,0.000000}%
\pgfsetstrokecolor{textcolor}%
\pgfsetfillcolor{textcolor}%
\pgftext[x=3.204959in,y=1.718348in,,]{\color{textcolor}\sffamily\fontsize{19.000000}{22.800000}\selectfont \(\displaystyle S_{1}\)}%
\end{pgfscope}%
\begin{pgfscope}%
\definecolor{textcolor}{rgb}{0.000000,0.000000,0.000000}%
\pgfsetstrokecolor{textcolor}%
\pgfsetfillcolor{textcolor}%
\pgftext[x=1.846458in,y=0.760385in,,]{\color{textcolor}\sffamily\fontsize{19.000000}{22.800000}\selectfont \(\displaystyle S_{2}\)}%
\end{pgfscope}%
\begin{pgfscope}%
\definecolor{textcolor}{rgb}{0.000000,0.000000,0.000000}%
\pgfsetstrokecolor{textcolor}%
\pgfsetfillcolor{textcolor}%
\pgftext[x=1.895146in,y=3.078085in,,]{\color{textcolor}\sffamily\fontsize{19.000000}{22.800000}\selectfont \(\displaystyle S_{3}\)}%
\end{pgfscope}%
\begin{pgfscope}%
\definecolor{textcolor}{rgb}{0.000000,0.000000,0.000000}%
\pgfsetstrokecolor{textcolor}%
\pgfsetfillcolor{textcolor}%
\pgftext[x=2.637146in,y=1.919015in,,]{\color{textcolor}\sffamily\fontsize{19.000000}{22.800000}\selectfont \(\displaystyle S_{4}\)}%
\end{pgfscope}%
\begin{pgfscope}%
\definecolor{textcolor}{rgb}{0.000000,0.000000,0.000000}%
\pgfsetstrokecolor{textcolor}%
\pgfsetfillcolor{textcolor}%
\pgftext[x=2.069616in,y=0.317777in,,]{\color{textcolor}\sffamily\fontsize{19.000000}{22.800000}\selectfont \(\displaystyle S_{5}\)}%
\end{pgfscope}%
\begin{pgfscope}%
\definecolor{textcolor}{rgb}{0.000000,0.000000,0.000000}%
\pgfsetstrokecolor{textcolor}%
\pgfsetfillcolor{textcolor}%
\pgftext[x=0.459600in,y=1.277404in,,]{\color{textcolor}\sffamily\fontsize{19.000000}{22.800000}\selectfont \(\displaystyle S_{6}\)}%
\end{pgfscope}%
\begin{pgfscope}%
\definecolor{textcolor}{rgb}{0.000000,0.000000,0.000000}%
\pgfsetstrokecolor{textcolor}%
\pgfsetfillcolor{textcolor}%
\pgftext[x=0.973595in,y=2.934979in,,]{\color{textcolor}\sffamily\fontsize{19.000000}{22.800000}\selectfont \(\displaystyle S_{7}\)}%
\end{pgfscope}%
\begin{pgfscope}%
\definecolor{textcolor}{rgb}{0.000000,0.000000,0.000000}%
\pgfsetstrokecolor{textcolor}%
\pgfsetfillcolor{textcolor}%
\pgftext[x=0.395041in,y=2.207640in,,]{\color{textcolor}\sffamily\fontsize{19.000000}{22.800000}\selectfont \(\displaystyle S_{8}\)}%
\end{pgfscope}%
\begin{pgfscope}%
\definecolor{textcolor}{rgb}{0.000000,0.000000,0.000000}%
\pgfsetstrokecolor{textcolor}%
\pgfsetfillcolor{textcolor}%
\pgftext[x=1.101336in,y=0.585310in,,]{\color{textcolor}\sffamily\fontsize{19.000000}{22.800000}\selectfont \(\displaystyle S_{9}\)}%
\end{pgfscope}%
\begin{pgfscope}%
\definecolor{textcolor}{rgb}{0.000000,0.000000,0.000000}%
\pgfsetstrokecolor{textcolor}%
\pgfsetfillcolor{textcolor}%
\pgftext[x=2.378843in,y=1.127935in,,]{\color{textcolor}\sffamily\fontsize{19.000000}{22.800000}\selectfont \(\displaystyle S_{10}\)}%
\end{pgfscope}%
\begin{pgfscope}%
\definecolor{textcolor}{rgb}{0.000000,0.000000,0.000000}%
\pgfsetstrokecolor{textcolor}%
\pgfsetfillcolor{textcolor}%
\pgftext[x=2.816129in,y=0.762370in,,]{\color{textcolor}\sffamily\fontsize{19.000000}{22.800000}\selectfont \(\displaystyle P_{11}\)}%
\end{pgfscope}%
\begin{pgfscope}%
\definecolor{textcolor}{rgb}{0.000000,0.000000,0.000000}%
\pgfsetstrokecolor{textcolor}%
\pgfsetfillcolor{textcolor}%
\pgftext[x=2.720431in,y=2.609016in,,]{\color{textcolor}\sffamily\fontsize{19.000000}{22.800000}\selectfont \(\displaystyle P_{12}\)}%
\end{pgfscope}%
\end{pgfpicture}%
\makeatother%
\endgroup%